\documentclass[12pt,fleqn,a4paper]{article} 
\usepackage{epsfig,graphics,graphicx}
\usepackage{clshan-math,clshan-dm-dd}
\usepackage{multirow}
\def \lsim {\:\raisebox{-0.7ex}{$\stackrel{\textstyle<}{\sim}$}\:}
\def \gsim {\:\raisebox{-0.7ex}{$\stackrel{\textstyle>}{\sim}$}\:}
\begin{document}
\thispagestyle{empty}
\begin{flushright}
 March 2018
\end{flushright}
\begin{center}
{\Large\bf
 Effects of Threshold Energy on                   \\
 Reconstructions of Properties of Low--Mass WIMPs \\ \vspace{0.2 cm}
 in Direct Dark Matter Detection Experiments}     \\
\vspace*{0.7cm}
 {\sc Yu Bai}$^{1, \ddagger}$,
 {\sc Weichao Sun}$^{2, \S}$,
 and {\sc Chung-Lin Shan}$^{2, \P}$       \\~\\
\vspace{0.5cm}
 ${}^1$%
 {\it School of Physics and Technology,
      Xinjiang University                 \\
      No.~666, Shengli Road,
      Urumqi, Xinjiang 830046, China}     \\~\\
 ${}^2$%
 {\it Xinjiang Astronomical Observatory,
      Chinese Academy of Sciences         \\
      No.~150, Science 1-Street,
      Urumqi, Xinjiang 830011, China}     \\~\\~\\
 ${}^{\ddagger}$%
 {\it E-mail:} {\tt baiyu@xao.ac.cn}      \\
\vspace{0.1cm}
 ${}^{\S}$%
 {\it E-mail:} {\tt sunweichao@xao.ac.cn} \\
\vspace{0.1cm}
 ${}^{\P}$%
 {\it E-mail:} {\tt clshan@xao.ac.cn}     \\~\\
\vspace{0.1cm}
\end{center}
\vspace{1cm}
\begin{abstract}
 In this paper,
 we revisit our model--independent methods
 developed for reconstructing properties of
 Weakly Interacting Massive Particles (WIMPs)
 by using measured recoil energies
 from direct Dark Matter detection experiments directly
 and take into account more realistically
 non--negligible threshold energy.
 All expressions for reconstructing the mass
 and the (ratios between the) spin--independent
 and the spin--dependent WIMP--nucleon couplings
 have been modified.
 We focus on low--mass ($\mchi \lsim\~15$ GeV) WIMPs
 and present the numerical results
 obtained by Monte Carlo simulations.
 Constraints caused by non--negligible threshold energy and
 technical treatments for improving reconstruction results
 will also be discussed.
\end{abstract}
\clearpage
\section{Introduction}

 So far
 Weakly Interacting Massive Particles (WIMPs)
 arising in several extensions of the Standard Model
 of particle physics
 are still one of the leading
 candidates for cosmological Dark Matter.
 In the last three decades,
 a large number of experiments have been built
 and are being planned
 to search for different WIMP candidates
 by direct detection of scattering recoil energies
 of ambient WIMPs off target nuclei
 in low--background underground laboratory detectors
 (see Refs.~\cite{%
  Goodman85, Wasserman86, Drukier86, Griest88,
  Smith90, SUSYDM96, Lewin96,
  Ramachers02, Jesus04, Gaitskell04,
  Cerdeno10, Saab12,
  Baudis12c, Baudis15, Drees16, JLiu17}).

 Using data
 from these direct Dark Matter detection experiments
 to reconstruct,
 e.g.,
 the mass and different couplings on nucleons
 is essential
 for understanding the nature of WIMPs
 and
 identifying them among new particles produced at colliders.
 Different methods have been purposed
 for reconstructing
 the WIMP mass $\mchi$
 \cite{Green-mchi07, Green-mchi08, Green-mchi12, Green-mchi13}
 and
 spin--independent (SI)/spin--dependent (SD) WIMP--nucleon cross sections
 $\sigma_{\chi {\rm (p, n)}}^{\rm SI, SD}$
 \cite{Cannoni10, Hoferichter16}.
 Recently,
 several applications of
 the maximum likelihood and Bayesian analyses
 have been developed,
 which treat
 the WIMP mass
 and different WIMP--nucleon couplings
 as well as
 the Solar and Earth's orbital velocities
 in the Galactic reference frame
 as fitting parameters simultaneously
 (see works by,
  e.g.,
  Y.~Akrami {\it et al.}
  \cite{Akrami10a, Akrami10b},
  M.~Pato {\it et al.}
  \cite{Pato10b, Pato11},
  C.~Arina {\it et al.}
  \cite{Arina11a, Arina12, Arina13a, Arina13b},
  D.~G.~Cerde$\rm \tilde{n}$o {\it et al.}
  \cite{Cerdeno13, Cerdeno14, Cerdeno18}
  and e.g.~\cite{McDermott11, Strege12, Newstead13, Savage15}).
 Furthermore,
 while
 some authors focus on studying
 effects of and/or constraints caused by uncertainties
 on the velocity and density distributions of Galactic Dark Matter
 \cite{Strigari09, APeter09, APeter11, Pato12},
 some model--independent methods
 have also been developed
 by P.~J.~Fox {\it et al.}
 \cite{Fox10a, Fox10b, Fox14},
 E.~Del Nobile {\it et al.}
 \cite{DelNobile13a, DelNobile13b, Cirelli13, NRopsDD, DelNobile14a, DelNobile14b},
 B.~Feldstein and F.~Kahlhoefer
 \cite{Feldstein14a, Feldstein14b, Kahlhoefer16},
 G.~B.~Gelmini {\it et al.}
 \cite{Gelmini14, Gelmini15b, Gelmini16, Gelmini17}
 and e.g.~\cite{Cherry14, Kahn14}.

 Besides these works,
 we started also in 2003 to study methods
 for reconstructing properties of WIMP particles
 by using (not a fitted recoil spectrum but)
 the measured recoil energies directly
 as model--independently as possible.
 As the first step,
 in Ref.~\cite{DMDDf1v}
 we introduced an exponential ansatz
 for reconstructing the measured recoil spectrum and in turn
 for reconstructing the (moments of the)
 time--averaged one--dimensional velocity distribution of halo WIMPs.
 This analysis requires
 no prior knowledge about the local WIMP density
 nor a WIMP scattering cross section on nucleus,
 the only required information
 is the mass of incident WIMPs.
 However,
 with a few hundreds or even thousands recorded events,
 only a few ($<$ 10) reconstructed points of
 the WIMP velocity distribution
 with pretty large statistical uncertainties
 could be obtained.
 In order to provide more detailed information
 about the WIMP velocity distribution
 as well as
 the characteristic Solar and Earth's Galactic velocities,
 we introduced therefore later the Bayesian analysis
 into our reconstruction procedure
 \cite{DMDDf1v-Bayesian}
 for fitting a functional form of
 the one--dimensional velocity distribution
 as well as
 for determining concretely,
 e.g.,
 the position of the peak of
 the fitted velocity distribution function
 and the values of
 the characteristic Solar and Earth's Galactic velocities.
 Moreover,
 based on the reconstruction of the moments of
 the one--dimensional WIMP velocity distribution function
 and the combinations of two or more experimental data sets
 with different target nuclei,
 we developed further the methods for model--independently
 determining the WIMP mass $\mchi$
 \cite{DMDDmchi},
 the (squared) SI scalar WIMP--proton coupling $|\frmp|^2$
 \cite{DMDDfp2}
 as well as
 the ratios of the SD axial--vector
 (to the SI scalar)
 WIMP--nucleon couplings/cross sections
 $\armn / \armp$ and $\sigma_{\chi ({\rm p, n})}^{\rm SD} / \sigmapSI$
 \cite{DMDDranap}.

 In these earlier works
 -- both of the theoretical derivations
 and numerical simulations --
 the minimal experimental cut--off energies
 of data sets to be analyzed
 are often assumed to be negligible.
 For experiments with heavy target nuclei,
 e.g.~Ge or Xe,
 and once WIMPs are heavy
 ($\gsim~100$ GeV),
 the systematic bias caused by this assumption
 could be neglected,
 compared with the pretty large statistical uncertainties.
 However,
 once WIMPs are light
 ($\lsim~50$ GeV)
 and a light target nucleus,
 e.g.~Si or Ar,
 is used,
 effects of non--negligible threshold energy
 has to be considered seriously and
 the expressions
 for the reconstructions of different WIMP properties
 and estimates of the statistical uncertainties
 would need to be modified properly.
 Meanwhile,
 some experimental collaborations
 have developed several detector techniques
 with different materials
 for searching for low--mass WIMPs.
 For instance,
 the CRESST experiment
 with their $\rm Al_2 O_3$ and $\rm Ca W O_4$ detectors
 \cite{Altmann01, Angloher14, Angloher15b, Petricca17, Strauss18},
 the CoGeNT and CDEX experiments
 with p--type point--contact Ge detectors
 \cite{Aalseth10, Aalseth12,
       WZhao13, QYue14, WZhao16, HJiang18},
 and the newest generation of the CDMS,
 the SuperCDMS, experiment
 with also Ge detectors
 \cite{Agnese13d, Agnese14a, Agnese15b, Agnese17a}.
 Recently,
 the PICO Collaboration
 with their
 PICO-2L $\rm C_3 F_8$ bubble chamber
 \cite{Amole15a, Amole16}
 and
 the DarkSide Collaboration
 with their DarkSide-50 Ar detector
 \cite{Agnes18a}
 have also announced the sensitivity on detecting low--mass WIMPs.

 Therefore,
 as a supplement of our earlier works,
 in \cite{DMDDf1v-calN}
 we have considered the needed modification of
 the normalization constant of
 the reconstructed one--dimensional WIMP velocity distribution function
 caused by
 non--negligible experimental threshold energy.
 And,
 in this paper,
 we revisit further our methods
 for the reconstructions of
 the WIMP mass
 as well as
 the (ratios between the) SI (scalar) and the SD (axial--vector)
 WIMP--nucleon couplings/cross sections
 by taking into account
 non--negligible experimental threshold energies of the analyzed data sets.
 All expressions
 for the reconstructions of $\mchi$ and
 (the ratios of) $\sigma_{\chi {\rm (p, n)}}^{\rm SI, SD}$
 have been checked and modified properly.
 We focus on effects of
 non--negligible threshold energy
 on the reconstructed WIMP properties
 for WIMP masses of $\mchi \lsim$ 15 GeV.

 The remainder of this paper is organized as follows.
 In Sec.~2,
 we first review our model--independent procedures
 for reconstructing different WIMP properties
 and modify our expressions.
 Then,
 in Sec.~3,
 we present numerical results of the reconstructed WIMP properties
 by using the modified expressions
 and discuss effects of
 non--negligible threshold energy
 for light WIMPs.
 We conclude in Sec.~4
 and give
 some technical details for our analysis
 in Appendix.

\section{Formalism}

 In this section,
 we first review briefly
 the modification of the normalization constant
 of the reconstructed one--dimensional WIMP velocity distribution.
 Then
 we derive the corresponding modifications of the expressions
 for our model--independent reconstructions
 of different WIMP properties.

\subsection{Modification of the normalization constant
            of the one--dimensional velocity distribution function}
\label{sec:f1v}

 The general expression for the differential event rate
 for elastic WIMP--nucleus scattering
 with both of
 the SI and the SD cross sections
 can be given by \cite{SUSYDM96, DMDDranap}:
\beq
     \dRdQ
  =  \frac{\rho_0}{2 \mchi \mrN^2}
     \bbigg{\sigmaSI \FSIQ + \sigmaSD \FSDQ}
     \intvmin \bfrac{f_1(v)}{v} dv
\~.
\label{eqn:dRdQ_SISD}
\eeq
 Here $R$ is the direct detection event rate,
 i.e.~the number of events
 per unit time and unit mass of detector material,
 $Q$ is the energy deposited in the detector,
 $\rho_0$ is the WIMP density near the Earth,
 $f_1(v)$ is the one--dimensional velocity distribution function
 of the WIMPs impinging on the detector,
 $v$ is the absolute value of the WIMP velocity
 in the laboratory frame.
 $\sigma_0^{\rm (SI, SD)}$ are the SI/SD total cross sections
 ignoring the form factor suppression
 and
 $F_{\rm (SI, SD)}(Q)$ indicate the elastic nuclear form factors
 corresponding to the SI/SD WIMP interactions,
 respectively.
 The reduced mass $\mrN$ is defined by
\(
         \mrN
 \equiv  \mchi \mN / \abrac{\mchi + \mN}
\),
 where $\mchi$ is the WIMP mass and
 $\mN$ that of the target nucleus.
 Finally,
 $\vmin$ is the minimal incoming velocity of incident WIMPs
 that can deposit the energy $Q$ in the detector:
\beq
     \vmin(Q)
  =  \alpha \sqrt{Q}
\label{eqn:vmin}
\eeq
 with the transformation constant
\(
         \alpha
 \equiv  \sqrt{\mN / 2 \mrN^2}
\).

 As the first step of our model--independent methods
 for reconstructing the one--dimensional velocity distribution
 as well as
 other particle properties of halo WIMPs,
 the entire experimental possible energy range
 between the minimal and maximal cut--offs $\Qmin$ and $\Qmax$
 of the analyzed data set
 needs to be divided into $B$ bins
 with central points $Q_n$ and widths $b_n$
 \cite{DMDDf1v}%
\footnote{
 Note that,
 due to the maximal cut--off
 on the incoming velocity of incident WIMPs,
 $\vmax$,
 which is related to
 the escape velocity from our Galaxy
 at the position of the Solar system,
 a kinematic maximal cut--off energy,
\beq
     Q_{\rm max, kin}
  =  \frac{\vmax^2}{\alpha^2}
\~,
\label{eqn:Qmax_kin}
\eeq
 has to be considered.
 For distinguishing two maximal cut--offs
 more clearly,
 we define and use hereafter
\mbox{\(
         \Qmax^{\ast}
 \equiv  {\rm min}\abrac{\Qmax,~Q_{\rm max, kin}}
\)},
 the smaller one between
 the experimental and kinematic maximal cut--off energies,
 as the upper bound of the recoil energy
 of the recorded events
 in this paper.
}:
\beq
      {\T Q_n - \frac{b_n}{2}}
 \le  \Qni
 \le  {\T Q_n + \frac{b_n}{2}}
\~,
     ~~~~ ~~~~ ~~~~ 
      i
  =   1,~2,~\cdots,~N_n,~
      n
  =   1,~2,~\cdots,~B,
\label{eqn:Qni}
\eeq
 where $\Qni$ denotes the measured recoil energy in the $n$th $Q-$bin
 and in each bin,
 $N_n$ events will be recorded.
 Since the recoil spectrum $dR / dQ$ is expected
 to be approximately exponential,
 in order to approximate the spectrum
 in a rather wider range,
 the following exponential ansatz
 for the measured recoil spectrum
 (before normalized by the exposure $\calE$)
 in the $n$th bin has been introduced
 \cite{DMDDf1v}:
\beq
          \adRdQ_{{\rm expt}, \~ n}
  \equiv  \adRdQ_{{\rm expt}, \~ Q \simeq Q_n}
  \equiv  r_n  \~ e^{k_n (Q - Q_{s, n})}
\~.
\label{eqn:dRdQn}
\eeq
 Here
\(
     r_n
  =  N_n / b_n
\)
 is the standard estimator
 for $(dR / dQ)_{\rm expt}$ at $Q = Q_n$,
 $k_n$ is the logarithmic slope of
 the recoil spectrum in the $n$th $Q-$bin,
 which can be computed numerically
 from the average value of the measured recoil energies
 in this bin
 \cite{DMDDf1v}:
\beq
     \bQn
 \equiv
     \frac{1}{N_n} \sumiNn \abrac{\Qni - Q_n}
  =  \afrac{b_n}{2} \coth\afrac{b_n k_n}{2}-\frac{1}{k_n}
\~.
\label{eqn:bQn}
\eeq
 Then the shifted point $Q_{s, n}$
 in the ansatz (\ref{eqn:dRdQn}),
 can be estimated by
 \cite{DMDDf1v}
\beq
     Q_{s, n}
  =  Q_n + \frac{1}{k_n} \ln \bfrac{\sinh(b_n k_n / 2)}{b_n k_n / 2}
\~.
\label{eqn:Qsn}
\eeq
 In Ref.~\cite{DMDDf1v},
 we derived that
 the functional form of
 the one--dimensional velocity distribution function
 can be given by the recoil spectrum as%
\footnote{
 Note that,
 originally and for so far most practical uses
 under the assumption that
 the SI WIMP--nucleus interaction dominates over the SD one,
 $F(Q)$ appearing in this
 and the next Sec.~\ref{sec:mchi}
 should be chosen as $F_{\rm SI}(Q)$.
 However,
 for light and strongly spin--sensitive target nuclei
 (namely,
  in the case that
  the SD WIMP--nucleus interaction dominates over the SI one)
 or the general case
 given in Eq.~(\ref{eqn:dRdQ_SISD}) of $dR / dQ$,
 one can apply all expressions given in this
 and the next Sec.~\ref{sec:mchi}
 straightforwardly
 (cf.~Sec.~\ref{sec:ranap}).
}
\beq
     f_1(v)
  =  \calN
     \cbrac{-2 Q \cdot \dd{Q} \bbrac{ \frac{1}{\FQ} \aDd{R}{Q} } }\Qva
\~.
\label{eqn:f1v_dRdQ}
\eeq
 By substituting the ansatz (\ref{eqn:dRdQn})
 into this functional form of $f_1(v)$
 and letting $Q = Q_{s, n}$,
 the reconstructed velocity distribution
 at points
\(
     v_{s, n}
  =  \alpha \sqrt{Q_{s, n}}
\)
 can thus be estimated by
 \cite{DMDDf1v}
\beq
     f_{1, {\rm rec}}(v_{s, n})
  =  \calN
     \bBigg{\frac{2 Q_{s, n} r_n}{F^2(Q_{s, n})}}
     \bbrac{\dd{Q} \ln \FQ \bigg|_{Q = Q_{s, n}} - k_n}
\~.
\label{eqn:f1v_Qsn}
\eeq
 In Ref.~\cite{DMDDf1v-calN},
 we considered further
 a minimal cut--off of the velocity distribution
 due to non--zero experimental threshold energy,
 $\vmin(\Qmin) = \alpha \sqrt{\Qmin} \equiv \vmin^{\ast}$,
 and introduced a model--independent trianglar estimator
 for the area under $f_1(v)$
 in the range of $0 \le v \le \vmin^{\ast}$.
 Then
 the normalization condition
 for the reconstructed velocity distribution function
 can be approximated by
 \cite{DMDDf1v-calN}
\beqn
     \intz f_1(v) \~ dv
 \eqnsimeq
     f_{1, {\rm rec}}(\vmin^{\ast}) \cdot \frac{\vmin^{\ast}}{2}
   + \calN
     \int_{\Qmin}^{\Qmax^{\ast}}
     \cbrac{-2 Q \cdot \ddRdQoFQdQ} \afrac{\alpha}{2 \sqrt{Q}} dQ
     \non\\
 \eqnsimeq
     f_{1, {\rm rec}}(\vmin^{\ast}) \cdot \frac{\alpha \sqrt{\Qmin}}{2}
   + \calN \afrac{\alpha}{2}
     \bbrac{  \frac{2 \Qmin^{1 / 2} r(\Qmin)}{\FQmin}
             + I_0(\Qmin, \Qmax^{\ast})}
     \non\\
 \=  1
\~.
\label{eqn:normalization_infty_mod}
\eeqn
 Here we have defined
\beq
         r(\Qmin)
 \equiv  \adRdQ_{{\rm expt},\~Q = \Qmin}
  =      r_1 \~ e^{k_1 (\Qmin - Q_{s, 1})}
\~,
\label{eqn:rmin}
\eeq
 with
\(
     r_1
  =  N_1 / b_1
\),
 is an estimated value
 of the measured recoil spectrum
 $(dR / dQ)_{\rm expt}$
 at $Q = \Qmin$,
 and $I_n(\Qmin, \Qmax^{\ast})$ can be estimated through the sum
 running over all events in the data set:
\beq
         I_n(\Qmin, \Qmax^{\ast})
 \equiv  \int_{\Qmin}^{\Qmax^{\ast}} Q^{(n - 1) / 2} \bdRdQoFQ dQ
 \to     \sum_a \frac{Q_a^{(n - 1) / 2}}{F^2(Q_a)}
\~.
\label{eqn:In_sum}
\eeq
 Note that,
 since the WIMP--nucleus scattering spectrum
 is expected to be exponential,
 the term of $\abrac{dR / dQ}_{{\rm expt},\~Q = \Qmax^{\ast}}$
 appearing in the second term of the second line
 in Eq.~(\ref{eqn:normalization_infty_mod})
 has been ignored.
 Moreover,
 by substituting Eq.~(\ref{eqn:dRdQn})
 into Eq.~(\ref{eqn:f1v_dRdQ})
 and setting $Q = \Qmin$,
 one can have
\beqn
      f_{1, {\rm rec}}(\vmin^{\ast})
 \=  \calN
     \bBigg{\frac{2 \Qmin r(\Qmin)}{\FQmin}}
     \bbrac{\dd{Q} \ln \FQ \bigg|_{Q = \Qmin} - k_1}
\~.
\label{eqn:f1v_Qmin}
\eeqn
 Hence,
 a model--independent approximation for
 the modified normalization constant $\calN$
 which can be estimated directly from the data
 is given by
 \cite{DMDDf1v-calN}
\beqn
     \calN
 \=  \frac{2}{\alpha}
     \cbrac{  \bBigg{\frac{2 \Qmin^{1 / 2} r(\Qmin)}{\FQmin}}
              \bbigg{  K_1(\Qmin) \~
                       \Qmin
                     + 1 }
            + I_0(\Qmin, \Qmax^{\ast})}^{-1}
\~,
\label{eqn:calN_sum_mod}
\eeqn
 where we have defined
\beq
         K_n(Q)
 \equiv  \dd{Q} \ln \FQ - k_n
\~.
\label{eqn:Kn}
\eeq
\subsection{Reconstructions of the WIMP mass
            and the SI WIMP--nucleon coupling}
\label{sec:mchi}

 Now we revisit our model--independent procedures
 for the determination of the WIMP mass $\mchi$
 and the (squared) SI scalar WIMP--nucleon coupling $|\frmp|^2$.
 The modified expressions corresponding to
 the modification of the normalization constant $\calN$
 given in Eq.~(\ref{eqn:calN_sum_mod})
 will be derived here.
 For more detailed discussions about these methods,
 please see Refs.~\cite{DMDDmchi, DMDDfp2}.

 From the functional form (\ref{eqn:f1v_dRdQ}) of $f_1(v)$,
 one can find that
\beq
     \int_{\vmin^{\ast}}^{\vmax} v^n f_1(v) \~ dv
 \simeq
     \calN \afrac{\alpha^{n + 1}}{2}
     \bbrac{  \frac{2 \Qmin^{(n + 1) / 2} r(\Qmin)}{\FQmin}
            + (n + 1) I_n(\Qmin, \Qmax^{\ast}) }
\~,
\label{eqn:int_v^n_vmin_ast_vmax}
\eeq
 where
 a term
 $2 \Qmax^{\ast \~ (n + 1) / 2} \abrac{dR / dQ}_{{\rm expt},\~Q = \Qmax^{\ast}} / F^2(\Qmax^{\ast})$
 has been ignored%
\footnote{
 Remind that,
 due to sizable contributions from large recoil energies
 \cite{DMDDf1v},
 this is not necessarily true for $n \ge 1$.
 Nevertheless,
 since we use usually only $n = -1$, 1, and 2,
 it has been found that
 Eq.~(\ref{eqn:int_v^n_vmin_ast_vmax}) and,
 in turn,
 Eq.~(\ref{eqn:moments}) can still be available
 for determining
 the WIMP mass
 as well as
 the (ratios between different) WIMP--nucleon couplings/cross sections
 (see Refs.~\cite{DMDDmchi, DMDDfp2, DMDDranap}
  and Sec.~3).
}$^{,\~}$%
\footnote{
 Remind here also that,
 without special remark
 the form factor $F(Q)$ appearing in this section
 could in principle also be the one
 for the SD WIMP--nucleus cross section
 or even for the general case
 with both of the SI and SD cross sections
 (cf.~Sec.~\ref{sec:ranap}).
 However,
 since
 our algorithmic procedure
 for the reconstruction of the WIMP mass
 includes also
 the second solution $\left. \mchi \right|_\sigma$
 given in Eq.~(\ref{eqn:mchi_Rsigma}),
 which is derived under the assumption of
 only the SI WIMP--nucleus interaction
 \cite{DMDDmchi},
 the form factor appearing in $\left. \mchi \right|_{\Expv{v^n}}$
 in Eq.~(\ref{eqn:mchi_Rn})
 has usually to be chosen for the SI cross section.
}.
 Then,
 similar to the calculation of the normalization condition
 given in Eq.~(\ref{eqn:normalization_infty_mod}),
 the moments of the one--dimensional WIMP velocity distribution function
 can be approximated by
\beqn
     \expv{v^n}
 \eqnequiv
     \intz v^n f_1(v) \~ dv
     \non\\
 \eqnsimeq
     f_{1, {\rm rec}}(\vmin^{\ast}) \cdot \frac{(\vmin^{\ast})^{n + 1}}{2}
   + \int_{\vmin^{\ast}}^{\vmax} v^n f_1(v) \~ dv
     \non\\
 \eqnsimeq
     \calN \afrac{\alpha^{n + 1}}{2}
     \cbrac{   \bBigg{\frac{2 \Qmin^{(n + 1) / 2} r(\Qmin)}{\FQmin}}
               \bbigg{  K_1(\Qmin) \~
                        \Qmin
                      + 1 }
             + (n + 1) I_n(\Qmin, \Qmax^{\ast}) }
     \non\\
 \=  \alpha^n
     \bfrac{  2 \Qmin^{(n + 1) / 2} r^{\ast}(\Qmin) / \FQmin
            + (n + 1) I_n(\Qmin, \Qmax^{\ast}) }
           {  2 \Qmin^{     1  / 2} r^{\ast}(\Qmin) / \FQmin
            +         I_0(\Qmin, \Qmax^{\ast}) }
\~.
\label{eqn:moments}
\eeqn
 Here
 the modified normalization constant $\calN$
 given in Eq.~(\ref{eqn:calN_sum_mod})
 has been used
 and we have defined
\beqn
     r^{\ast}(\Qmin)
 \eqnequiv
     r(\Qmin)
     \bBig{K_1(\Qmin) \~ \Qmin + 1}
\~.
\label{eqn:rmin_ast}
\eeqn
 In our earlier work
 on the determination of the WIMP mass $\mchi$,
 it has already been found that
 by requiring that
 the values of a given moment of $f_1(v)$
 estimated by Eq.~(\ref{eqn:moments})
 from two experiments
 with different target nuclei, $X$ and $Y$, agree,
 $\mchi$ appearing in the prefactor $\alpha^n$
 on the right--hand side of Eq.~(\ref{eqn:moments})
 can be solved analytically as
 \cite{DMDDmchi}%
\beq
     \left. \mchi \right|_{\Expv{v^n}}
  =  \frac{\sqrt{\mX \mY} - \mX (\calR_{n, X} / \calR_{n, Y})}
          {\calR_{n, X} / \calR_{n, Y} - \sqrt{\mX / \mY}}
\~,
\label{eqn:mchi_Rn}
\eeq
 with $\calR_{n, (X, Y)}$
 now modified directly as
\beq
     \calR_{n, X}
 \equiv
     \bfrac{  2 \QminX^{(n + 1) / 2} r_X^{\ast}(\QminX) / \FQminX
            + (n + 1) \InX }
           {  2 \QminX^{     1  / 2} r_X^{\ast}(\QminX) / \FQminX
            +         \IzX }^{1 / n}
\~,
\label{eqn:RnX_min}
\eeq
 and $\calR_{n, Y}$ can be defined analogously%
\footnote{
 Hereafter,
 without special remark
 all notations defined for the target $X$
 can be defined analogously for the targets $Y$ and $Z$.
}.
 Here $n \ne 0$,
 $m_{(X, Y)}$ and $F_{(X, Y)}(Q)$
 are the masses and the form factors of the nucleus $X$ and $Y$,
 respectively,
 and $r_{(X, Y)}^{\ast}(Q_{{\rm min}, (X, Y)})$
 refer to the counting rates
 (modified by Eq.~(\ref{eqn:rmin_ast}))
 for the target $X$ and $Y$
 at the relatively lowest recoil energies included in the analysis.

 On the other hand,
 assuming the SI scalar WIMP interaction dominates,
 the zero--momentum--transfer cross section in Eq.~(\ref{eqn:dRdQ_SISD})
 can be expressed as
 \cite{SUSYDM96}
\beq
     \sigmaSI
  =  \afrac{4}{\pi} \mrN^2 \bBig{Z f_{\rm p} + (A - Z) f_{\rm n}}^2
 \simeq
     A^2 \afrac{\mrN}{\mrp}^2 \sigmapSI
\~.
\label{eqn:sigma0SI}
\eeq
 Here $Z$ is the atomic number of the target nucleus,
 i.e.~the number of protons,
 $A$ is the atomic mass number,
 $A - Z$ is then the number of neutrons,
 $f_{\rm (p, n)}$ are the effective scalar couplings of WIMPs
 on protons p and on neutrons n,
 respectively;
 the theoretical prediction that
 the scalar couplings
 on protons and on neutrons
 are approximately equal:
\(
         f_{\rm n}
 \simeq  f_{\rm p}
\)
 has been adopted,
 the tiny mass difference between a proton and a neutron
 has been neglected,
 and
\beq
     \sigmapSI
  =  \afrac{4}{\pi} \mrp^2 |f_{\rm p}|^2
\label{eqn:sigmapSI}
\eeq
 is the SI WIMP--nucleon cross section.
 Now,
 by applying Eq.~(\ref{eqn:int_v^n_vmin_ast_vmax})
 with $n = -1$
 and substituting the expression (\ref{eqn:sigma0SI}) for $\sigmaSI$
 into Eq.~(\ref{eqn:dRdQ_SISD}),
 we can have
\beqn
     \adRdQ_{{\rm expt}, \~ Q = \Qmin}
  =  \calE A^2
     \afrac{\rho_0}{2 \mchi}
     \bbrac{\afrac{4}{\pi} |f_{\rm p}|^2}
     \FSIQmin
     \cbrac{ \calN
             \bfrac{r(\Qmin)}{\FSIQmin} }
\~.
\label{eqn:dRdQ_Qmin}
\eeqn
 It can therefore obtain that
 \cite{DMDDfp2}
\beq
     |\frmp|^2
  =  \frac{1}{\rho_0}
     \afrac{\pi}{4 \sqrt{2}}
     \afrac{1}{\calE A^2 \sqrt{\mN}}
     \bbrac{\frac{2 \Qmin^{1 / 2} r^{\ast}(\Qmin)}{\FSIQmin} + I_0}
     \abrac{\mchi + \mN}
\~.
\label{eqn:fp2}
\eeq
 Note that,
 similar to the expression (\ref{eqn:moments})
 for the moments of
 the one--dimensional WIMP velocity distribution,
 instead of $r(\Qmin)$,
 the first term in the bracket
 on the right--hand side of the estimator
 of the SI WIMP--nucleon coupling
 given here
 is now proportional to $r^{\ast}(\Qmin)$.
 Since
 $|\frmp|^2$ is identical for different targets,
 it leads to the second expression for determining $\mchi$
 \cite{DMDDmchi}:
\beq
     \left. \mchi \right|_\sigma
  =  \frac{\abrac{\mX / \mY}^{5 / 2} \mY - \mX (\calR_{\sigma, X} / \calR_{\sigma, Y})}
          {\calR_{\sigma, X} / \calR_{\sigma, Y} - \abrac{\mX / \mY}^{5 / 2}}
\~.
\label{eqn:mchi_Rsigma}
\eeq
 Here $m_{(X, Y)} \propto A_{(X, Y)}$ has been assumed,
 and $\calR_{\sigma, (X, Y)}$
 are now modified to
\beq
     \calR_{\sigma, X}
 \equiv
     \frac{1}{\calE_X}
     \bbrac{\frac{2 \QminX^{1 / 2} r_X^{\ast}(\QminX)}{\FSIQminX} + \IzX}
\~,
\label{eqn:RsigmaX_min}
\eeq
 and similarly for $\calR_{\sigma, Y}$;
 $\calE_{(X, Y)}$ here are the experimental exposures
 with the target $X$ and $Y$.

 In order to yield the best--fit WIMP mass
 as well as to minimize its statistical uncertainty,
 the $\chi^2$ function
 combining the estimators for different $n$
 in Eq.~(\ref{eqn:mchi_Rn}) with each other
 and with the estimator in Eq.~(\ref{eqn:mchi_Rsigma})
 has been introduced
 \cite{DMDDmchi}
\beq
     \chi^2(\mchi)
  =  \sum_{i, j}
     \abrac{f_{i, X} - f_{i, Y}} {\cal C}^{-1}_{ij} \abrac{f_{j, X} - f_{j, Y}}
\~,
\label{eqn:chi2}
\eeq
 where
\cheqna
\beq
     f_{i, X}
 \equiv
     \alpha_X^i
     \bfrac{  2 \QminX^{(i + 1) / 2} r_X^{\ast}(\Qmin) / \FSIQminX
            + (i + 1) I_{i, X}}
           {  2 \QminX^{     1  / 2} r_X^{\ast}(\Qmin) / \FSIQminX
            +         \IzX    }
     \afrac{1}{300~{\rm km/s}}^i
\~,
\label{eqn:fiXa}
\eeq
 for $i = -1,~1,~2,~\dots,~n_{\rm max}$,
 and
\cheqnb
\beq
     f_{n_{\rm max} + 1, X}
 \equiv
     \calE_X
     \bfrac{A_X^2}
           {  2 \QminX^{1  / 2} r_X^{\ast}(\Qmin) / \FSIQminX
            +         \IzX }
     \afrac{\sqrt{\mX}}{\mchi + \mX}
\~;
\label{eqn:fiXb}
\eeq
\cheqn
 the other $n_{\rm max} + 2$ functions $f_{i, Y}$
 can be defined analogously.
 Here $n_{\rm max}$ determines the highest moment of $f_1(v)$
 that is included in the fit.
 Since the $X$ and $Y$ quantities
 are statistically completely independent,
 the total covariance matrix
 $\cal C$ can be written as a sum of two terms:
\beq
     {\cal C}_{ij}
  =  {\rm cov}\abrac{f_{i, X}, f_{j, X}} + {\rm cov}\abrac{f_{i, Y}, f_{j, Y}}
\~,
\label{eqn:Cij}
\eeq
 with
\beqn
     {\rm cov}\abrac{f_i, f_j}
 \=  \calN_{\rm m}^2
     \bbiggl{  f_i \~ f_j \~ {\rm cov}(I_0, I_0)
             + \Td{\alpha}^{i + j} (i + 1) (j + 1) {\rm cov}(I_i, I_j) }
     \non\\
 \conti ~~~~ ~~ 
             - \Td{\alpha}^j (j + 1) f_i \~ {\rm cov}(I_0, I_j)
             - \Td{\alpha}^i (i + 1) f_j \~ {\rm cov}(I_0, I_i)\bigg.
     \non\\
 \conti ~~~~ ~~~~ ~~ 
             + D_i D_j \sigma^2(r^{\ast}(\Qmin))
             - \abrac{D_i f_j + D_j f_i} {\rm cov}(r^{\ast}(\Qmin), I_0)\Bigg.
     \non\\
 \conti ~~~~ ~~~~ ~~~~ ~~ 
     \bbiggr{+ \Td{\alpha}^j (j + 1) D_i \~ {\rm cov}(r^{\ast}(\Qmin), I_j)
             + \Td{\alpha}^i (i + 1) D_j \~ {\rm cov}(r^{\ast}(\Qmin), I_i)}
\~.
        \non\\
\label{eqn:cov_fi}
\eeqn
 Here we have defined
\beq
         \calN_{\rm m}
 \equiv  \afrac{\alpha}{2} \calN
  =      \bbrac{\frac{2 \Qmin^{1 / 2} r^{\ast}(\Qmin)}{\FSIQmin} + I_0}^{-1}
\~,
\label{eqn:calNm}
\eeq
 and
\(
         \Td{\alpha}
 \equiv  \alpha / 300~{\rm km~s^{-1}}
\);
 as the definitions of $f_{i, (X, Y)}$,
 while
\cheqna
\beq
         D_i
 \equiv  \frac{1}{\calN_{\rm m}} \bPp{f_i}{r^{\ast}(\Qmin)}
 =       \frac{2}{\FSIQmin}
         \abigg{\Td{\alpha}^i \Qmin^{(i + 1) / 2} - \Qmin^{1 / 2} \~ f_i}
\~,
\label{eqn:Dia}
\eeq
 for $i = -1,~1,~2,~\dots,~n_{\rm max}$,
\cheqnb
\beq
     D_{n_{\rm max} + 1}
  =  \frac{2}{\FSIQmin} \abrac{-\Qmin^{1 / 2} f_{n_{\rm max} + 1}}
\~.
\label{eqn:Dib}
\eeq
\cheqn

 With the modified definitions of
 $\calR_{n, (X, Y)}$ and $\calR_{\sigma, (X, Y)}$
 given by Eqs.~(\ref{eqn:RnX_min}) and (\ref{eqn:RsigmaX_min}),
 one can follow the procedure developed in Ref.~\cite{DMDDmchi}
 to reconstruct the WIMP mass $\mchi$ straightforwardly.
 Remind only that,
 while,
 as discussed in Ref.~\cite{DMDDmchi},
 the upper cuts on $f_1(v)$ in two data sets
 should be (approximately) equal
 and it in turn requires that
\beq
     \QmaxY^{\ast}
  =  \afrac{\alpha_X}{\alpha_Y}^2 \QmaxX^{\ast}
\~,
\label{eqn:Qmax_match}
\eeq
 a similar correction between $\QminX$ and $\QminY$
 is not necessary,
 since the estimator of the moments of $f_1(v)$
 given by Eq.~(\ref{eqn:moments})
 has already taken into account the integral
 below the non--zero experimental threshold energy
 $v \le \alpha_{(X, Y)} \sqrt{Q_{{\rm min}, (X, Y)}}$.

\subsection{Reconstructions of the ratios between
            the SD and SI WIMP--nucleon cross sections}
\label{sec:ranap}

 Finally,
 we come back to consider
 the scattering event rate estimated by
 the general combination of
 the SI and SD
 WIMP--nucleus interactions
 in Eq.~(\ref{eqn:dRdQ_SISD})
 and derive the modified expressions
 for the reconstructions of the ratios between
 the SD WIMP coupling on neutrons to that on protons,
 $\armn / \armp$,
 and between the SD and SI WIMP--nucleon cross sections,
 $\sigma_{\chi ({\rm p, n})}^{\rm SD} / \sigmapSI$.
 For more detailed discussions about these reconstructions,
 please see Ref.~\cite{DMDDranap}.

 While the expression for the SI WIMP--nucleus cross section
 has been given in Eq.~(\ref{eqn:sigma0SI}),
 the SD WIMP--nucleus cross section
 can be expressed as
 \cite{SUSYDM96}
\beqn
     \sigmaSD
 \=  \afrac{32}{\pi} G_F^2 \~ \mrN^2
     \afrac{J + 1}{J} \bBig{\Srmp \armp + \Srmn \armn}^2
     \non\\
 \=  \frac{4}{3} \afrac{J + 1}{J} \afrac{\mrN}{\mrp}^2
     \bbrac{\Srmp + \Srmn \afrac{\armn}{\armp}}^2
     \sigmapSD
\~.
\label{eqn:sigma0SD}
\eeqn
 Here $G_F$ is the Fermi constant,
 $J$ is the total spin of the target nucleus,
 $\expv{S_{\rm (p, n)}}$ are the expectation values of
 the proton and neutron group spins
 (see Table \ref{tab:Sp/n}),
 $a_{\rm (p, n)}$ are the effective SD WIMP couplings
 on protons and on neutrons,
 respectively,
 and
 the SD WIMP cross section on protons or on neutrons
 can be given as
\beq
     \sigma_{\chi {\rm (p, n)}}^{\rm SD}
  =  \afrac{24}{\pi} G_F^2 \~ m_{\rm r, (p, n)}^2 |a_{\rm (p, n)}|^2
\~.
\label{eqn:sigmap/nSD}
\eeq
\begin{table}[t!]
\small
\begin{center}
\renewcommand{\arraystretch}{1.5}
 \begin{tabular}{|| c   c   c   c   c   c   c   c ||}
\hline
\hline
 \makebox[1.3cm][c]{Isotope}        &
 \makebox[0.9cm][c]{$Z$}            & \makebox[0.9cm][c]{$J$}     &
 \makebox[1.5cm][c]{$\Srmp$}        & \makebox[1.5cm][c]{$\Srmn$} &
 \makebox[1.8cm][c]{$-\Srmp/\Srmn$} & \makebox[1.8cm][c]{$\Srmn/\Srmp$} &
 \makebox[3.7cm][c]{Natural abundance (\%)} \\
\hline
\hline
 $\rmXA{F}{19}$   &  9 & 1/2 &                   0.441  & \hspace{-1.8ex}$-$0.109 &
      4.05  &  $-$0.25   &       100   \\
\hline
 $\rmXA{Na}{23}$  & 11 & 3/2 &                   0.248  &                   0.020 &
  $-$12.40  &     0.08   &       100   \\
\hline
 $\rmXA{I}{127}$  & 53 & 5/2 &                   0.309  &                   0.075 &
   $-$4.12  &     0.24   &       100   \\
\hline
 $\rmXA{Xe}{131}$ & 54 & 3/2 & \hspace{-1.8ex}$-$0.009  & \hspace{-1.8ex}$-$0.227 &
   $-$0.04  &    25.2    &        21   \\
\hline
\hline
\end{tabular}
\end{center}
\caption{
 List of the relevant spin values of the nuclei
 used in our simulations presented in this paper.
 More details can be found in
 e.g.~Refs.~\cite{SUSYDM96, Tovey00, Giuliani05, Girard05}.
}
\label{tab:Sp/n}
\end{table}

 Consider at first the case that
 the SD WIMP--nucleus interaction dominates over the SI one
 and thus
 the first SI term, $\sigmaSI \FSIQ$,
 in the bracket on the right--hand side of Eq.~(\ref{eqn:dRdQ_SISD})
 can be neglected.
 Similar to the calculation given in Eq.~(\ref{eqn:dRdQ_Qmin}),
 one can obtain an expression
 for the SD WIMP--nucleus cross section
 straightforwardly as
\beqn
     \sigmaSD
 \=  \frac{1}{\rho_0}
     \bbrac{\frac{1}{\calE\sqrt{2}} \afrac{\mchi^2 \mN^{3 / 2}}{\mchi + \mN}}
     \bbrac{\frac{2 \Qmin^{1 / 2} r^{\ast}(\Qmin)}{\FSDQmin} + I_0}
\~.
\label{eqn:sigmaSD}
\eeqn
 By combining two target nuclei,
 $X$ and $Y$,
 and substituting the expression (\ref{eqn:sigma0SD}) for $\sigmaSD$
 into Eq.~(\ref{eqn:sigmaSD}),
 the ratio between two SD WIMP--nucleon couplings
 has been solved analytically as
 \cite{DMDDranap}
\beqn
     \afrac{\armn}{\armp}_{\pm, n}^{\rm SD}
 \=- \frac{\SpX \pm \SpY \abrac{\calR_{J, n, X} / \calR_{J, n, Y}} }
          {\SnX \pm \SnY \abrac{\calR_{J, n, X} / \calR_{J, n, Y}} }
\~,
     ~~~~ ~~~~ ~~~~ ~~~~ 
    \forall \~ n \ne 0
.
\label{eqn:ranapSD}
\eeqn
 Here we have defined%
\footnote{
 Note that
 the form factors involved in both of
 $\calR_{n, (X, Y)}$ and $\calR_{\sigma, (X, Y)}$
 defined in Eqs.~(\ref{eqn:RnX_min}) and (\ref{eqn:RsigmaX_min})
 as well as
 in the estimator for $I_{n, (X, Y)}$ given by Eq.~(\ref{eqn:In_sum})
 must be chosen as for the SD WIMP--nucleus interaction.
}
\beq
     \calR_{J, n, X}
 \equiv
     \bbrac{\Afrac{J_X}{J_X + 1}
     \frac{\calR_{\sigma, X}}{\calR_{n, X}}}^{1 / 2}
\~.
\label{eqn:RJnX}
\eeq
 Note that,
 as discussed in Ref.~\cite{DMDDranap},
 because the couplings in Eq.~(\ref{eqn:sigma0SD}) are squared,
 we have two solutions for $\armn / \armp$ here,
 which depends simply on the signs of $\SnX$ and $\SnY$:
 if both $\SnX$ and $\SnY$ are positive or negative,
 the ``$+$ (plus)'' solution $(\armn / \armp)^{\rm SD}_{+, n}$
 will be the solution to be taken;
 in contrast,
 if the signs of $\SnX$ and $\SnY$ are opposite,
 the ``$-$ (minus)'' solution
 will be the suitable one.

 Now we consider the general combination of
 both of the SI and SD cross sections.
 By dividing Eq.~(\ref{eqn:sigma0SD})
 by Eq.~(\ref{eqn:sigma0SI}),
 the ratio between the SD and SI WIMP--proton cross section
 can be expressed as
\beq
     \frac{\sigmaSD}{\sigmaSI}
  =  \afrac{32}{\pi} G_F^2 \~ \mrp^2 \Afrac{J + 1}{J}
     \bfrac{\Srmp + \Srmn (\armn / \armp)}{A}^2 \frac{|\armp|^2}{\sigmapSI}
  =  \calCp \afrac{\sigmapSD}{\sigmapSI}
\~,
\label{eqn:rsigmaSDSI}
\eeq
 where we defined
\beq
         \calCp
 \equiv  \frac{4}{3} \afrac{J + 1}{J}
         \bfrac{\Srmp + \Srmn (\armn / \armp)}{A}^2
\~.
\label{eqn:Cp}
\eeq
 Then,
 by substituting Eq.~(\ref{eqn:rsigmaSDSI})
 and then Eq.~(\ref{eqn:sigma0SI})
 into Eq.~(\ref{eqn:dRdQ_SISD}),
 the general expression
 for the differential event rate
 can be rewritten as
\beq
     \adRdQ_{\rm expt}
  =  \calE
     A^2 \! \afrac{\rho_0 \sigmapSI}{2 \mchi \mrp^2} \!\!
     \bbrac{\FSIQ + \calCp \FSDQ \afrac{\sigmapSD}{\sigmapSI}}
     \intvmin \bfrac{f_1(v)}{v} dv
\~.
\label{eqn:dRdQ_SISD_expt}
\eeq
 This implies that
 one can use Eq.~(\ref{eqn:dRdQ_Qmin})
 with the replacement of $\FQmin$ by
\(
         F_{\rm SI}'^2(\Qmin)
 \equiv  F_{\rm SI} ^2(\Qmin)
       + \calCp F_{\rm SD}^2(\Qmin) \abrac{\sigmapSD / \sigmapSI}
\).
 By combining two targets $X$ and $Y$,
 the ratio of the SD WIMP--proton cross section
 to the SI one can be solved analytically as
 \cite{DMDDranap}%
\footnote{
 Similarly,
 the ratio of the SD WIMP--neutron cross section
 to the SI one can be solved as
\beq
     \frac{\sigmanSD}{\sigmapSI}
  =  \frac{\FSIQminY (\calR_{m, X} / \calR_{m, Y}) - \FSIQminX}
          {\calCnX \FSDQminX               - \calCnY \FSDQminY (\calR_{m, X} / \calR_{m, Y})}
\~,
\label{eqn:rsigmaSDnSI}
\eeq
 with
\beq
         \calCn
 \equiv  \frac{4}{3} \Afrac{J + 1}{J}
         \bfrac{\Srmp (\armp / \armn) + \Srmn}{A}^2
\~.
\label{eqn:Cn}
\eeq
 Note hereafter that,
 except a few special expressions given explicitly,
 all formulae for protons can be applied for neutrons straightforwardly
 by replacing p $\to$ n.
}
\beq
     \frac{\sigmapSD}{\sigmapSI}
  =  \frac{\FSIQminY (\calR_{m, X} / \calR_{m, Y})  - \FSIQminX}
          {\calCpX \FSDQminX                - \calCpY \FSDQminY (\calR_{m, X} / \calR_{m, Y})}
\~,
\label{eqn:rsigmaSDpSI}
\eeq
 where we have defined
\beq
     \calR_{m, X}
 \equiv
     \frac{r_X^{\ast}(\QminX)}{\calE_X A_X^2}
\~,
\label{eqn:RmX}
\eeq
 and similar to $\calR_{m, Y}$.
 Remind that,
 as the estimator (\ref{eqn:ranapSD}) for $\armn / \armp$,
 one can use Eq.~(\ref{eqn:rsigmaSDpSI})
 to estimate $\sigma_{\chi {\rm p}}^{\rm SD} / \sigmapSI$
 without a prior knowledge of the WIMP mass $\mchi$.
 Moreover,
 since ${\cal C}_{{\rm p}, (X, Y)}$
 depend only on the nature of the detector materials,
 $\sigma_{\chi {\rm p}}^{\rm SD} / \sigmapSI$
 is practically only a function of $\calR_{m, (X, Y)}$,
 which can be estimated by using events
 in the lowest available energy ranges.

 Meanwhile,
 for the general combination of
 the SI and SD WIMP--nucleus cross sections,
 the $\armn / \armp$ ratio
 appearing in Eq.~(\ref{eqn:Cp})
 can also be solved analytically
 by introducing a third nucleus
 with only an SI sensitivity:
\(
     \Srmp_Z
  =  \Srmn_Z
  =  0
\),
 i.e.~%
\(
     {\cal C}_{{\rm p}, Z}
  =  0
\)
 \cite{DMDDranap}
\beqn
     \afrac{\armn}{\armp}_{\pm}^{\rm SI + SD}
 \=  \frac{-\abrac{\cpX \snpX - \cpY \snpY}
           \pm \sqrt{\cpX \cpY} \vbrac{\snpX - \snpY}}
          {\cpX \snpX^2 - \cpY \snpY^2}
     \non\\
 \=  \cleft{\renewcommand{\arraystretch}{0.5}
            \begin{array}{l l l}
             \\
             \D -\frac{\sqrt{\cpX} \mp \sqrt{\cpY}}{\sqrt{\cpX} \snpX \mp \sqrt{\cpY} \snpY}\~, &
             ~~~~ ~~~~ & {\rm for}~\snpX > \snpY, \\~\\~\\ 
             \D -\frac{\sqrt{\cpX} \pm \sqrt{\cpY}}{\sqrt{\cpX} \snpX \pm \sqrt{\cpY} \snpY}\~, &
                       & {\rm for}~\snpX < \snpY. \\~\\
            \end{array}}
\label{eqn:ranapSISD}
\eeqn
 Here we have defined
\cheqna
\beq
     \cpX
 \equiv
     \frac{4}{3} \Afrac{J_X + 1}{J_X} \bfrac{\SpX}{A_X}^2
     \bbrac{  \FSIQminZ \afrac{\calR_{m, Y}}{\calR_{m, Z}} \!
            - \FSIQminY} \!
     \FSDQminX
\~,
\label{eqn:cpX}
\eeq
\cheqnb
\beq
         \cpY
 \equiv  \frac{4}{3} \Afrac{J_Y + 1}{J_Y} \bfrac{\SpY}{A_Y}^2
         \bbrac{  \FSIQminZ \afrac{\calR_{m, X}}{\calR_{m, Z}} \!
                - \FSIQminX} \!
         \FSDQminY
\~,
\label{eqn:cpY}
\eeq
\cheqn
 and
\beq
         \snpX
 \equiv  \frac{\SnX}{\SpX}
\~.
\label{eqn:snpX}
\eeq
 Remind that,
 first,
 $(\armn / \armp)_{\pm}^{\rm SI + SD}$ and $c_{{\rm p}, (X, Y)}$
 given in Eqs.~(\ref{eqn:ranapSISD}), (\ref{eqn:cpX}), and (\ref{eqn:cpY})
 are functions of only
 $r_{(X, Y, Z)}^{\ast}(Q_{{\rm min}, (X, Y, Z)})$,
 which can be estimated with events
 in the lowest energy ranges.
 Second,
 while the decision of the suitable solution of
 $(\armn / \armp)_{\pm, n}^{\rm SD}$
 depends on the signs of $\SnX$ and $\SnY$,
 the decision for $(\armn / \armp)_{\pm}^{\rm SI + SD}$
 depends not only on the signs of
 $s_{{\rm n / p}, (X, Y)}$,
 but also on the order of the two targets.
 For instance,
 for an F + I combination used in our simulations
 presented in the paper,
 since \mbox{$ s_{{\rm n/p}, \rmXA{F}{19}}  = -0.247
             < s_{{\rm n/p}, \rmXA{I}{127}} = 0.243$}
 and since $s_{{\rm n/p}, \rmXA{F}{19}}$
 and $s_{{\rm n/p}, \rmXA{I}{127}}$
 have the opposite signs,
 the ``$-$ (minus)'' solution of the lower expression
 in the second line of Eq.~(\ref{eqn:ranapSISD})
 (or the ``$-$ (minus)'' solution of the expression in the first line)
 is then the suitable solution
 (see Ref.~\cite{DMDDranap}
  for detailed discussions).

 Furthermore,
 one can also choose at first a nucleus
 with only an SI sensitivity
 as the second target:
\(
     \SpY
  =  \SnY
  =  0
\),
 i.e.~%
\(
     \calCpY
  =  0
\).
 The expression in Eq.~(\ref{eqn:rsigmaSDpSI})
 can thus be reduced to
\beq
     \frac{\sigmapSD}{\sigmapSI}
  =  \frac{\FSIQminY (\calR_{m, X} / \calR_{m, Y}) - \FSIQminX}
          {\calCpX \FSDQminX}
\~.
\label{eqn:rsigmaSDpSI_even}
\eeq
 Then,
 by choosing a nucleus with (much) larger
 proton (or neutron) group spin
 as the first target:
\(
         \SpX
 \gg     \SnX
 \simeq  0
\),
 the $\armn / \armp$ dependence of $\calCpX$
 given in Eq.~(\ref{eqn:Cp})
 can be eliminated as%
\footnote{
 Analogously,
 from the definition (\ref{eqn:Cn}) of $\calCn$,
 one can choose $\SnX \gg \SpX \simeq 0$
 to eliminate its $\armp / \armn$ dependence
 and get
\beq
         \calCnX
 \simeq  \frac{4}{3} \afrac{J_X + 1}{J_X} \bfrac{\SnX}{A_X}^2
\~.
\label{eqn:CnX_n}
\eeq
 Then we can have
\beq
     \frac{\sigmanSD}{\sigmapSI}
  =  \frac{\FSIQminY (\calR_{m, X} / \calR_{m, Y}) - \FSIQminX}
          {\calCnX \FSDQminX}
\~.
\label{eqn:rsigmaSDnSI_even}
\eeq
}
\beq
         \calCpX
 \simeq  \frac{4}{3} \Afrac{J_X + 1}{J_X} \bfrac{\SpX}{A_X}^2
\~.
\label{eqn:CpX_p}
\eeq
\section{Numerical results}

 In this section,
 we present Monte Carlo simulation results
 of the reconstructions of different WIMP properties%
\footnote{
 Note that
 all of the (uncertainty bounds on the)
 reconstructed WIMP properties
 presented in this paper
 are as usual the median values
 of the simulated results.
}
 by using the modified expressions
 given in the previous section
 with non--zero threshold energy.

 First of all,
 since the lighter the WIMP mass,
 the more problematic
 the non--negligible experimental threshold energy,
 we present simulation results
 with input WIMP masses less than 200 GeV and
 focus our discussions
 with $m_{\chi, {\rm in}} \le 15$ GeV.
 For the input one--dimensional velocity distribution function
 of halo WIMPs,
 we,
 as usual,
 take into account
 the orbital motion of the Solar system around our Galaxy
 as well as
 that of the Earth around the Sun
 and thus use
 the shifted Maxwellian velocity distribution
 given by \cite{Lewin96}:
\beqn
     f_{1, \sh}(v)
 \=  \cleft{\renewcommand{\arraystretch}{1.75}
            \begin{array}{l l}
             \D
             N_{\sh}
             \afrac{v}{v_0 \ve}
             \bBig{  e^{-(v - \ve)^2 / v_0^2}
                   - e^{-(v + \ve)^2 / v_0^2} } \~, &
             {\rm for}~v \le \vesc - \ve \~,                 \\
             \D
             N_{\sh}
             \afrac{v}{v_0 \ve}
             \bBig{  e^{-(v - \ve)^2 / v_0^2}
                   - e^{-\vesc^2     / v_0^2} } \~, &
             {\rm for}~\vesc - \ve \le v \le \vesc + \ve \~, \\
             0 \~,                                  &
             {\rm for}~v \ge \vesc + \ve \~ \equiv \vmax \~,
            \end{array}}
\label{eqn:f1v_sh_vesc}
\eeqn
 with the normalization constant
\beq
     N_{\sh}
  =  \bbrac{  \sqrt{\pi} \~
              \erf{\D \afrac{\vesc}{v_0}}
            - \afrac{2 \vesc}{v_0}
              e^{-\vesc^2 / v_0^2}        }^{-1}
\~.
\label{eqn:N_sh_vesc}
\eeq
 Here
 $v_0 \simeq 220$ km/s
 is the Solar orbital speed around the Galactic center
 and
 $\ve$ is the time--dependent
 Earth's velocity in the Galactic frame
 \cite{Freese88, SUSYDM96}:
\beq
     \ve(t)
  =  v_0 \bbrac{1.05 + 0.07 \cos\afrac{2 \pi (t - t_{\rm p})}{1~{\rm yr}}}
\~,
\label{eqn:ve}
\eeq
 with $t_{\rm p} \simeq$ June 2nd,
 the date
 on which the velocity of the Earth relative to the WIMP halo is maximal%
\footnote{
 As usual,
 in all our simulations
 the time dependence of the Earth's velocity in the Galactic frame,
 the second term of $\ve(t)$,
 will be ignored,
 i.e.~$\ve = 1.05 \~ v_0$ is used.
}.
 In addition,
 the escape velocity from our Galaxy
 in the position of the Solar system
 has been set as $\vesc = 500$ km/s
 \cite{RPP16AP}.

 Meanwhile,
 for the elastic scattering form factors
 for the SI and SD WIMP--nucleus cross sections,
 we use
\beq
     F_{\rm SI}^2(Q)
  =  \bfrac{3 j_1(q R_1)}{q R_1}^2 e^{-(q s)^2}
\~,
\label{eqn:FQ_SI_WS}
\eeq
 and
\beqn
     F_{\rm SD}^2(Q)
 \=  \cleft{\renewcommand{\arraystretch}{1.5}
            \begin{array}{l l l}
             j_0^2(q R_1)                      \~, & ~~~~ ~~~~ & 
             {\rm for}~q R_1 \le 2.55~{\rm or}~q R_1 \ge 4.5 \~, \\
             {\rm const.} \simeq 0.047         \~, &           &
             {\rm for}~2.5 5 \le q R_1 \le 4.5 \~.
            \end{array}}
\label{eqn:FQ_SD_TS}
\eeqn
 Here $Q$ is the recoil energy
 transferred from the incident WIMP to the target nucleus,
 $j_1(x)$ and $j_0(x)$ are the spherical Bessel functions,
\(
     q
  =  \sqrt{2 \mN Q}
\)
 is the transferred 3--momentum,
 for the effective nuclear radius we use
\(
     R_1
  =  \sqrt{R_A^2 - 5 s^2}
\)
 with
\(
     R_A
 \simeq
     1.2 \~ A^{1 / 3} \~ {\rm fm}
\)
 and a nuclear skin thickness
\(
     s
 \simeq
     1 \~ {\rm fm}
\).
 Additionally,
 the SI WIMP--nucleus cross section
 has been fixed to be $\sigmapSI = 10^{-9}$ pb
 in {\em all} of our simulations.

 Finally,
 we assumed that
 all experimental systematic uncertainties
 as well as
 the uncertainty on the measurement of the recoil energy
 could be ignored.
 Instead of all relevant isotopes of an element,
 we have considered only a single isotope at a time,
 since the uncertainty caused by using different isotopes
 has been estimated to be negligible.
 5,000 experiments with 50 total events on average
 (Poisson--distributed,
  before cuts on $\Qmax^{\ast}$
  determined by Eq.~(\ref{eqn:Qmax_match}))
 in one experiment have been simulated.

\subsection{Small, but non--zero threshold energy}

 As a warm up,
 we consider at first
 a small,
 but non--zero experimental threshold energy
 of $\Qmin = 0.25$ keV.

\subsubsection{Reconstruction of the WIMP mass}
\label{sec:mchi-500-0025}

 We first present our simulation results of the reconstruction of
 the most important WIMP property:
 the WIMP mass $\mchi$.
 The algorithmic procedure
 introduced in Ref.~\cite{DMDDmchi}
 by minimizing the $\chi^2(\mchi)$ function
 defined in Eq.~(\ref{eqn:chi2})
 have been used
 with
 $\rmXA{Si}{28}$ and $\rmXA{Ge}{76}$
 as two target nuclei
 \cite{DMDDmchi}.
 The input WIMP masses
 between 2 and 200 GeV
 have been simulated.

\begin{figure}[t!]
\begin{center}
 \includegraphics[width = 15 cm]{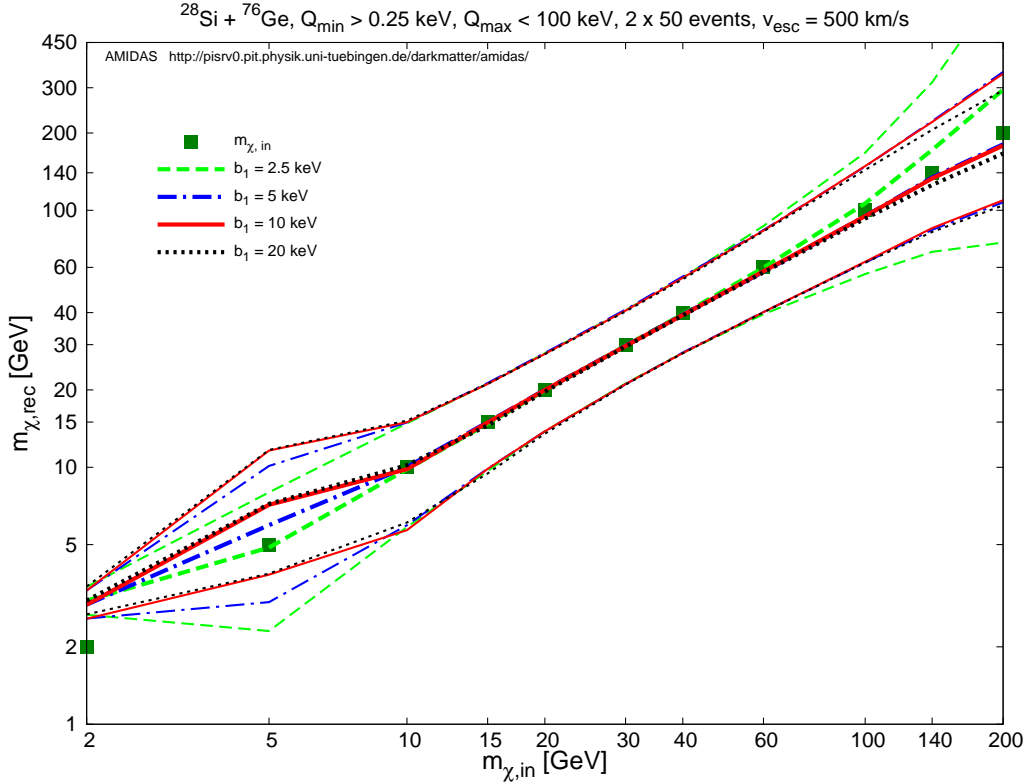}
\end{center}
\caption{
 The reconstructed WIMP masses
 and the lower and upper bounds of
 the 1$\sigma$ statistical uncertainties
 as functions of the input WIMP mass
 with a $\rmXA{Si}{28}$ + $\rmXA{Ge}{76}$ target combination
 for input WIMP masses between 2 and 200 GeV.
 A small experimental threshold energy
 of $\Qmin = 0.25$ keV
 has been simulated.
 Four different widths of the first energy bin $b_1$
 have been used:
 2.5 keV (dashed green),
 5 keV (dash--dotteded blue),
 10 keV (solid red),
 and 20 keV (dotted black).
 The squared dark--green points
 indicate the true (input) WIMP masses
 in our simulations.
 Each experiment contains 50 events on average
 (before cuts on $\Qmax^{\ast}$
  determined by Eq.~(\ref{eqn:Qmax_match})).
 See the text for further details.
}
\label{fig:mchi-SiGe-500-0025}
\end{figure}

 In Fig.~\ref{fig:mchi-SiGe-500-0025},
 we show
 the reconstructed WIMP masses
 and the lower and upper bounds of
 the 1$\sigma$ statistical uncertainties
 as functions of the input WIMP mass.
 Four different widths of the first energy bin $b_1$
 have been used:
 2.5 keV (dashed green),
 5 keV (dash--dotteded blue),
 \mbox{10 keV} (solid red),
 and 20 keV (dotted black).
 Note however that,
 although in our simulations
 a fixed width of the first energy bin
 has been set initially,
 for the three lightest input WIMP masses,
 $b_1$ has been tuned to be
 the {\em analyzed} energy range,
 i.e.,~$b_1 = \Qmax^{\ast} - \Qmin$,
 since this is less than the initial setup of $b_1$
 in our simulations.
 For example,
 for the input WIMP mass of \mbox{$m_{\chi, {\rm in}} = 2$ GeV},
 $b_1$ has been tuned to be 1.33 keV
 for the Si target
 and only 0.39 keV
 for the Ge target
 (remind that $\Qmin = 0.25$ keV)
 for all four initially different $b_1$.
 It is thus that
 the reconstructed WIMP masses
 as well as
 the statistical uncertainty bounds
 are the same
 for these four cases.
 Furthermore,
 for our simulations
 with the input WIMP mass of $m_{\chi, {\rm in}} = 5$ GeV,
 $b_1$ has been tuned to be 7.79 keV
 for the Si target (for only the cases of $b_1 = 10$ and 20 keV)
 as well as
 3.42 keV (for however the cases of $b_1 = 5$, 10 and 20 keV)
 for the Ge target.
 In Table \ref{tab:Qmax_kin},
 we list
 the kinematic maximal cut--off energies
 of the nuclei used in our simulations presented in this paper
 for different WIMP masses
 between 2 and 20 GeV
 for reference%
\footnote{
 Remind that
 the given values
 depend on the maximal cut--off of
 the WIMP velocity distribution
 and in turn
 the escape velocity from our Galaxy
 and the Earth's velocity in the Galactic frame.
 In our simulations
 we have set $\ve = 1.05 \~ v_0 = 231$ km/s
 and $\vesc = 500$ km/s
 (i.e., $\vmax = 731$ km/s).
 For reference,
 we give also the values
 corresponding to $\vesc = 600$ km/s
 ($\vmax = 831$ km/s)
 in the parentheses.
}.
\begin{table}[t!]
\small
\begin{center}
\renewcommand{\arraystretch}{1.5}
\begin{tabular}{|| c | c   c | c   c   c   c   c ||}
\hline
\hline
 \makebox[1.3 cm][c]{Isotope}          &
 \makebox[0.8 cm][c]{$Z$}              & \makebox[0.8 cm][c]{$A$}              &
 \makebox[2.1 cm][c]{$\mchi =  2$ GeV} & \makebox[2.1 cm][c]{$\mchi =  5$ GeV} &
 \makebox[2.1 cm][c]{$\mchi = 10$ GeV} & \makebox[2.1 cm][c]{$\mchi = 15$ GeV} &
 \makebox[2.1 cm][c]{$\mchi = 20$ GeV} \\
\hline
\hline
\hline
 \multirow{2}{*}{$\rmXA{F}{19}$}    & \multirow{2}{*}{9}  & \multirow{2}{*}{19}  &
          2.17  &  10.23  &  27.45  &  44.30  &  59.22    \\
 &  &  & (2.81) & (13.22) & (35.48) & (57.25) & (76.54)   \\
\hline
 \multirow{2}{*}{$\rmXA{Na}{23}$}   & \multirow{2}{*}{11} & \multirow{2}{*}{23}  &
          1.86  &   9.14  &  25.82  &  43.23  &  59.39    \\
 &  &  & (2.41) & (11.81) & (33.37) & (55.86) & (76.75)   \\
\hline
 \multirow{2}{*}{$\rmXA{Si}{28}$}   & \multirow{2}{*}{14} & \multirow{2}{*}{28}  &
          1.58  &   8.04  &  23.85  &  41.38  &  58.44    \\
 &  &  & (2.04) & (10.39) & (30.82) & (53.47) & (75.52)   \\
\hline
 \multirow{2}{*}{$\rmXA{Ar}{40}$}   & \multirow{2}{*}{18} & \multirow{2}{*}{40}  &
          1.15  &   6.22  &  19.87  &  36.55  &  54.10    \\
 &  &  & (1.49) &  (8.03) & (25.68) & (47.23) & (69.91)   \\
\hline
 \multirow{2}{*}{$\rmXA{Ge}{76}$}   & \multirow{2}{*}{32} & \multirow{2}{*}{76}  &
          0.64  &   3.67  &  12.92  &  25.78  &  40.91    \\
 &  &  & (0.82) &  (4.75) & (16.70) & (33.32) & (52.87)   \\
\hline
 \multirow{2}{*}{$\rmXA{I}{127}$}   & \multirow{2}{*}{53} & \multirow{2}{*}{127} &
          0.39  &   2.32  &   8.57  &  17.85  &  29.48    \\
 &  &  & (0.50) &  (3.00) & (11.07) & (23.07) & (38.09)   \\
\hline
 \multirow{2}{*}{$\rmXA{Xe}{131}$}  & \multirow{2}{*}{54} & \multirow{2}{*}{131} &
          0.38  &   2.25  &   8.34  &  17.43  &  28.83    \\
 &  &  & (0.49) &  (2.91) & (10.78) & (22.52) & (37.26)   \\
\hline
 \multirow{2}{*}{$\rmXA{Xe}{136}$}  & \multirow{2}{*}{54} & \multirow{2}{*}{136} &
          0.36  &   2.18  &   8.08  &  16.92  &  28.06    \\
 &  &  & (0.47) &  (2.81) & (10.45) & (21.87) & (36.27)   \\
\hline
\hline
\end{tabular}
\end{center}
\caption{
 The kinematic maximal cut--off energies
 (in unit of keV)
 of the nuclei used in our simulations presented in this paper
 for different WIMP masses
 between 2 and 20 GeV.
 Remind that
 we have set $\ve = 1.05 \~ v_0 = 231$ km/s
 and the values given here
 (in the parentheses)
 are corresponding to
 $\vesc = 500$ km/s
 (i.e., $\vmax = 731$ km/s)
 and
 $\vesc = 600$ km/s
 (\mbox{$\vmax = 831$ km/s}),
 respectively.
}
\label{tab:Qmax_kin}
\end{table}

 It can be found
 in Fig.~\ref{fig:mchi-SiGe-500-0025}
 clearly that,
 for the input WIMP mass of $m_{\chi, {\rm in}} = 2$ GeV,
 the reconstructed mass
 is a bit overestimated.
 The main reason should be the following:
 the corresponding kinematic maximal cut--off energies
 are very low
 (see Table \ref{tab:Qmax_kin}):
 $Q_{\rm max, kin, Si} = 1.58$ keV and
 $Q_{\rm max, kin, Ge} = 0.64$ keV.
 Thus,
 for the germanium target,
 the threshold energy of \mbox{$\Qmin = 0.25$ keV}
 cuts almost 40\% of
 the theoretically analyzable energy range.
 According to the transformation
 given by Eq.~(\ref{eqn:vmin}),
 this means a minimal cut--off velocity of
 $\vmin^{\ast} = \vmin(0.25~{\rm keV}) \simeq 460$ km/s,
 which not only cuts $\sim$ 63\% of the considered velocity range
 ($0 \le v \le \vmax = 731$ km/s),
 but also far beyond the peak of our (input) velocity distribution function
 at $\simeq$ 310 km/s.
 Hence,
 our trianglar approximation
 for the velocity range of $v \le \vmin^{\ast}$
 for estimating
 the normalization constant $\calN$
 as well as
 the moments of the WIMP velocity distribution,
 $\expv{v^n}$,
 in Eqs.~(\ref{eqn:calN_sum_mod}) and (\ref{eqn:moments})
 can obviously not hold anymore.

 Similarly,
 due to the pretty small kinematic cut--off energy
 and the relatively high threshold energy,
 for the input WIMP mass of $m_{\chi, {\rm in}} = 5$ GeV,
 the reconstructed WIMP mass
 could also be somehow overestimated.
 However,
 it should be caused by
 the use of the logarithmically linear ansatz
 (\ref{eqn:dRdQn})
 for reconstructing $dR / dQ$:
 in all cases with the tuned (reduced) bin width
 discussed before,
 we have used in fact
 the whole analyzed energy range
 as the first bin
 to reconstruct the recoil spectrum,
 or,
 equivalently,
 to estimate the logarithmic slope $k_1$.
 This would be too wide
 and $k_1$ could thus be {\em underestimated}
 (see Fig.~1 of Ref.~\cite{DMDDf1v}).
 Consequently,
 $K_1(\Qmin)$ and $r^{\ast}(\Qmin)$
 defined in Eqs.~(\ref{eqn:Kn}) and (\ref{eqn:rmin_ast})
 would in turn be {\em overestimated}.
 Moreover,
 by increasing the initial bin width,
 only the used one for the Si target increases,
 whereas for the Ge target
 the used bin width has been kept as 3.42 keV,
 except of the smallest case of $b_1 = 2.5$ keV;
 it would therefore be the reason that
 the WIMP mass could still be reconstructed pretty precisely
 with $b_1 = 2.5$ keV
 but show a $b_1$--dependent overestimate.

 On the other hand,
 for the input WIMP masses $m_{\chi, {\rm in}} \gsim 10$ GeV,
 our simulations show that
 the algorithmic procedure with the modified
 $\calR_{n, (X, Y)}$ and $\calR_{\sigma, (X, Y)}$
 could reconstruct the WIMP masses pretty precisely
 with statistical uncertainties of $\sim$ 30\% to 45\%,
 regardless of the width of the first energy bin;
 except that,
 once WIMPs are heavier than $\sim 60$ GeV,
 the use of a small bin width of $\sim 2.5$ keV
 would overestimate the reconstructed masses and
 enlarge the statistical uncertainties.
 We will discuss about effects of taking
 different widths of the first energy bin
 in more detail
 in Sec.~\ref{sec:mchi-500-100}.

\subsubsection{Reconstruction of the SI WIMP--nucleon coupling}
\label{sec:fp2-500-0025}

 We continue to present our simulation results
 of the reconstruction of the (squared)
 SI (scalar) WIMP--nucleon coupling $|\frmp|^2$
 estimated by Eq.~(\ref{eqn:fp2})
 for
 the input WIMP masses
 between 2 and 200 GeV.
 Note that,
 since the WIMP mass could (in principle) be reconstructed pretty well,
 we show here
 only the reconstructed $|\frmp|^2$ with the true (input) WIMP masses.

\begin{figure}[p!]
\begin{center}
\vspace{-0.5cm}
 \includegraphics[width = 15 cm]{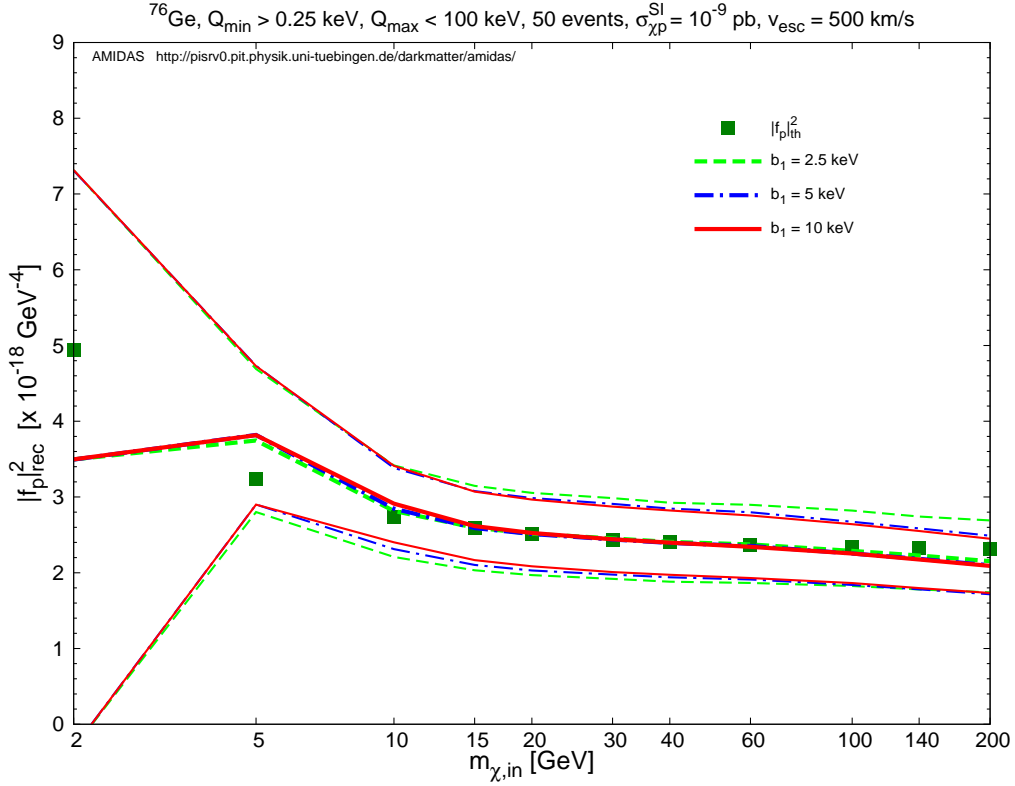} \\ \vspace{ 0.25cm}
 \includegraphics[width = 15 cm]{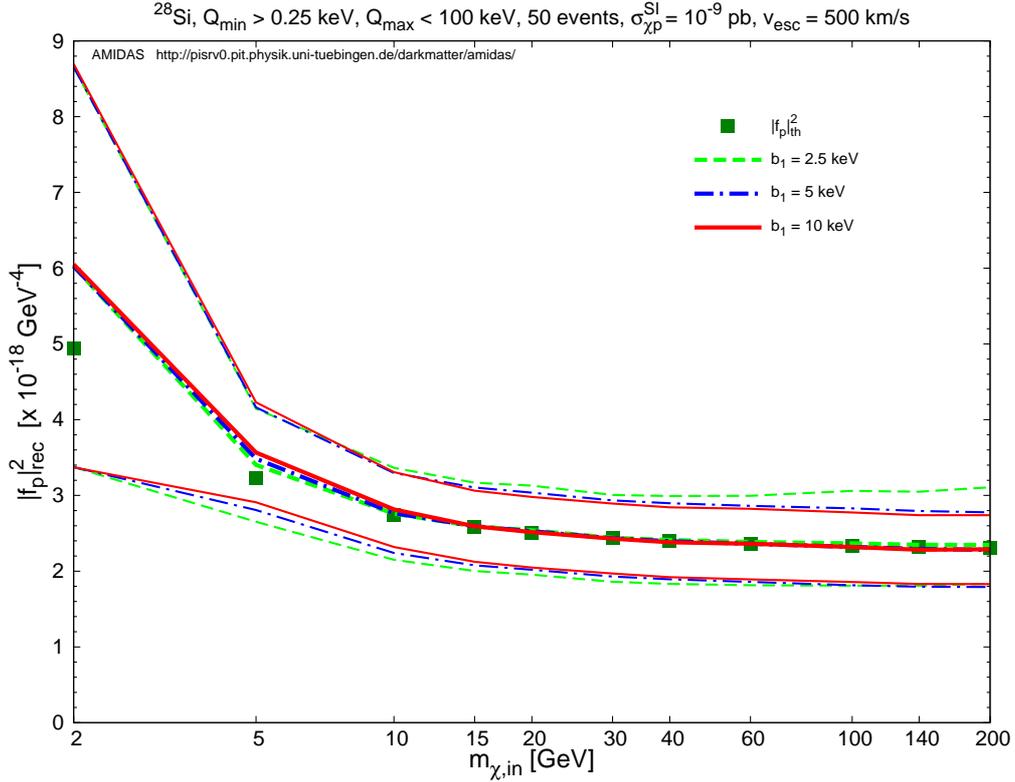} \\ \vspace{-0.5 cm}
\end{center}
\caption{
 The reconstructed squared SI (scalar) WIMP--nucleon couplings
 and the lower and upper bounds of
 the 1$\sigma$ statistical uncertainties
 as functions of the input WIMP mass
 with a $\rmXA{Ge}{76}$ (upper)
 or a $\rmXA{Si}{28}$ (lower) target
 for input WIMP masses between 2 and 200 GeV.
 The required WIMP mass has been assumed
 to be the true (input) values.
 {\em Three} different widths of the first energy bin
 $b_1$
 (before tuning)
 have been used:
 2.5 keV (dashed green),
 5 keV (dash--dotteded blue),
 and 10 keV (solid red).
 The squared dark--green points
 indicate the true (input) WIMP masses
 and the theoretically estimated $|\frmp|^2$ values.
 Each experiment contains 50 events on average.
 The SI WIMP--nucleus cross section
 has been fixed to be
 $\sigmapSI = 10^{-9}$ pb.
}
\label{fig:fp2-500-0025-input}
\end{figure}

 In Figs.~\ref{fig:fp2-500-0025-input},
 we show
 the reconstructed squared SI WIMP--nucleon couplings
 and the lower and upper bounds of
 the 1$\sigma$ statistical uncertainties
 as functions of the input WIMP mass
 with a $\rmXA{Ge}{76}$ (upper)
 and a $\rmXA{Si}{28}$ (lower) target separately.
 Three different widths of the first energy bin
 $b_1$
 (before tuning)
 have been used:
 2.5 keV (dashed green),
 5 keV (dash--dotteded blue),
 and 10 keV (solid red).

 As discussed previously,
 for the input WIMP mass of $m_{\chi, {\rm in}} = 2$ GeV,
 the corresponding kinematic maximal cut--off energies are
 $Q_{\rm max, kin, Si} = 1.58$ keV and
 $Q_{\rm max, kin, Ge} = 0.64$ keV,
 respectively.
 This means that,
 while for the germanium target
 the threshold energy of \mbox{$\Qmin = 0.25$ keV}
 cuts almost 40\% of the
 theoretically analyzable energy range,
 for the silicon target,
 only 16\% of the energy range
 has been cut.
 Consequently,
 our simulations shown in Figs.~\ref{fig:fp2-500-0025-input}
 demonstrate clearly that,
 once the priorly known or reconstructed WIMP mass is pretty light,
 $|\frmp|^2$ could be (strongly) underestimated
 by using data with {\em heavy} target nuclei (Ge or Xe),
 but (much) better reconstructed
 with {\em light} target nuclei (Si and Ar).

 On the other hand,
 for the input WIMP masses $m_{\chi, {\rm in}} \gsim 10$ GeV,
 the SI WIMP--nucleon coupling
 could be reconstructed very precisely
 with statistical uncertainties of $\sim$ 20\% to 30\%,
 by using not only the Ge and Si targets,
 but in fact also other available nuclei,
 e.g.~Ar or Xe.
 However,
 as already shown in Ref.~\cite{DMDDfp2},
 for input WIMP masses of $m_{\chi, {\rm in}} \gsim 100$ GeV,
 $|\frmp|^2$ could be a bit underestimated
 with heavy (Ge and Xe) nuclei.
 More comparisons between results
 reconstructed with light (Si) and heavy (Ge) target nuclei
 when the threshold energy increases
 will be given in Sec.~\ref{sec:fp2-500-100}.

\subsubsection{Reconstruction of the ratio between the SD WIMP--nucleon couplings}
\label{sec:ranap-500-0025}

 Now
 we consider the ratio between two SD WIMP--nucleon couplings
 $\armn / \armp$
 with
 both of the estimators given in
 Eqs.~(\ref{eqn:ranapSD}) and (\ref{eqn:ranapSISD}).
 Only the combination of the $\rmXA{F}{19}$ +  $\rmXA{I}{127}$ nuclei
 has been used in our simulations,
 since a much wider range of the $\armn / \armp$ ratio
 (between $\pm 4$)
 can (in principle) be well reconstructed
 \cite{DMDDranap}.
 Remind that,
 as discussed in Ref.~\cite{DMDDranap} and Sec.~\ref{sec:ranap},
 for the F + I target combination,
 one should take
 the ``$-$ (minus)'' solution
 for both of Eqs.~(\ref{eqn:ranapSD}) and Eqs.~(\ref{eqn:ranapSISD}).
 Note also that,
 in all simulations presented
 in this paper,
 a non--zero SI WIMP--nucleon cross section
 $\sigmapSI = 10^{-9}$ pb
 has always been taken into account%
\footnote{
 Remind here that
 Eq.~(\ref{eqn:ranapSD}) has however been derived
 by considering only
 the SD WIMP--nucleus cross section.
}.
\begin{figure}[p!]
\begin{center}
\vspace{-0.5cm}
 \includegraphics[width = 15 cm]{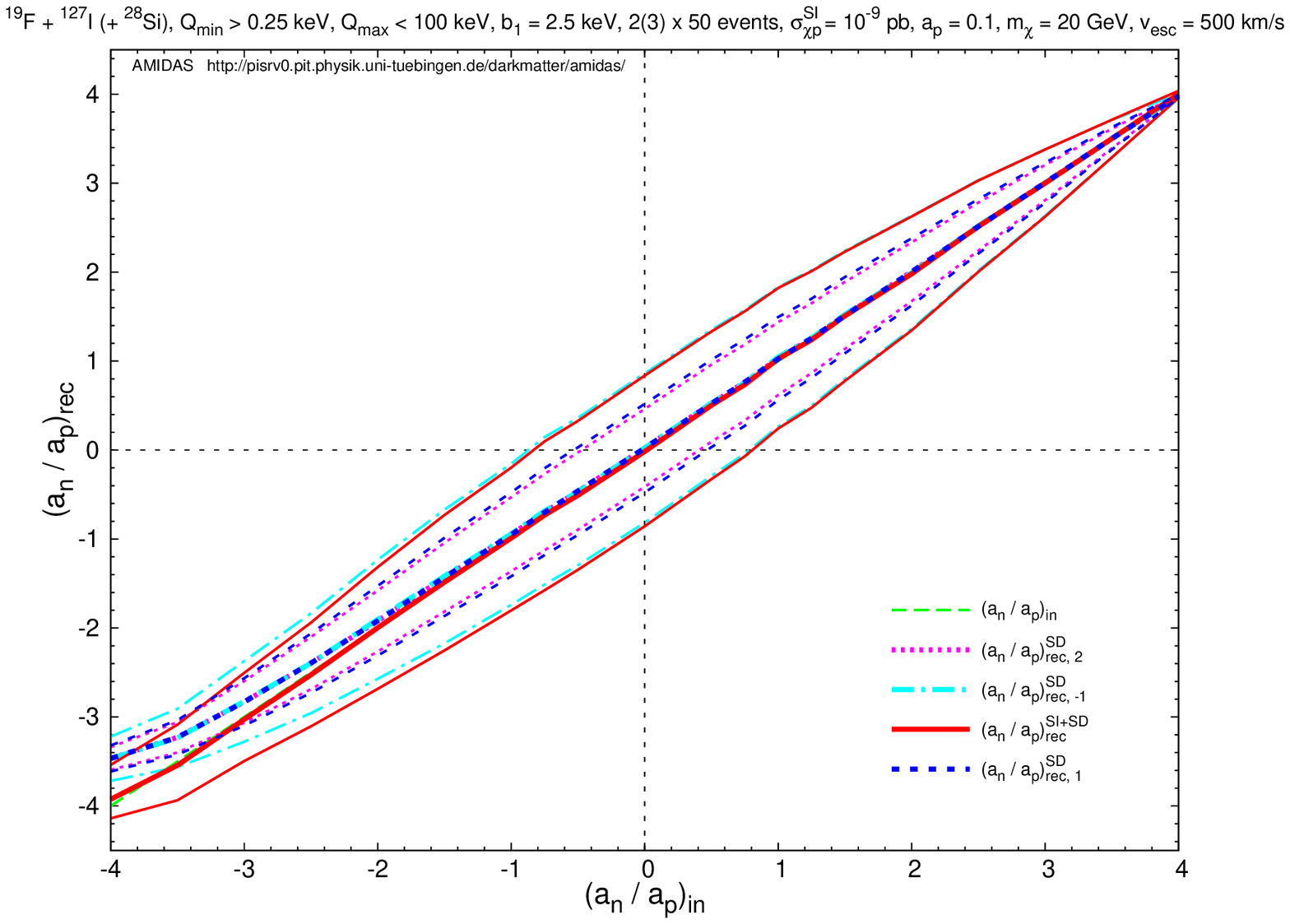} \\ \vspace{ 0.25cm}
 \includegraphics[width = 15 cm]{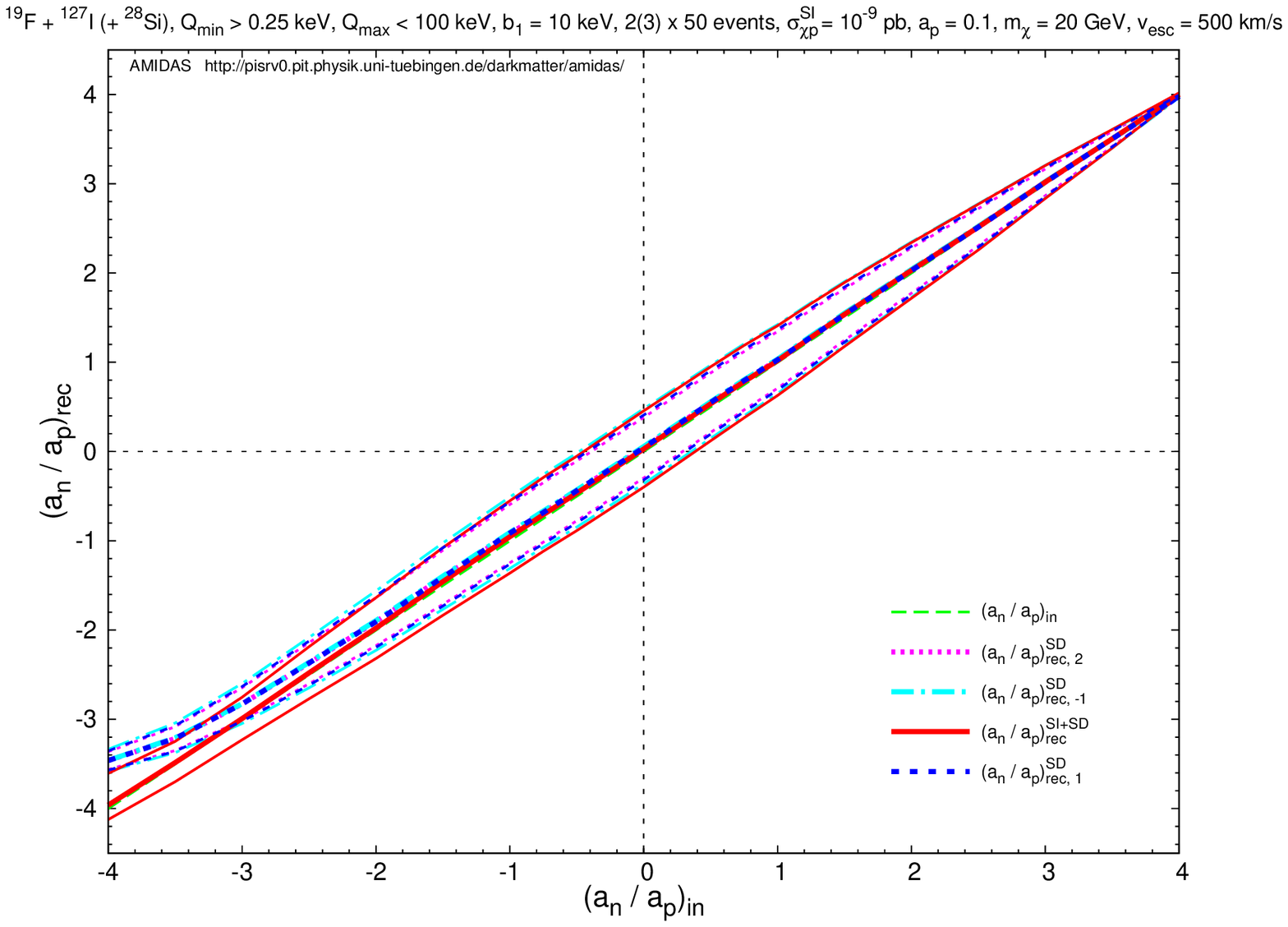} \\ \vspace{-0.5 cm}
\end{center}
\caption{
 The reconstructed $\armn / \armp$ ratios
 estimated by Eq.~(\ref{eqn:ranapSD})
 and their lower and upper bounds of
 the 1$\sigma$ statistical uncertainties
 with $n = -1$ (dash--dotted cyan),
 1 (long--dotted blue),
 and 2 (dotted magenta)
 as well as
 those estimated by Eq.~(\ref{eqn:ranapSISD})
 (solid red)
 as functions of the input $\armn / \armp$ ratio
 with the $\rmXA{F}{19}$ + $\rmXA{I}{127}$ target combination
 and $\rmXA{Si}{28}$ as the third spinless target
 for input $\armn / \armp$ ratios between $\pm 4$.
 The mass of incident WIMPs
 has been set as 20 GeV.
 Remind that
 a non--zero SI WIMP--nucleon cross section
 $\sigmapSI = 10^{-9}$ pb
 has been included.
 A relatively small bin width of $b_1 = 2.5$ keV (upper)
 and a wider width of $b_1 = 10$ keV (lower)
 (before tuning)
 have been used.
 Each experiment contains 50 events on average.
}
\label{fig:ranap-ranap-FI-500-0025}
\end{figure}
\begin{figure}[p!]
\begin{center}
 \includegraphics[width = 15 cm]{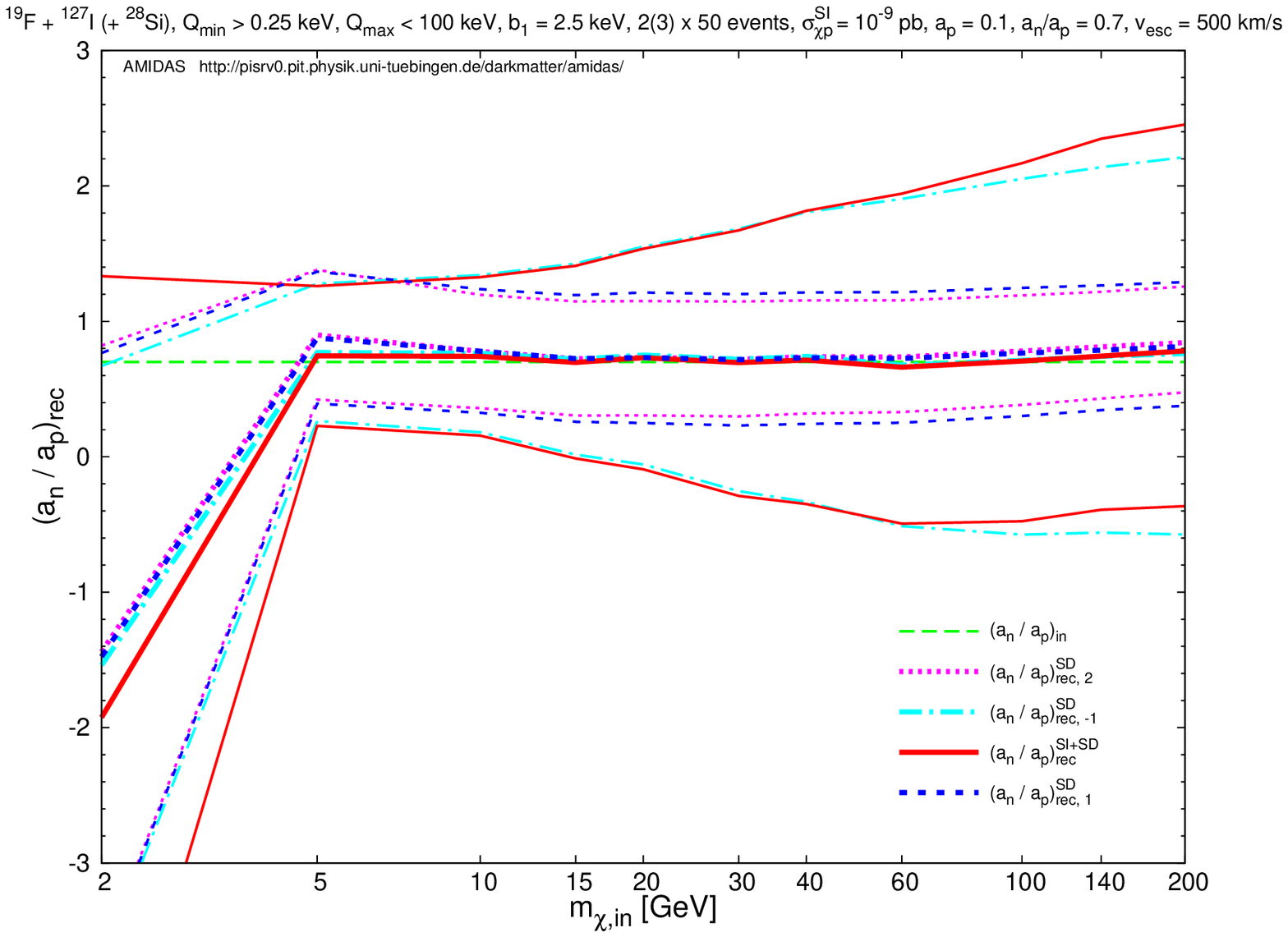} \\ \vspace{ 0.75cm}
 \includegraphics[width = 15 cm]{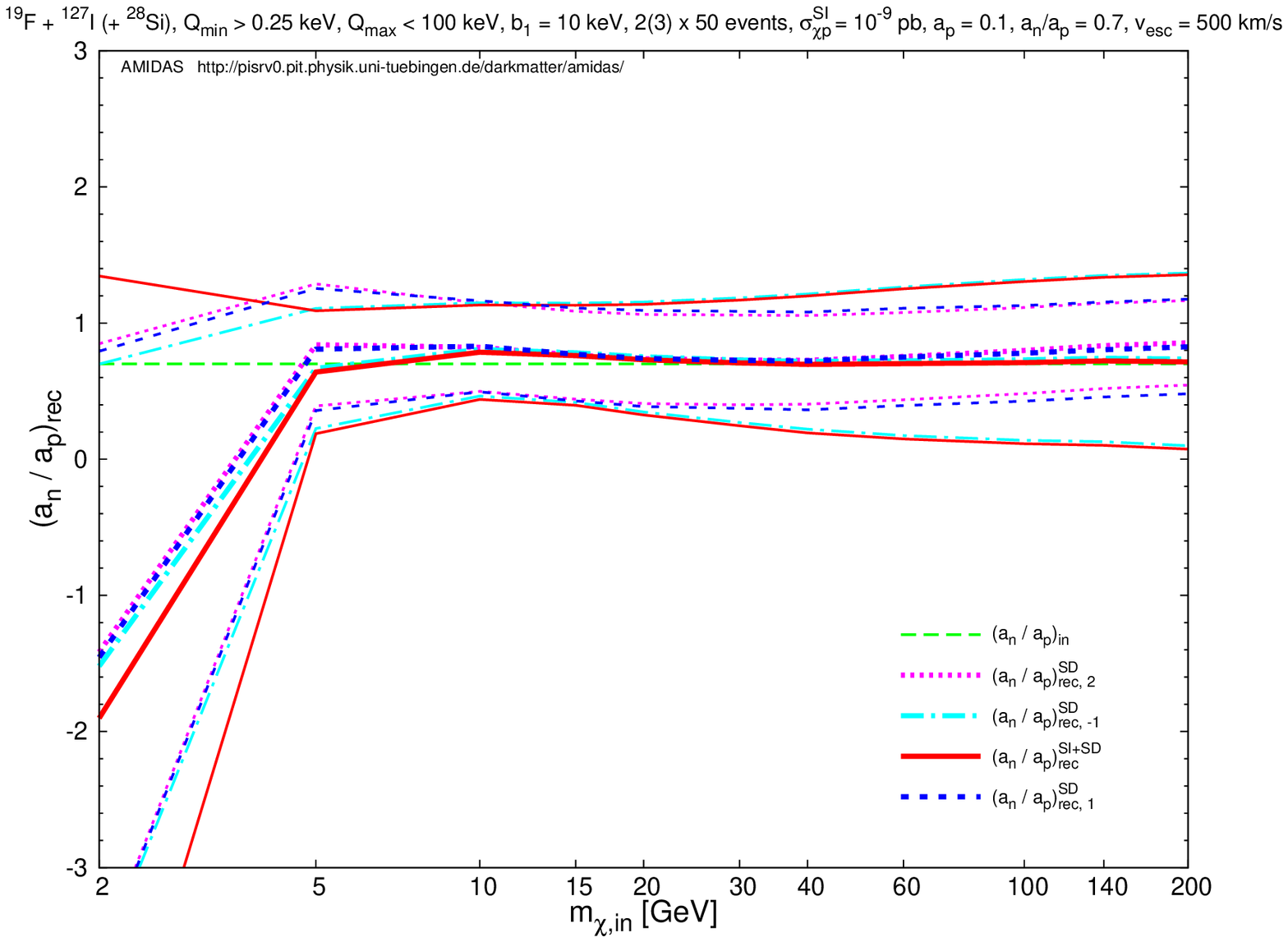}
\end{center}
\caption{
 The reconstructed $\armn / \armp$ ratios
 and their lower and upper bounds of
 the 1$\sigma$ statistical uncertainties
 as functions of the input WIMP mass
 with the $\rmXA{F}{19}$ + $\rmXA{I}{127}$ target combination
 and $\rmXA{Si}{28}$ as the third spinless target
 for input WIMP masses between 2 and 200 GeV.
 The input $\armn / \armp$ ratio
 has been fixed as 0.7.
 Other parameters and all notations are
 the same as in Figs.~\ref{fig:ranap-ranap-FI-500-0025}.
}
\label{fig:ranap-mchi-FI-500-0025}
\end{figure}

 In Figs.~\ref{fig:ranap-ranap-FI-500-0025},
 we show at first
 the reconstructed $\armn / \armp$ ratios
 estimated by Eq.~(\ref{eqn:ranapSD})
 and their lower and upper bounds of
 the 1$\sigma$ statistical uncertainties
 with $n = -1$ (dash--dotted cyan),
 1 (long--dotted blue),
 and 2 (dotted magenta)
 as well as
 those estimated by Eq.~(\ref{eqn:ranapSISD})
 (solid red)
 as functions of the input $\armn / \armp$ ratio
 with a $\rmXA{F}{19}$ + $\rmXA{I}{127}$ target combination
 and $\rmXA{Si}{28}$ as the third spinless target
 for input $\armn / \armp$ ratios between $\pm 4$.
 The mass of incident WIMPs
 has been set as 20 GeV.
 A relatively small bin width of $b_1 = 2.5$ keV (upper)
 and a wider width of $b_1 = 10$ keV (lower)
 (before tuning)
 have been used.

 It can be found
 in the upper frame of Figs.~\ref{fig:ranap-ranap-FI-500-0025}
 that,
 although
 the reconstructed $\armn / \armp$ ratios
 given by Eq.~(\ref{eqn:ranapSD})
 with three different $n$
 are slightly overestimated
 at the end closed to $\Srmp_{\rm I} / \Srmn_{\rm I} = -4.12$
 (due to including the non--zero SI WIMP--nucleus cross section
  \cite{DMDDranap}),
 not surprisingly,
 the ratio
 given by Eq.~(\ref{eqn:ranapSISD})
 could precisely match all the true (input) values.
 However,
 the 1$\sigma$ statistical uncertainties
 estimated with $n = -1$
 and from the general Eq.~(\ref{eqn:ranapSISD})
 are almost twice as large as
 those with $n = 1$ and 2.
 Nevertheless,
 by increasing the width of the first energy bin,
 the (difference between the) 1$\sigma$ statistical uncertainties
 by using both estimators
 could be reduced (significantly).

 On the other hand,
 in Figs.~\ref{fig:ranap-mchi-FI-500-0025},
 we show
 the reconstructed $\armn / \armp$ ratios
 and their 1$\sigma$ statistical uncertainties
 as functions of the input WIMP mass
 with the F + I (+ Si) target combination
 for input WIMP masses between 2 and 200 GeV.
 The input $\armn / \armp$ ratio
 has been fixed as 0.7.

 First,
 as discussed before,
 for the lightest input WIMP mass of $m_{\chi, {\rm in}} = 2$ GeV,
 the kinematic maximal cut--off energy of the iodine target
 is only 0.39 keV
 and
 the threshold energy of \mbox{$\Qmin = 0.25$ keV}
 cuts thus more than 60\% of
 the theoretically analyzable energy range.
 Consequently,
 the reconstructed $\armn / \armp$ ratios
 are strongly underestimated.
 Nevertheless,
 for all larger input WIMP masses $m_{\chi, {\rm in}} \gsim 5$ GeV,
 the reconstructions of the $\armn / \armp$ ratio
 become to match the true (input) values very well.
 Furthermore,
 two plots in Figs.~\ref{fig:ranap-mchi-FI-500-0025}
 show more clearly that,
 with the use of a small bin width of $b_1 = 2.5$ keV,
 the 1$\sigma$ statistical uncertainties
 with $n = 1$ and 2
 are approximately equal
 and (much smaller) than
 the uncertainties estimated with $n = -1$
 and from the general Eq.~(\ref{eqn:ranapSISD}),
 especially
 for the large input WIMP masses.
 And similarly,
 by increasing the width of the first energy bin,
 the (difference between the) uncertainties
 (with different $n$)
 could be reduced (significantly);
 the larger the true (input) WIMP mass,
 the more significantly the uncertainty could be reduced
 and,
 (with a bin width as large as $b_1 \simeq 10$ keV),
 the uncertainties would (almost) be
 independent of the true (input) WIMP mass.

\subsubsection{Reconstructions of the ratios between the SD and SI WIMP--nucleon cross sections}
\label{sec:rsigma-500-0025}

 We present further
 the simulation results of the reconstructions of
 the ratios of the SD and SI WIMP cross sections
 on protons and on neutrons,
 $\sigma_{\chi ({\rm p, n})}^{\rm SD} / \sigmapSI$.
 Note that,
 first,
 for using Eqs.~(\ref{eqn:rsigmaSDpSI}) and (\ref{eqn:rsigmaSDnSI}),
 the needed $\armn / \armp$ ratio
 has been reconstructed
 only by Eq.~(\ref{eqn:ranapSISD}).
 Second,
 in all plots
 the width of the first energy bin
 (before tuning)
 has been fixed as $b_1 = 10$ keV.

\begin{figure}[p!]
\begin{center}
\vspace{-0.5cm}
 \includegraphics[width = 15 cm]{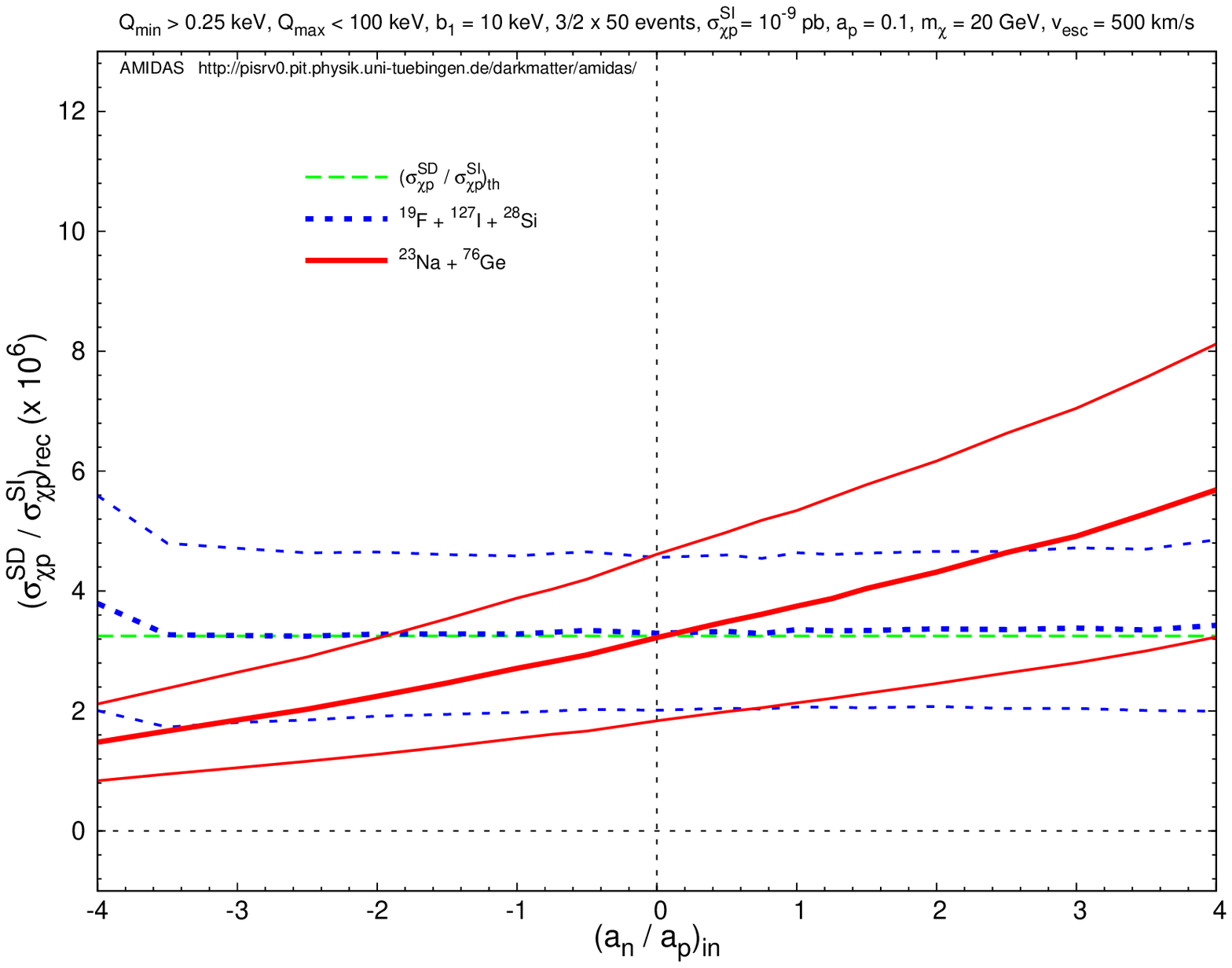} \\ \vspace{ 0.25cm}
 \includegraphics[width = 15 cm]{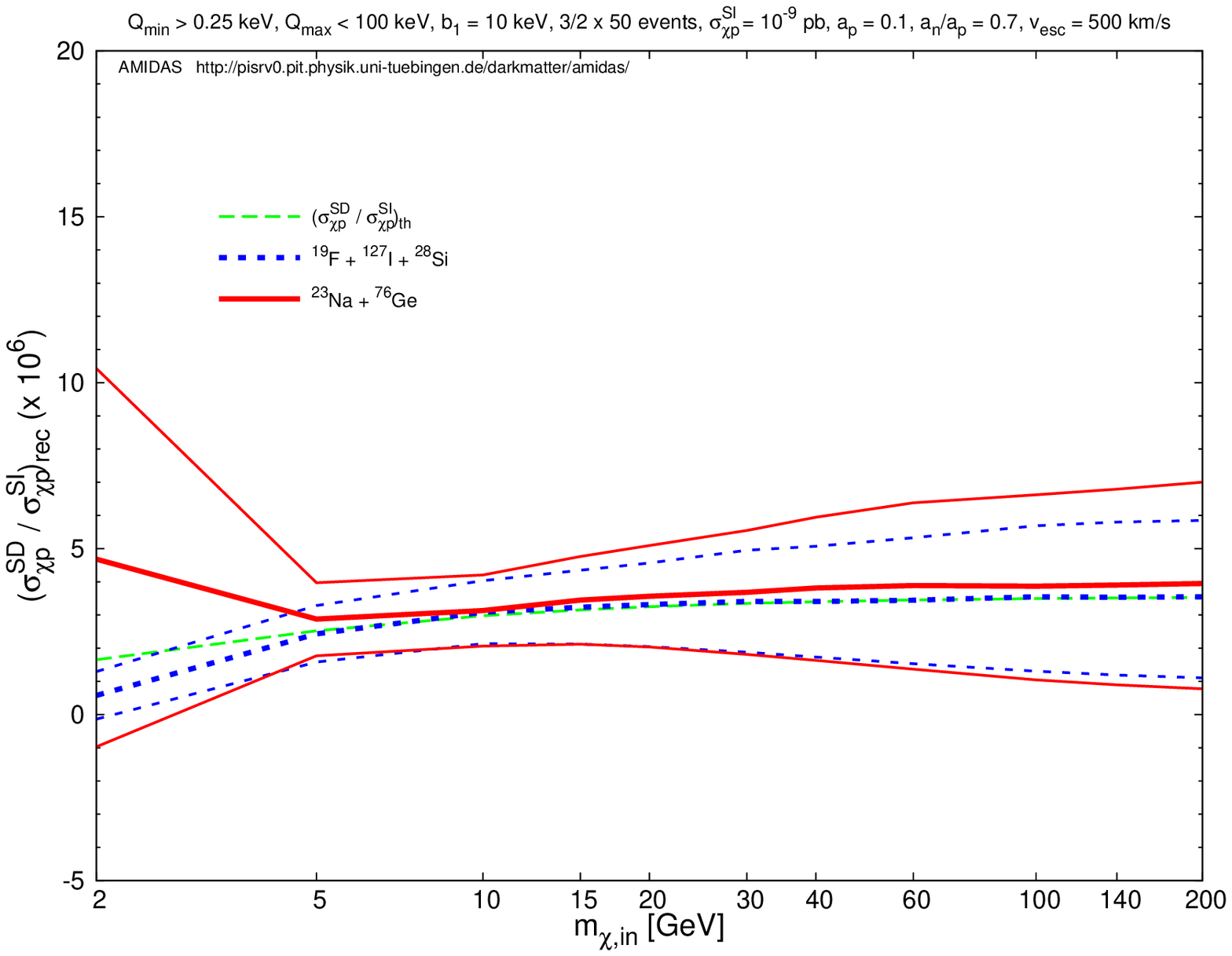}  \\ \vspace{-0.25cm}
\end{center}
\caption{
 The reconstructed $\sigmapSD / \sigmapSI$ ratios
 estimated by Eq.~(\ref{eqn:rsigmaSDpSI})
 and their lower and upper bounds of
 the 1$\sigma$ statistical uncertainties
 with the $\rmXA{F}{19}$ + $\rmXA{I}{127}$ + $\rmXA{Si}{28}$ target combination
 (long--dotted blue)
 as well as
 by Eq.~(\ref{eqn:rsigmaSDpSI_even})
 with the $\rmXA{Na}{23}$ + $\rmXA{Ge}{76}$ target combination
 (solid red).
 Upper:
 with a fixed input WIMP mass of $m_{\chi, {\rm in}} = 20$ GeV
 as functions of the input $\armn / \armp$ ratio
 between $\pm 4$;
 lower:
 with a fixed input ratio of $\armn / \armp = 0.7$
 as functions of the input WIMP mass
 between 2 and 200 GeV.
 Only the $\armn / \armp$ ratio
 reconstructed by Eq.~(\ref{eqn:ranapSISD})
 has been included
 and the width of the first energy bin
 (before tuning)
 has been fixed as $b_1 = 10$ keV.
 Other parameters are
 the same as in Figs.~\ref{fig:ranap-ranap-FI-500-0025}
 and \ref{fig:ranap-mchi-FI-500-0025}.
}
\label{fig:rsigmaSDpSI-FI-500-0025-100}
\end{figure}
\begin{figure}[p!]
\begin{center}
\vspace{-0.5cm}
 \includegraphics[width = 15 cm]{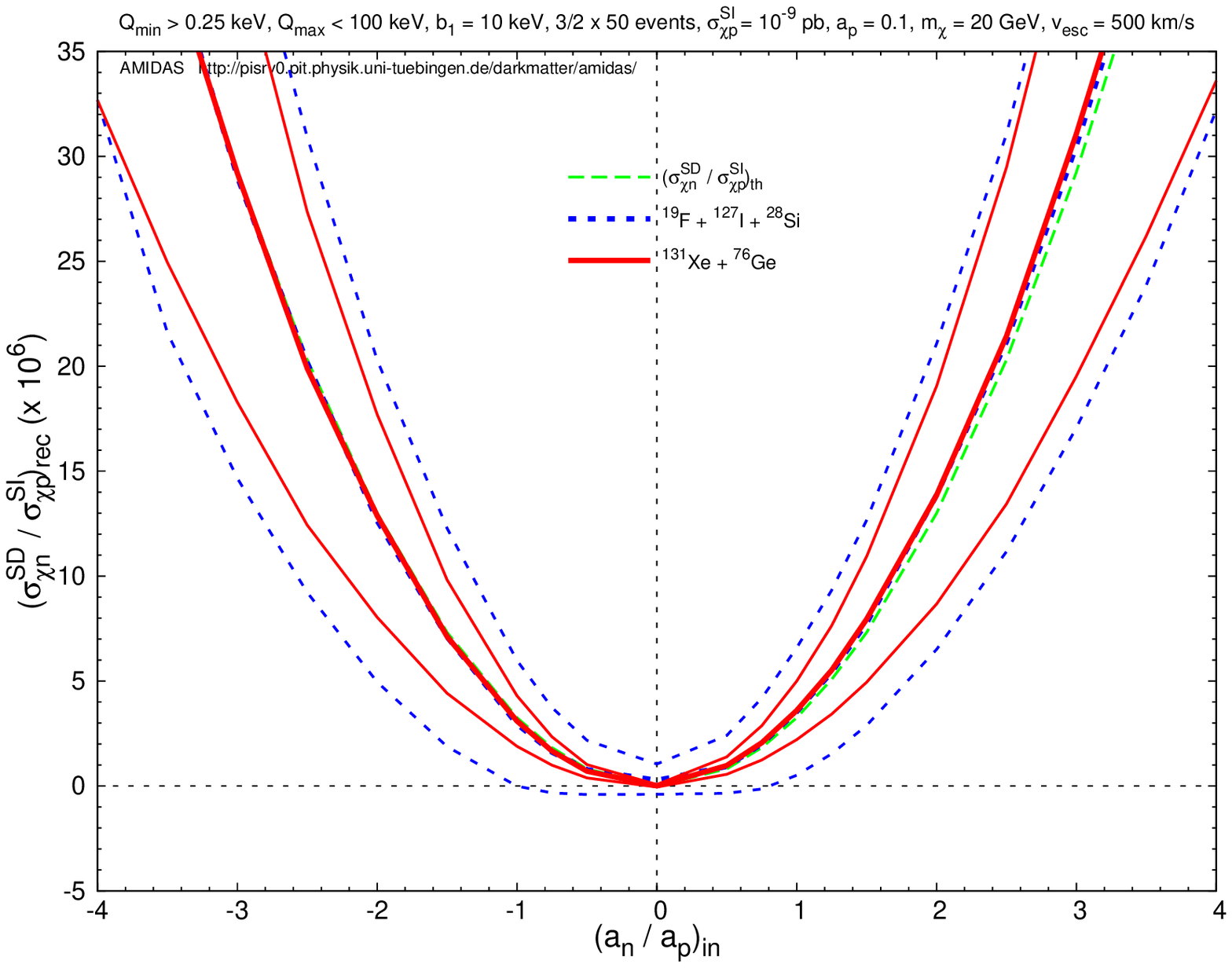} \\ \vspace{ 0.25cm}
 \includegraphics[width = 15 cm]{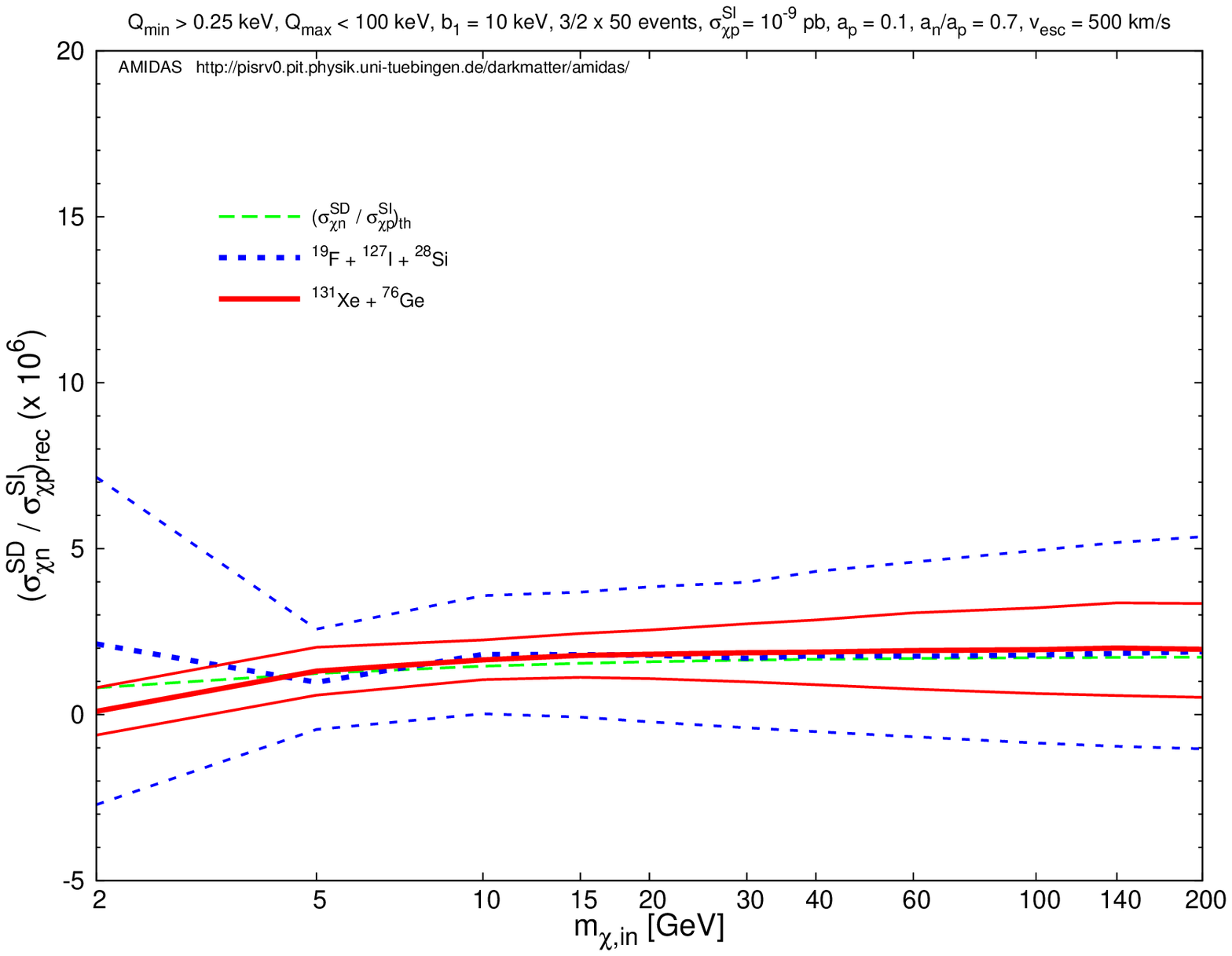}  \\ \vspace{-0.25cm}
\end{center}
\caption{
 The reconstructed $\sigmanSD / \sigmapSI$ ratios
 estimated by Eq.~(\ref{eqn:rsigmaSDnSI})
 and their lower and upper bounds of
 the 1$\sigma$ statistical uncertainties
 with the $\rmXA{F}{19}$ + $\rmXA{I}{127}$ + $\rmXA{Si}{28}$ target combination
 (long--dotted blue)
 as well as
 by Eq.~(\ref{eqn:rsigmaSDnSI_even})
 with the $\rmXA{Xe}{131}$ + $\rmXA{Ge}{76}$ target combination
 (solid red).
 Upper:
 with a fixed input WIMP mass of $m_{\chi, {\rm in}} = 20$ GeV
 as functions of the input $\armn / \armp$ ratio
 between $\pm 4$;
 lower:
 with a fixed input ratio of $\armn / \armp = 0.7$
 as functions of the input WIMP mass
 between 2 and 200 GeV.
 Only the $\armn / \armp$ ratio
 reconstructed by Eq.~(\ref{eqn:ranapSISD})
 has been included
 and the width of the first energy bin
 (before tuning)
 has been fixed as $b_1 = 10$ keV.
 Other parameters are
 the same as in Figs.~\ref{fig:ranap-ranap-FI-500-0025}
 and \ref{fig:ranap-mchi-FI-500-0025}.
}
\label{fig:rsigmaSDnSI-FI-500-0025-100}
\end{figure}

 In Figs.~\ref{fig:rsigmaSDpSI-FI-500-0025-100},
 we show
 the reconstructed $\sigmapSD / \sigmapSI$ ratios
 estimated by Eq.~(\ref{eqn:rsigmaSDpSI})
 and their lower and upper bounds of
 the 1$\sigma$ statistical uncertainties
 with the $\rmXA{F}{19}$ + $\rmXA{I}{127}$ + $\rmXA{Si}{28}$ target combination
 (long--dotted blue)
 as well as
 by Eq.~(\ref{eqn:rsigmaSDpSI_even})
 with a $\rmXA{Na}{23}$ + $\rmXA{Ge}{76}$ target combination
 (solid red).
 While
 in the upper frame,
 the $\sigmapSD / \sigmapSI$ ratios
 and their statistical uncertainties
 are reconstructed
 with a fixed input WIMP mass of $m_{\chi, {\rm in}} = 20$ GeV
 and given
 as functions of the input $\armn / \armp$ ratio
 between $\pm 4$,
 in the lower frame,
 the input ratio of $\armn / \armp$ is fixed as 0.7
 and the results are given
 as functions of the input WIMP mass
 between 2 and 200 GeV.

 It can be seen clearly that,
 in the wide range of interests of $|\armn / \armp| \le 4$,
 the $\sigmapSD / \sigmapSI$ ratios
 could be reconstructed by
 Eqs.~(\ref{eqn:rsigmaSDpSI})
 with the $\rmXA{F}{19}$ + $\rmXA{I}{127}$ + $\rmXA{Si}{28}$ target combination
 (much) better:
 except of the a--bit--overestimated one
 at the end of $(\armn / \armp)_{\rm in} = -4$,
 not only the $\sigmapSD / \sigmapSI$ ratios
 can be estimated very precisely,
 but also their statistical uncertainties
 are independent of the input $\armn / \armp$ ratios.
 On the other hand,
 except of
 the lightest input WIMP mass of \mbox{$m_{\chi, {\rm in}} = 2$ GeV},
 with a fixed $\armn / \armp$ ratio (lower frame),
 for the input WIMP masses of \mbox{$m_{\chi, {\rm in}} \gsim 5$ GeV},
 the $\sigmapSD / \sigmapSI$ ratios
 could be reconstructed by both combinations very well.
 Note however that,
 although the reconstructed results
 with the $\rmXA{Na}{23}$ + $\rmXA{Ge}{76}$ target combination
 shown in the lower frame
 look also pretty nice,
 as discussed above,
 this is only because that
 the input $\armn / \armp$ ratio is fixed as 0.7.
 As shown in the upper frame,
 once the true $\armn / \armp$ ratio
 exceeds the range of $|\armn / \armp| \lsim 1$,
 the reconstructions
 by the $\rmXA{Na}{23}$ + $\rmXA{Ge}{76}$ target combination
 would be (strongly) over-- or underestimated!

 On the other hand,
 in Figs.~\ref{fig:rsigmaSDnSI-FI-500-0025-100},
 we show
 the reconstructed $\sigmanSD / \sigmapSI$ ratios
 estimated by Eq.~(\ref{eqn:rsigmaSDnSI})
 and their lower and upper bounds of
 the 1$\sigma$ statistical uncertainties
 with the $\rmXA{F}{19}$ + $\rmXA{I}{127}$ + $\rmXA{Si}{28}$ target combination
 (long--dotted blue)
 as well as by Eq.~(\ref{eqn:rsigmaSDnSI_even})
 with a $\rmXA{Xe}{131}$ + $\rmXA{Ge}{76}$ target combination
 (solid red).

 Our results show that,
 although
 both combinations
 could offer well reconstructed $\sigmanSD / \sigmapSI$ ratios,
 with either a fixed WIMP mass or a fixed $\armn / \armp$ ratio,
 but,
 in contrast to the results
 shown in Figs.~\ref{fig:rsigmaSDpSI-FI-500-0025-100},
 the statistical uncertainties
 estimated with the $\rmXA{Xe}{131}$ + $\rmXA{Ge}{76}$ targets
 could now be much smaller (the half of) than
 those with $\rmXA{F}{19}$ + $\rmXA{I}{127}$ + $\rmXA{Si}{28}$ combination.
 Moreover,
 for the lightest input WIMP mass of $m_{\chi, {\rm in}} = 2$ GeV,
 while
 the reconstructed ratio
 with the $\rmXA{F}{19}$ + $\rmXA{I}{127}$ + $\rmXA{Si}{28}$ combination
 could now be overestimated,
 the one
 with the $\rmXA{Xe}{131}$ + $\rmXA{Ge}{76}$ targets
 could be underestimated.
 Nevertheless,
 as usual,
 for the input WIMP masses of $m_{\chi, {\rm in}} \gsim 5$ GeV,
 the $\sigmanSD / \sigmapSI$ ratio
 could be reconstructed very well
 with uncertainty depending only slightly
 on the true (input) WIMP mass
 by using both target combinations.

\subsection{Raising the threshold energy}

 In this section,
 we raise the threshold energy to
 $\Qmin = 2.5$ and even $\Qmin = 5$ keV
 and then
 repeat our simulations
 in order to confirm our observations
 discussed previously.

\subsubsection{Reconstruction of the WIMP mass}
\label{sec:mchi-500-100}
\begin{figure}[t!]
\begin{center}
 \includegraphics[width = 15 cm]{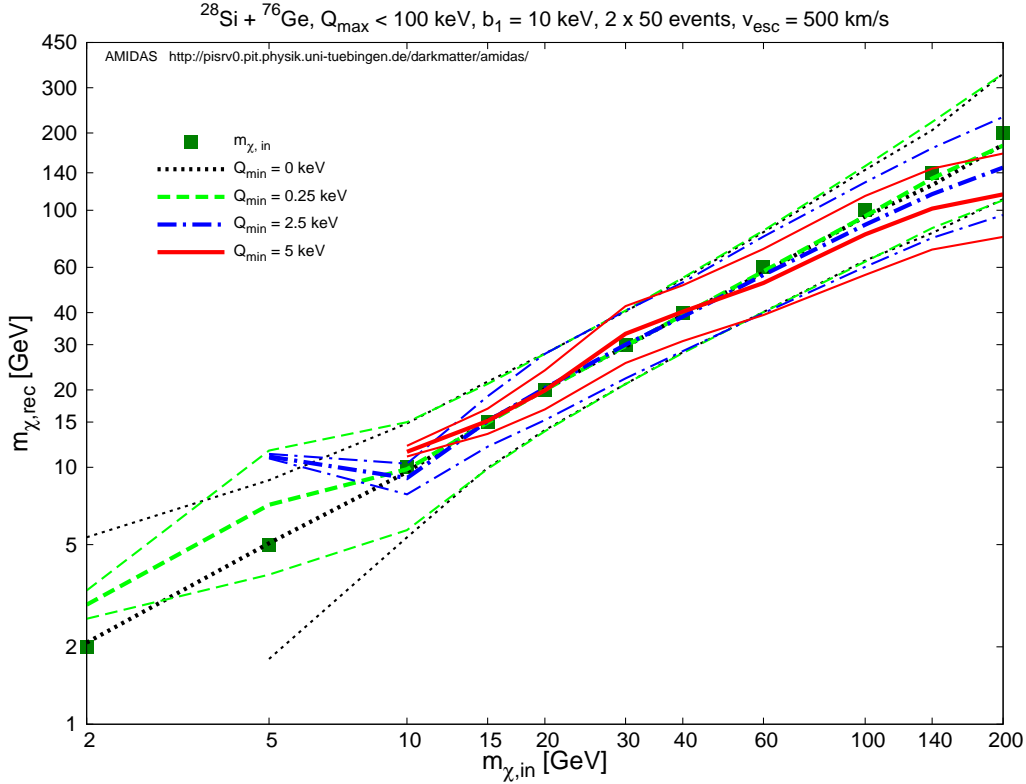}
\end{center}
\caption{
 As in Fig.~\ref{fig:mchi-SiGe-500-0025},
 except that
 three different minimal cut--off energies
 have been considered:
 $\Qmin = 0.25$ keV (dashed green),
 2.5 keV (dash--dotteded blue),
 and 5 keV (solid red).
 As a reference,
 the reconstructed masses
 and the statistical uncertainties
 with zero minimal cut--off energy $\Qmin = 0$
 (dotted black)
 have also been given here.
 The width of the first energy bin
 (before tuning)
 has been fixed as
 $b_1 = 10$ keV.
}
\label{fig:mchi-SiGe-500-100}
\end{figure}
\begin{figure}[p!]
\begin{center}
 \includegraphics[width = 15 cm]{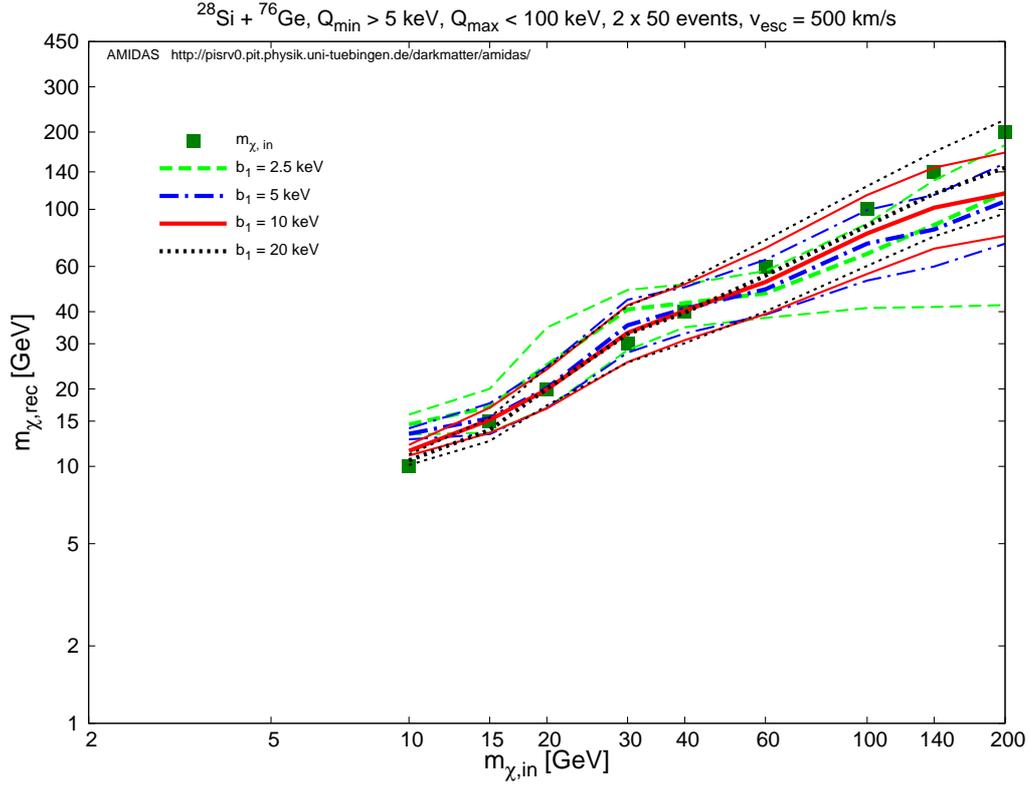} \\ \vspace{ 1 cm}
 \includegraphics[width = 15 cm]{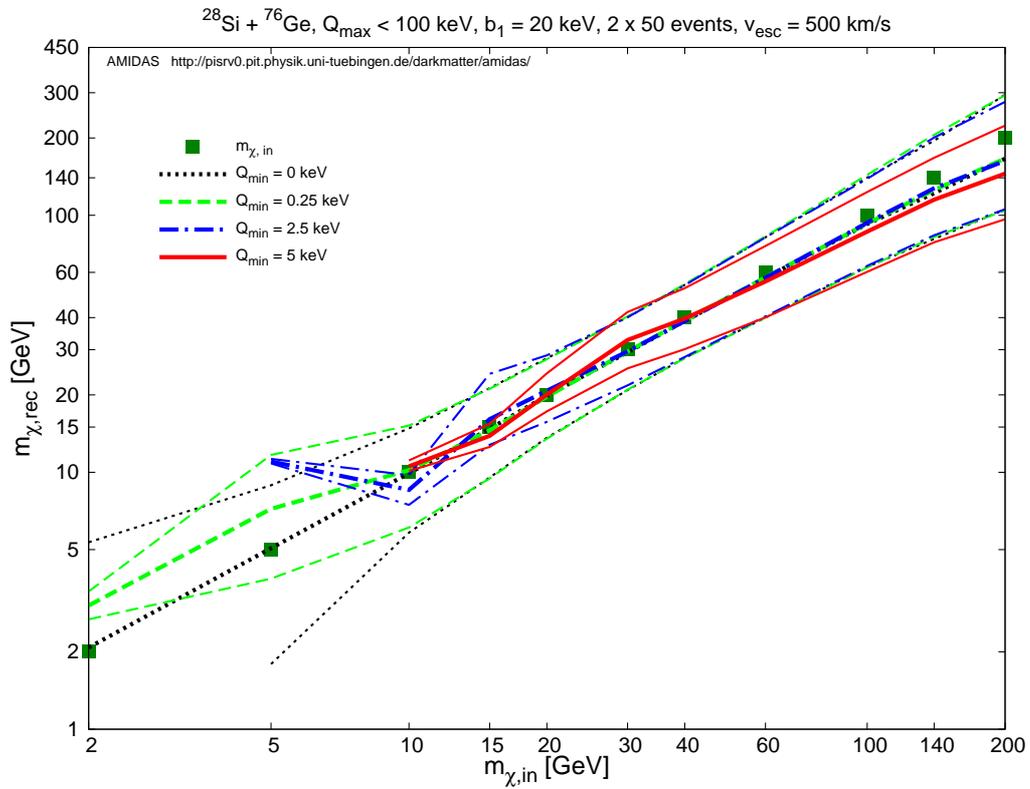}
\end{center}
\caption{
 Upper:
 as in Fig.~\ref{fig:mchi-SiGe-500-0025},
 except that
 the minimal cut--off energy
 has been raised to
 \mbox{$\Qmin = 5$ keV}.
 Lower:
 as in Fig.~\ref{fig:mchi-SiGe-500-100},
 except that
 the width of the first energy bin
 (before tuning)
 has been enlarged to
 $b_1 = 20$ keV.
}
\label{fig:mchi-SiGe-500-200}
\end{figure}

 In Fig.~\ref{fig:mchi-SiGe-500-100},
 we show
 the reconstructed WIMP masses
 and the 1$\sigma$ statistical uncertainty bounds
 with three different minimal cut--off energies together:
 $\Qmin = 0.25$ keV (dashed green),
 2.5 keV (dash--dotteded blue),
 and 5 keV (solid red).
 As a reference,
 the reconstructed masses and the statistical uncertainties
 with zero minimal cut--off energy $\Qmin = 0$
 (dotted black)
 have also been given.
 The width of the first energy bin
 (before tuning)
 has been fixed as
 $b_1 = 10$ keV.

 It can be found here that,
 as discussed in Sec.~\ref{sec:mchi-500-0025},
 for the input WIMP mass of $m_{\chi, {\rm in}} = 2$ or 5 GeV,
 the corresponding kinematic maximal cut--off energies
 for the Si and Ge targets
 are only
 $Q_{\rm max, kin, Si} = 1.58$ or 8.04 keV and
 $Q_{\rm max, kin, Ge} = 0.64$ or 3.67 keV.
 It is thus obviously that
 no WIMP events can be observed above
 an experimental threshold energy
 of $\Qmin = 2.5$ (or 5) keV
 with a Si or Ge target.
 Additionally,
 for the larger input WIMP masses of $m_{\chi, {\rm in}} = 5$ (and 10) GeV,
 the corresponding kinematic maximal cut--off energies
 for the Ge target
 are only $Q_{\rm max, kin, Ge} = 3.67$ (and 12.92) keV.
 Hence,
 the $\Qmin = 2.5$ (and 5) keV threshold energies
 cut 68\% (and 39\%) of
 the theoretically analyzable energy ranges.
 This results in overestimates of
 the reconstructed WIMP mass
 for input WIMP masses $m_{\chi, {\rm in}} \lsim 10$ GeV.

 On the other hand,
 for the input WIMP masses of $m_{\chi, {\rm in}} \gsim 50$ GeV,
 our simulations show that
 the higher the minimal cut--off energy,
 the more underestimate the reconstructed WIMP mass.
 In order to alleviate this underestimate,
 we consider then a much wider width of $b_1 = 20$ keV
 for the first energy bin
 in Figs.~\ref{fig:mchi-SiGe-500-200}.
 The upper frame of Figs.~\ref{fig:mchi-SiGe-500-200} shows
 the reconstructed WIMP masses
 by using data sets
 with a common minimal cut--off energy of $\Qmin = 5$ keV
 but four different widths of the first energy bin:
 2.5 keV (dashed green),
 5 keV (dash--dotteded blue),
 10 keV (solid red),
 and 20 keV (dotted black).
 It can be found obviously that,
 with a fixed threshold energy as high as $\Qmin = 5$ keV,
 the larger the width of the first energy bin,
 the higher (preciser) the reconstructed WIMP masses;
 the underestimate of the reconstructed WIMP masses
 between 60 and 200 GeV
 could indeed be alleviated.
 Meanwhile,
 we show
 in the lower frame of Figs.~\ref{fig:mchi-SiGe-500-200}
 the reconstructed WIMP masses
 with a larger fixed width of the first energy bin of $b_1 = 20$ keV
 for
 four different minimal cut--off energies:
 $\Qmin = 0$ (dotted black),
 0.25 keV (dashed green),
 \mbox{2.5 keV} (dash--dotteded blue),
 and 5 keV (solid red).
 Comparing with the results shown
 in Fig.~\ref{fig:mchi-SiGe-500-100},
 one can see clearly
 the improvement of the WIMP mass reconstruction.
 Additionally and more importantly,
 our simulations indicates that,
 the larger the minimal cut--off energy,
 the larger the alleviation
 by using a larger $b_1$
 could be.

 It would be reasonable to expect that
 the reconstruction of the WIMP mass
 (as well as
  the other properties)
 could be somehow (or even strongly) improved,
 once the true escape velocity of our Galaxy
 would be larger
 (than the value used in our presented simulations).
 However,
 according to our further simulations
 with a larger escape velocity of $\vesc = 600$ km/s,
 this expectation would unfortunately not be certainly true.
 Firstly,
 with a larger escape velocity and
 thus a larger kinematic cut--off
 on the velocity distribution
 as well as
 on the recoil spectra,
 the ``truncation'' problem
 discussed in Sec.~3.1 and, especially, above
 could indeed be alleviated a little bit
 and then
 the systematic deviations of the reconstructed results
 from the true (input) values
 could be reduced a little bit,
 but only a very little bit,
 much smaller than what one would expect.
 Actually,
 due to the exponential--like shape of the predicted recoil spectrum,
 in a data set with a few tens of total events
 at most {\em only one} extra event could ``occasionally'' be recorded
 in the extended high--energy range
 (between,
  e.g.,
  3.67 and 4.75 keV for the Ge target
  and the WIMP mass of 5 GeV,
  see Table \ref{tab:Qmax_kin}).
 On the other hand,
 as shown in Table \ref{tab:Qmax_kin},
 for e.g.~the I and Xe nuclei,
 the cut--offs of the kinematic energy
 could become (from lower to) higher than
 the experimental threshold energies
 and some WIMP events could thus be observed.
 However,
 as discussed earlier,
 since large parts of the theoretically analyzable energy ranges
 would still be cut by the relatively pretty high threshold energies,
 the reconstructed results would be (strongly) deviated
 from the true (input) values
 and not (very) reliable.

\subsubsection{Reconstruction of the SI WIMP--nucleon coupling}
\label{sec:fp2-500-100}
\begin{figure}[p!]
\begin{center}
 \includegraphics[width = 15 cm]{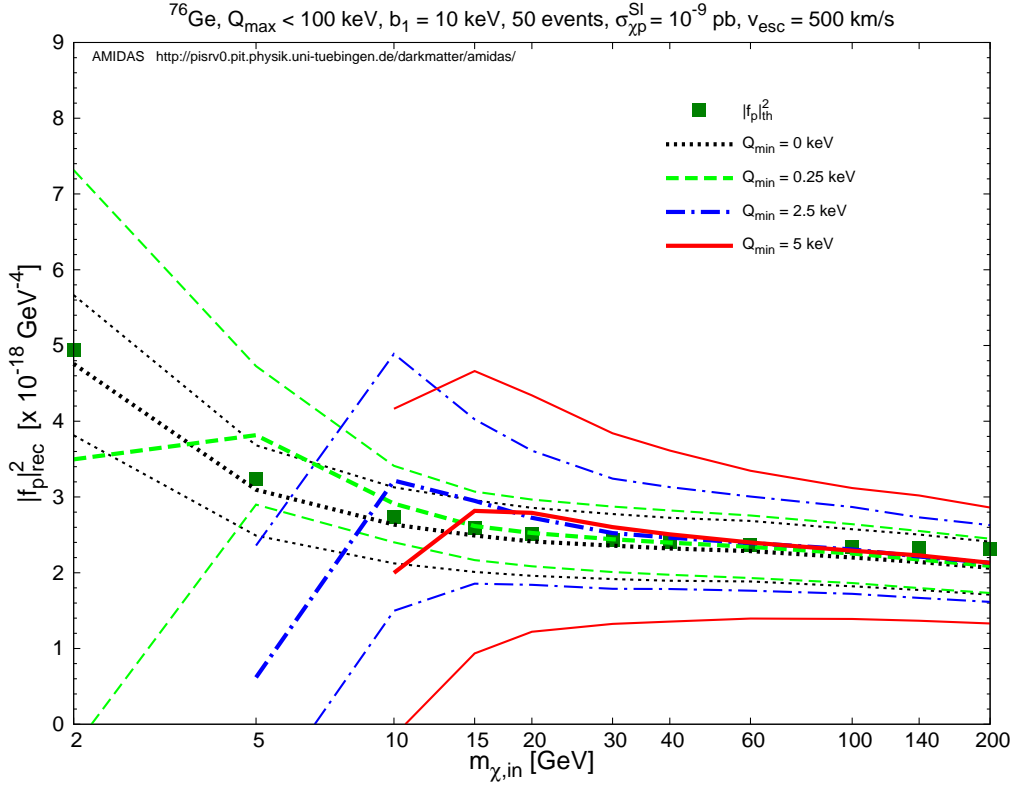} \\ \vspace{0.75cm}
 \includegraphics[width = 15 cm]{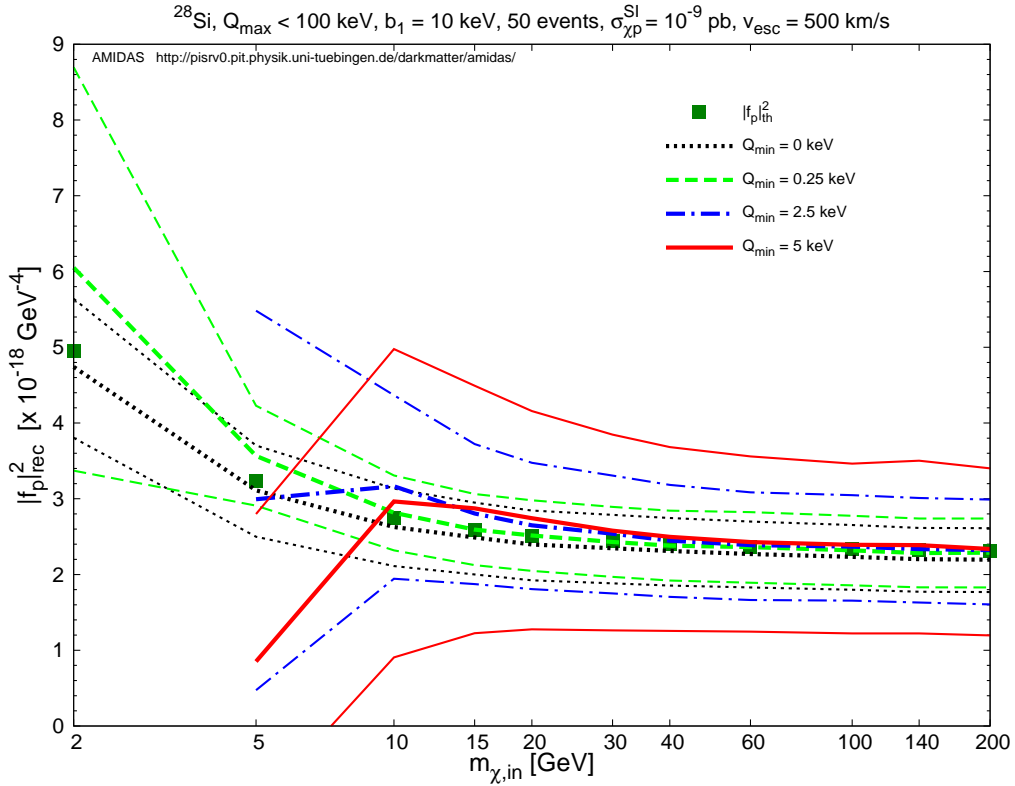}
\end{center}
\caption{
 As in Figs.~\ref{fig:fp2-500-0025-input},
 except that
 three different minimal cut--off energies
 have been considered:
 $\Qmin = 0.25$ keV (dashed green),
 2.5 keV (dash--dotteded blue),
 and 5 keV (solid red).
 As a reference,
 the reconstructed couplings
 and the statistical uncertainties
 with zero minimal cut--off energy $\Qmin = 0$
 (dotted black)
 have also been given here.
 Note that
 the width of the first energy bin
 (before tuning)
 has been fixed as
 $b_1 = 10$ keV.
}
\label{fig:fp2-500-100-input}
\end{figure}

 In Fig.~\ref{fig:fp2-500-100-input},
 we show
 the reconstructed (squared) SI WIMP--nucleon couplings
 and the 1$\sigma$ statistical uncertainty bounds
 with three different minimal cut--off energies together:
 $\Qmin = 0.25$ keV (dashed green),
 2.5 keV (dash--dotteded blue),
 and 5 keV (solid red).
 As a reference,
 the reconstructed couplings
 and the statistical uncertainties
 with zero minimal cut--off energy $\Qmin = 0$
 have also been given.
 Note that
 the width of the first energy bin
 (before tuning)
 has been fixed as
 $b_1 = 10$ keV.

 As discussed in Secs.~\ref{sec:fp2-500-0025} and \ref{sec:mchi-500-100},
 for the input WIMP masses of $m_{\chi, {\rm in}} = 2$ (and 5) GeV,
 no WIMP events can be observed above
 $\Qmin = 5$ (or 2.5) keV experimental cut--off energy
 by using a germanium (or silicon) target.
 Also,
 for the larger input WIMP masses of $m_{\chi, {\rm in}} = 5$ (and 10) GeV,
 large parts of the theoretically analyzable energy ranges
 would be cut by a threshold energy
 of $\Qmin = 2.5$ (or 5) keV
 and
 this results in the underestimates of
 the reconstructed SI coupling
 for input WIMP masses $m_{\chi, {\rm in}} \lsim 10$ GeV
 (for Ge, upper)
 or 5 GeV (for Si, lower).
 Nevertheless,
 for the input WIMP masses of $m_{\chi, {\rm in}} \gsim 10$ (15) GeV,
 the SI WIMP--proton couplings
 could always be reconstructed pretty precisely
 by using both of the Si (light) and Ge (heavy) targets.
 By increasing the minimal cut--off energy,
 one would only obtain larger statistical uncertainties
 on the reconstructed couplings.
 Moreover,
 as observed before,
 for the input WIMP masses of $m_{\chi, {\rm in}} \gsim 100$ GeV,
 the $|\frmp|^2$ couplings
 could be reconstructed more precisely
 with light (Si and Ar) targets
 then with heavy (Ge and Xe) targets,
 with however slightly larger statistical uncertainties.

\subsubsection{Reconstruction of the ratio between the SD WIMP--nucleon couplings}
\label{sec:ranap-500-100}
\begin{figure}[p!]
\begin{center}
\vspace{-0.75cm}
 \includegraphics[width = 15 cm]{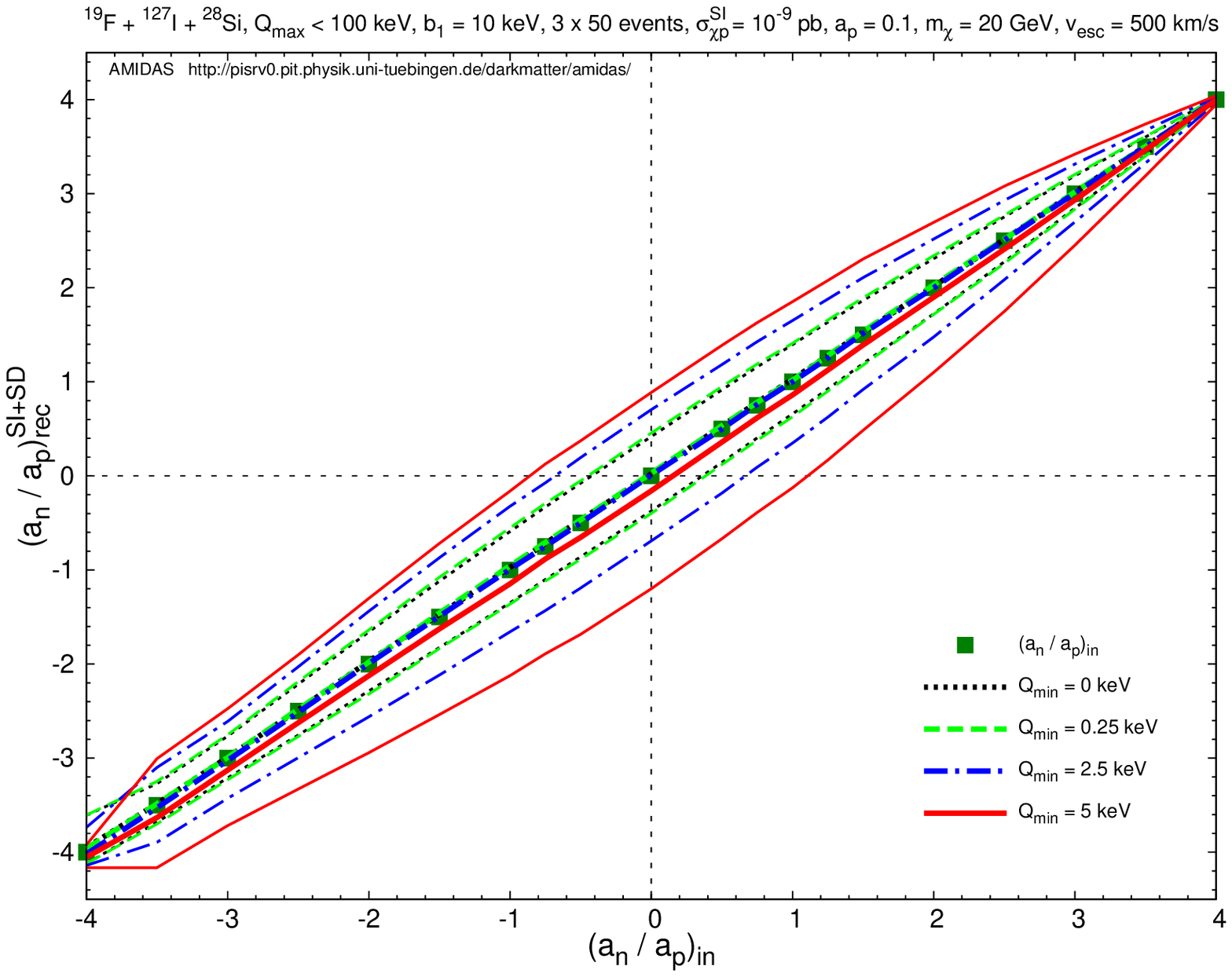} \\ \vspace{ 0.25cm}
 \includegraphics[width = 15 cm]{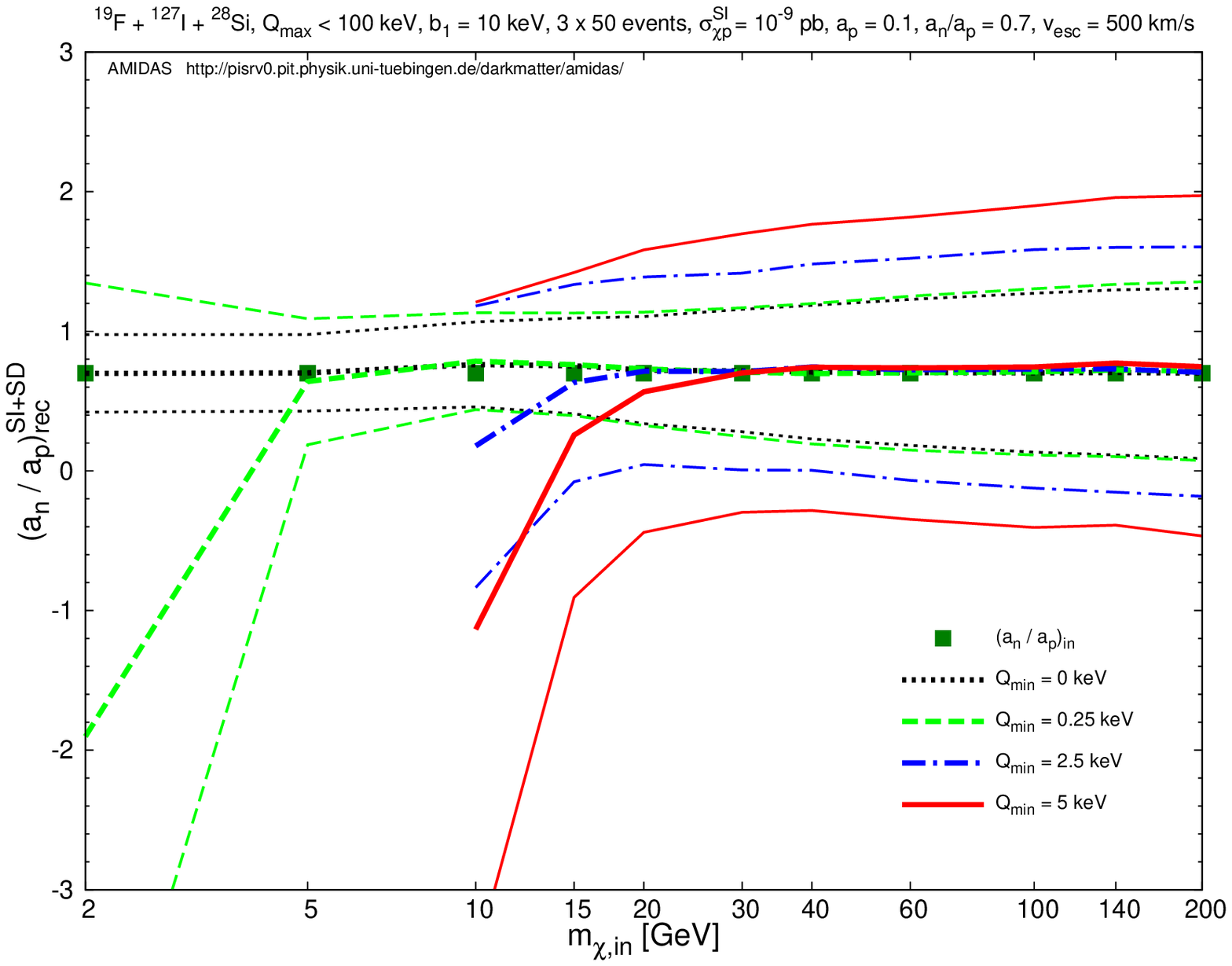}  \\ \vspace{-0.75cm}
\end{center}
\caption{
 The reconstructed $\armn / \armp$ ratios
 estimated only by Eq.~(\ref{eqn:ranapSISD})
 and the 1$\sigma$ statistical uncertainty bounds
 with the $\rmXA{F}{19}$ + $\rmXA{I}{127}$ + $\rmXA{Si}{28}$ target combination.
 Upper:
 the reconstructed $\armn / \armp$ ratios
 with a fixed input WIMP mass of $m_{\chi, {\rm in}} = 20$ GeV
 as functions of the input $\armn / \armp$ ratio
 between $\pm 4$;
 lower:
 with a fixed input ratio of $\armn / \armp = 0.7$
 as functions of the input WIMP mass
 between 2 and 200 GeV.
 The squared dark--green points
 indicate the true (input) (WIMP masses
 and) $\armn / \armp$ values.
 Four different minimal cut--off energies
 have been considered:
 $\Qmin = 0$ (dotted black),
 0.25 keV (dashed green),
 2.5 keV (dash--dotteded blue),
 and 5 keV (solid red).
 The width of the first energy bin
 (before tuning)
 has been fixed as
 $b_1 = 10$ keV.
 Other parameters are
 the same as in Figs.~\ref{fig:ranap-ranap-FI-500-0025}
 and \ref{fig:ranap-mchi-FI-500-0025}.
}
\label{fig:ranapSISD-FI-500-100}
\end{figure}

 In Figs.~\ref{fig:ranapSISD-FI-500-100},
 we show
 the reconstructed $\armn / \armp$ ratios
 estimated only by Eq.~(\ref{eqn:ranapSISD})
 and the 1$\sigma$ statistical uncertainty bounds
 with the $\rmXA{F}{19}$ + $\rmXA{I}{127}$ + $\rmXA{Si}{28}$ target combination.
 Three different minimal cut--off energies:
 $\Qmin = 0.25$ keV (dashed green),
 2.5 keV (dash--dotteded blue),
 and 5 keV (solid red)
 have been presented
 with the results of zero minimal cut--off energy $\Qmin = 0$
 (dotted black)
 as a reference.
 The width of the first energy bin
 (before tuning)
 has been fixed as
 $b_1 = 10$ keV.

 In the lower frame of Figs.~\ref{fig:ranapSISD-FI-500-100},
 the reconstructed $\armn / \armp$ ratios
 and the uncertainty bounds
 with a fixed input ratio of $\armn / \armp = 0.7$
 has been given
 as functions of the input WIMP mass
 between 2 and 200 GeV.
 As discussed before,
 this plot shows again that,
 for the input WIMP masses of $m_{\chi, {\rm in}} = 2$ and 5 GeV,
 the corresponding kinematic maximal cut--off energies
 for the I target
 are only
 $Q_{\rm max, kin, I}  = 0.39$ and 2.32 keV,
 and thus
 no WIMP events can be observed above
 $\Qmin = 2.5$ and 5 keV experimental cut--off energies.
 Furthermore,
 for the larger input WIMP mass of $m_{\chi, {\rm in}} = 10$ or 15 GeV,
 the corresponding kinematic maximal cut--off
 for iodine
 is only
 $Q_{\rm max, kin, I} = 8.57$ or 17.85 keV.
 Hence,
 the $\Qmin = 2.5$ and 5 keV threshold energies
 cuts 29\% and 58\%
 (for the input WIMP mass of $m_{\chi, {\rm in}} = 10$ GeV)
 of the theoretically analyzable energy range.
 This results again in the underestimates of
 the reconstructed $\armn / \armp$ ratios
 for $m_{\chi, {\rm in}} \lsim 10$ (and 15) GeV.

 Nevertheless,
 and importantly,
 comparing with the results presented in Ref.~\cite{DMDDranap}
 (with the unmodified estimators),
 the reconstructions of the $\armn / \armp$ ratios
 have been strongly improved here,
 once some (real) WIMP events can be observed
 above the experimental threshold energies.
 Moreover,
 if the WIMP mass is larger than ${\cal O}(20)$ GeV,
 even with a threshold energy of $\sim$ 5 keV,
 we could in principle reconstruct the $\armn / \armp$ ratio
 very precisely.
 Additionally,
 as also shown
 in the upper frame of Figs.~\ref{fig:ranapSISD-FI-500-100},
 the plot of
 the reconstructed $\armn / \armp$ ratios
 and statistical uncertainty bounds
 with a fixed input WIMP mass of $m_{\chi, {\rm in}} = 20$ GeV
 between $\armn / \armp = \pm 4$,
 increasing the minimal cut--off energy
 would only enlarge the statistical uncertainties
 on the reconstructed $\armn / \armp$ ratios,
 which would almost be independent of
 the true (input) WIMP mass.

\subsubsection{Reconstructions of the ratios between the SD and SI WIMP--nucleon cross sections}
\label{sec:rsigma-500-100}
\begin{figure}[p!]
\begin{center}
 \includegraphics[width = 15 cm]{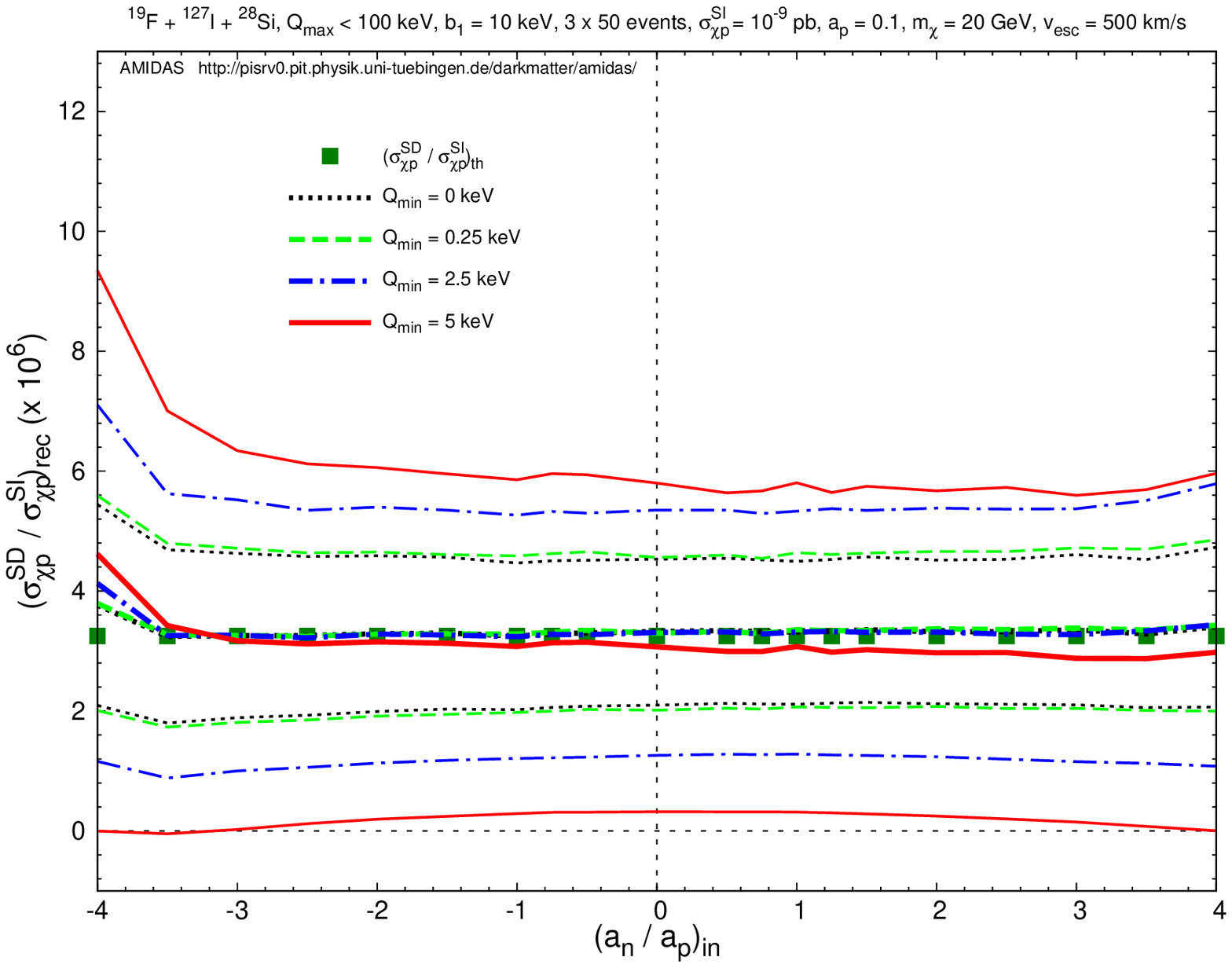} \\ \vspace{ 0.75cm}
 \includegraphics[width = 15 cm]{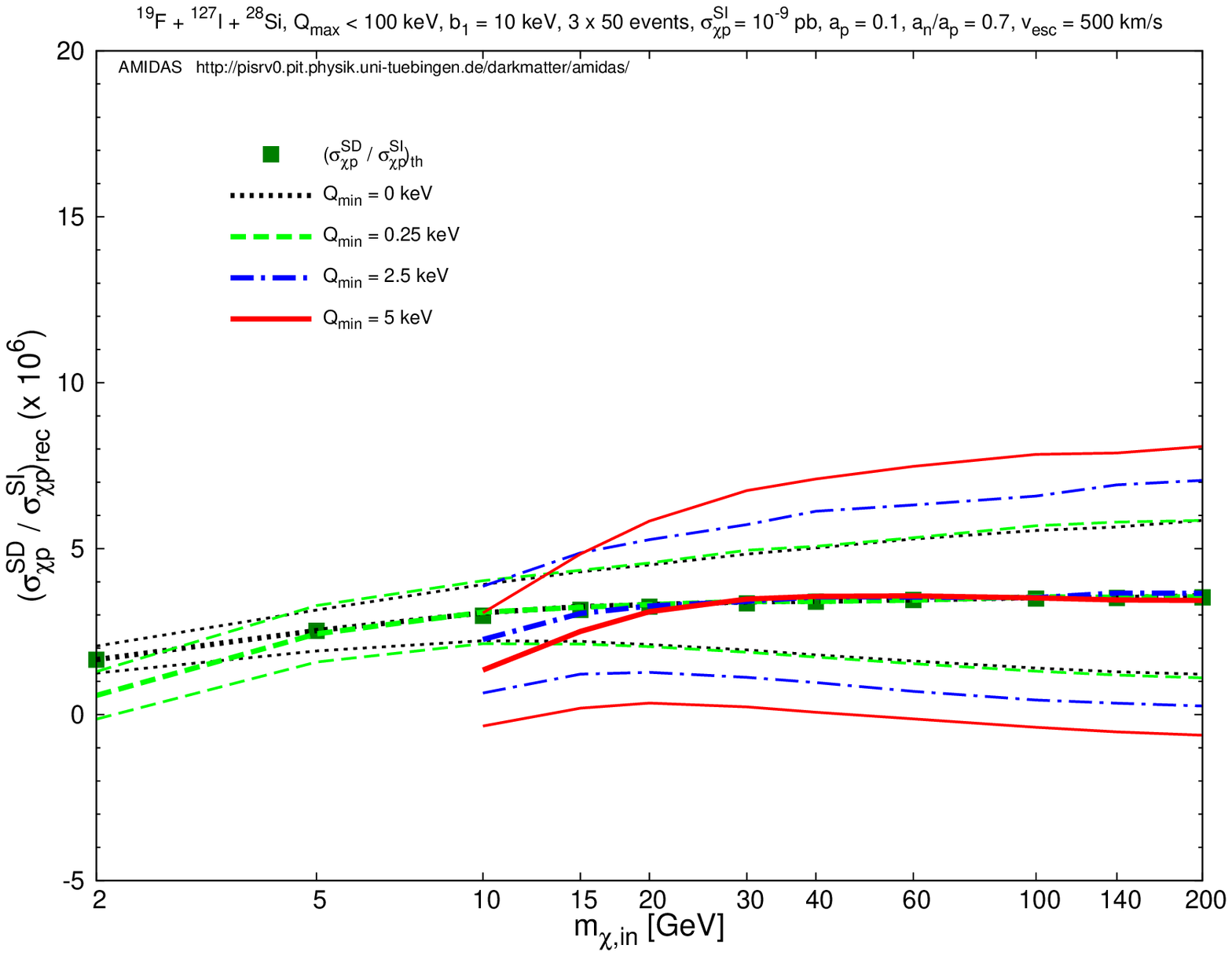}
\end{center}
\caption{
 The reconstructed $\sigmapSD / \sigmapSI$ ratios
 estimated only by Eq.~(\ref{eqn:rsigmaSDpSI})
 and the 1$\sigma$ statistical uncertainty bounds
 with the $\rmXA{F}{19}$ + $\rmXA{I}{127}$ + $\rmXA{Si}{28}$ target combination
 for input $\armn / \armp$ ratios between $\pm 4$ (upper)
 and input WIMP masses between 2 and 200 GeV (lower).
 Other parameters and notations are
 the same as in Figs.~\ref{fig:ranapSISD-FI-500-100}.
}
\label{fig:rsigmaSDpSIXYZ-FI-500-100}
\end{figure}
\begin{figure}[p!]
\begin{center}
 \includegraphics[width = 15 cm]{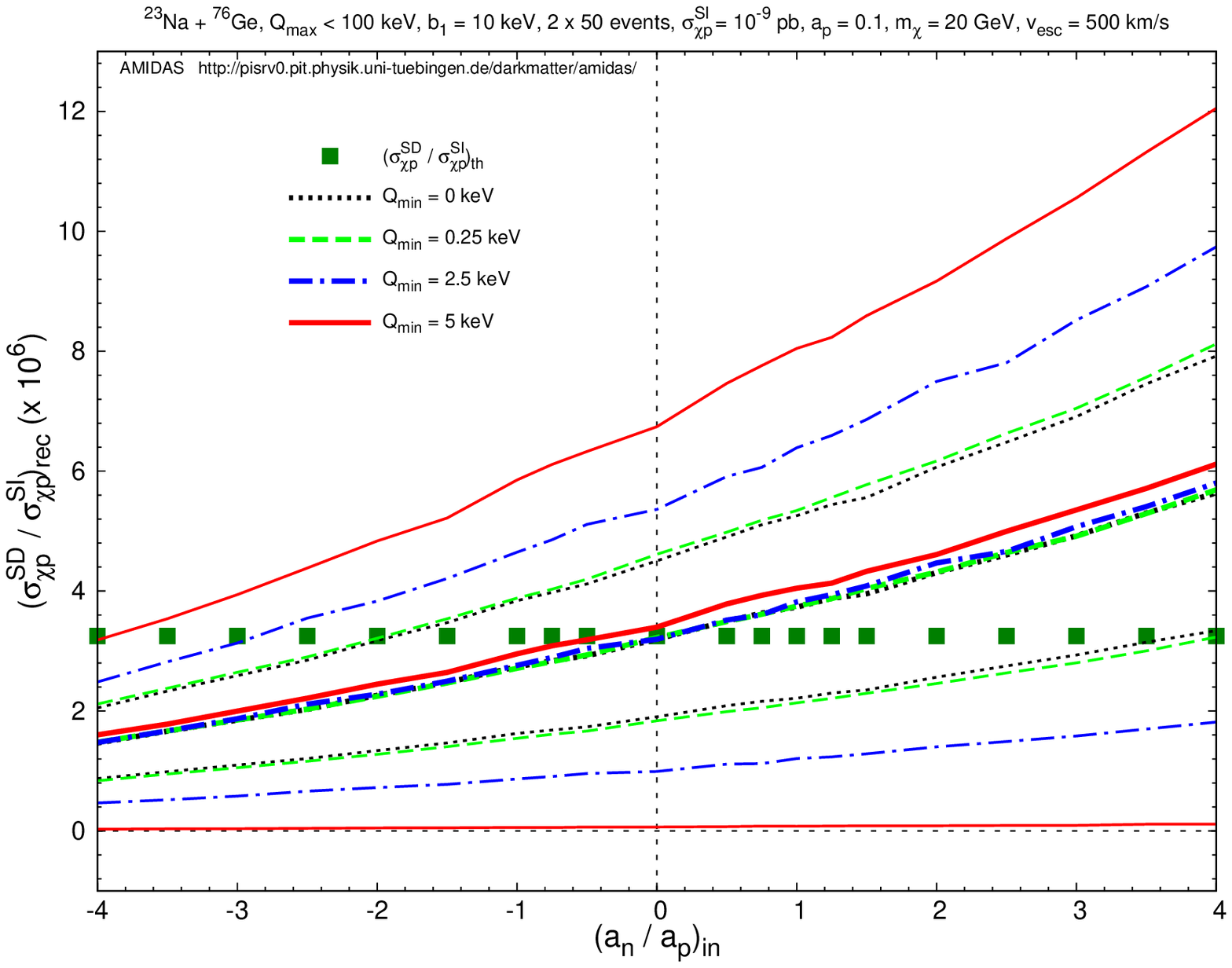} \\ \vspace{ 0.75cm}
 \includegraphics[width = 15 cm]{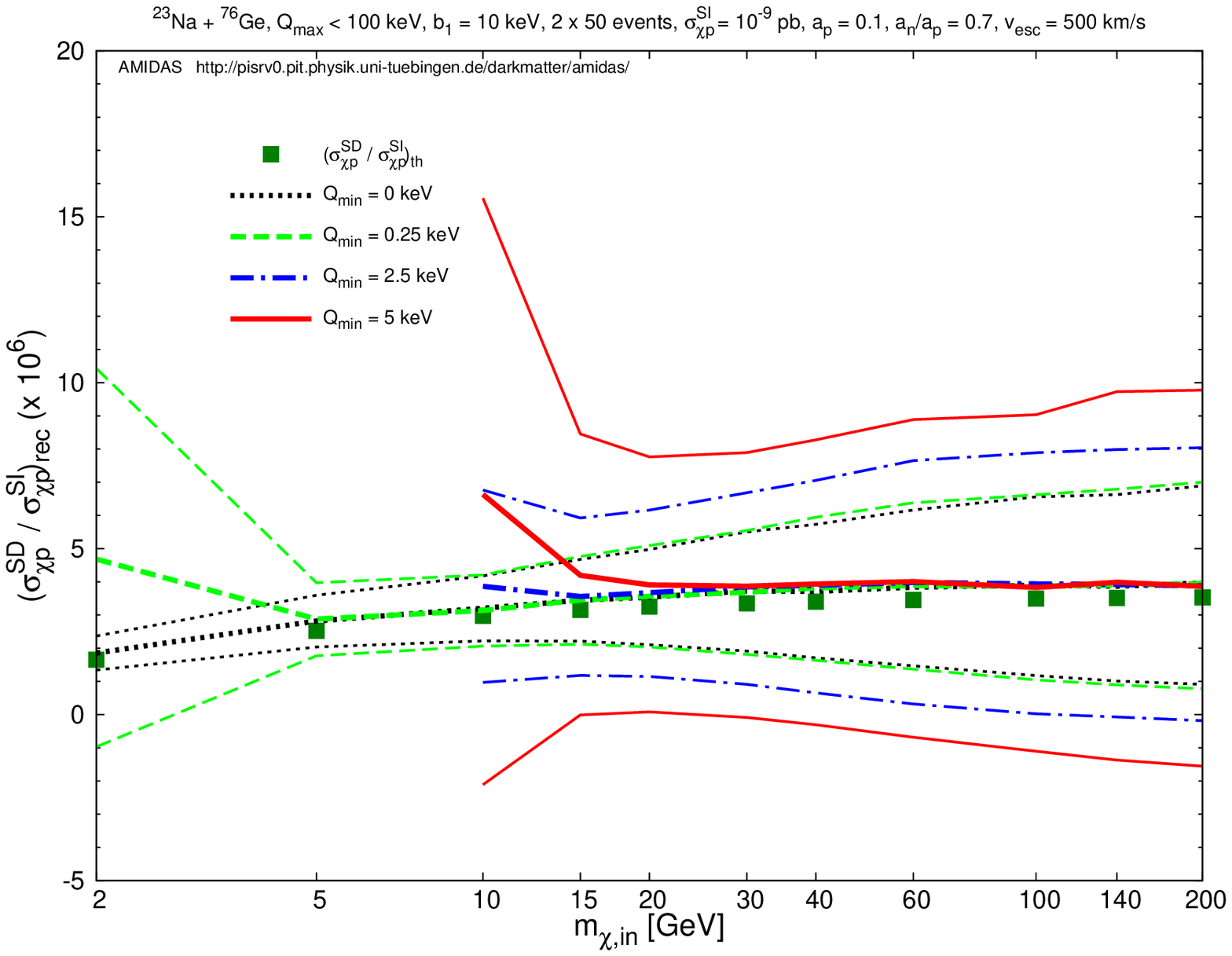}
\end{center}
\caption{
 As in Figs.~\ref{fig:rsigmaSDpSIXYZ-FI-500-100},
 except that
 $\sigmapSD / \sigmapSI$ and $\sigma\abrac{\sigmapSD / \sigmapSI}$
 have been estimated by
 Eq.~(\ref{eqn:rsigmaSDpSI_even})
 with the $\rmXA{Na}{23}$ + $\rmXA{Ge}{76}$ target combination.
}
\label{fig:rsigmaSDpSIXY-FI-500-100}
\end{figure}
\begin{figure}[p!]
\begin{center}
 \includegraphics[width = 15 cm]{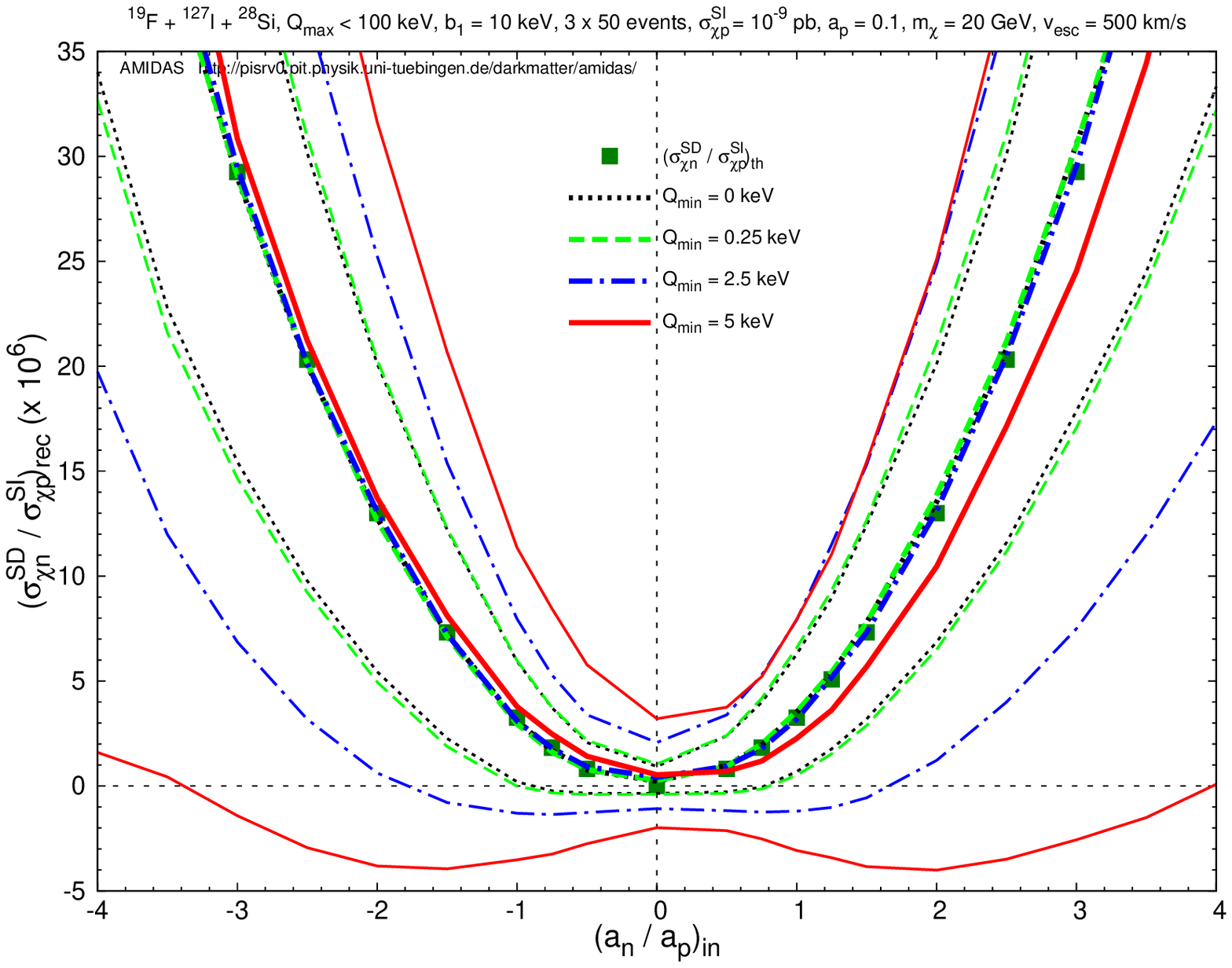} \\ \vspace{ 0.75cm}
 \includegraphics[width = 15 cm]{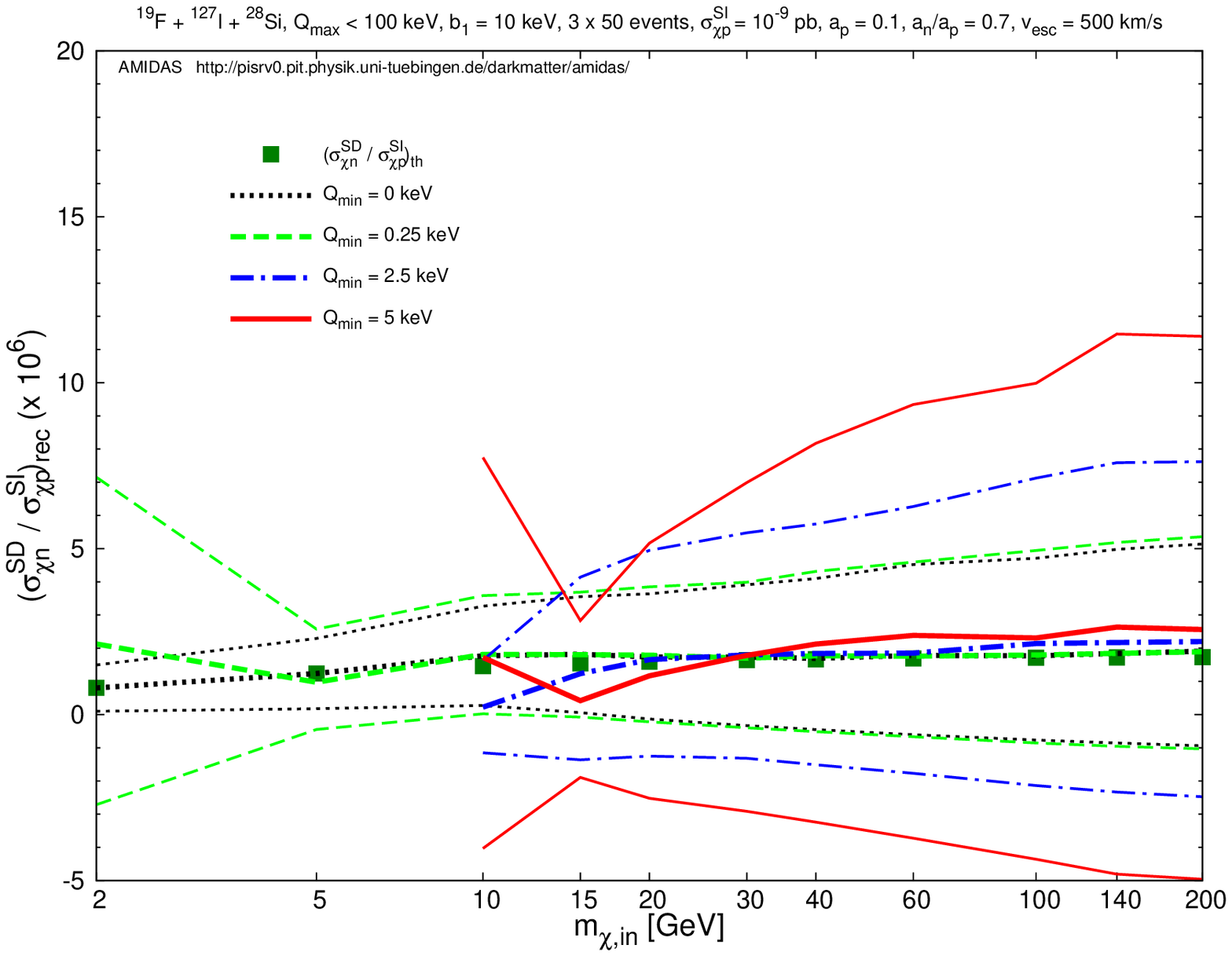}
\end{center}
\caption{
 The reconstructed $\sigmanSD / \sigmapSI$ ratios
 estimated only by Eq.~(\ref{eqn:rsigmaSDnSI})
 and the 1$\sigma$ statistical uncertainty bounds
 with the $\rmXA{F}{19}$ + $\rmXA{I}{127}$ + $\rmXA{Si}{28}$ target combination
 for input $\armn / \armp$ ratios between $\pm 4$ (upper)
 and input WIMP masses between 2 and 200 GeV (lower).
 Other parameters and all notations are
 the same as in Figs.~\ref{fig:ranapSISD-FI-500-100}.
}
\label{fig:rsigmaSDnSIXYZ-FI-500-100}
\end{figure}
\begin{figure}[p!]
\begin{center}
 \includegraphics[width = 15 cm]{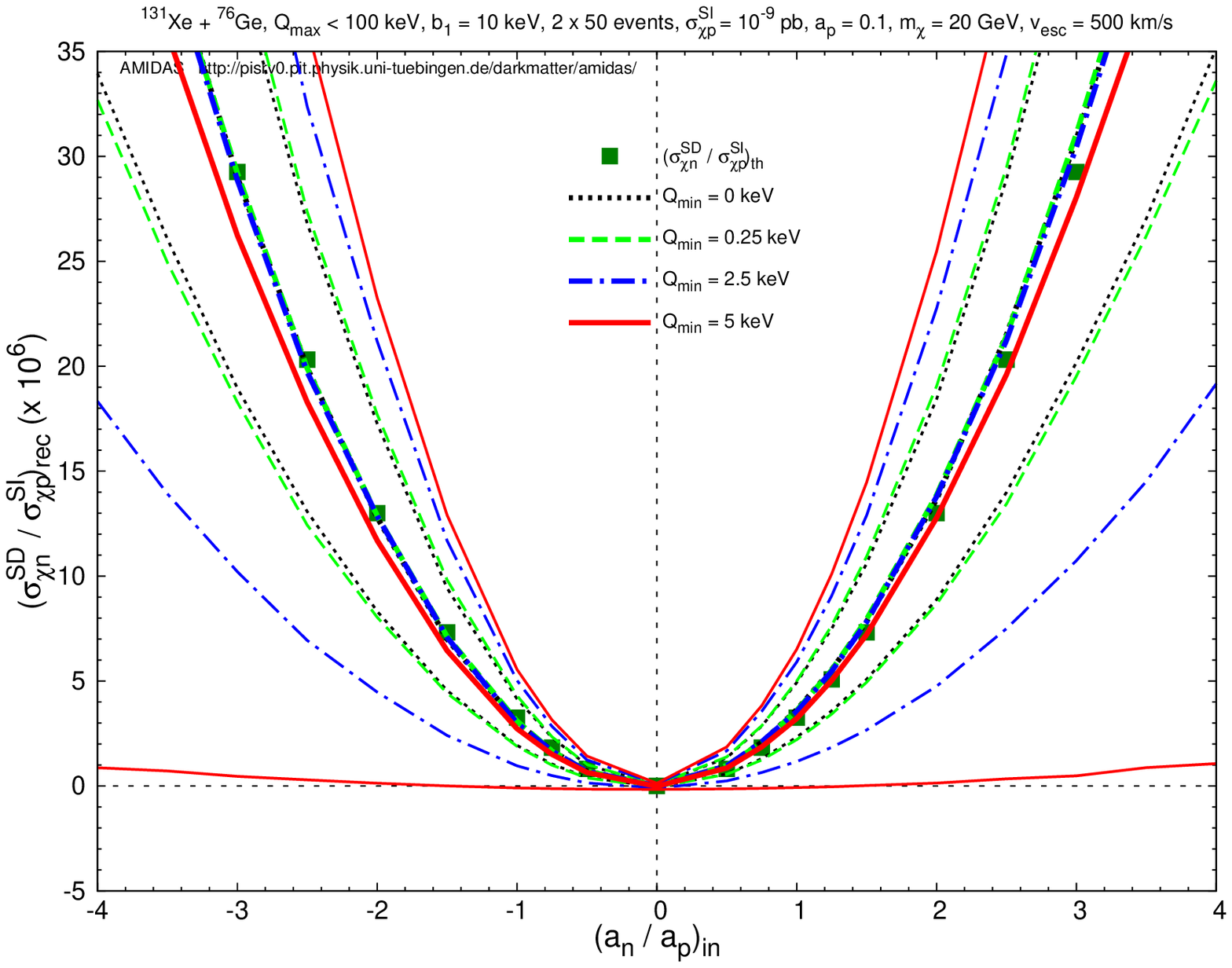} \\ \vspace{ 0.75cm}
 \includegraphics[width = 15 cm]{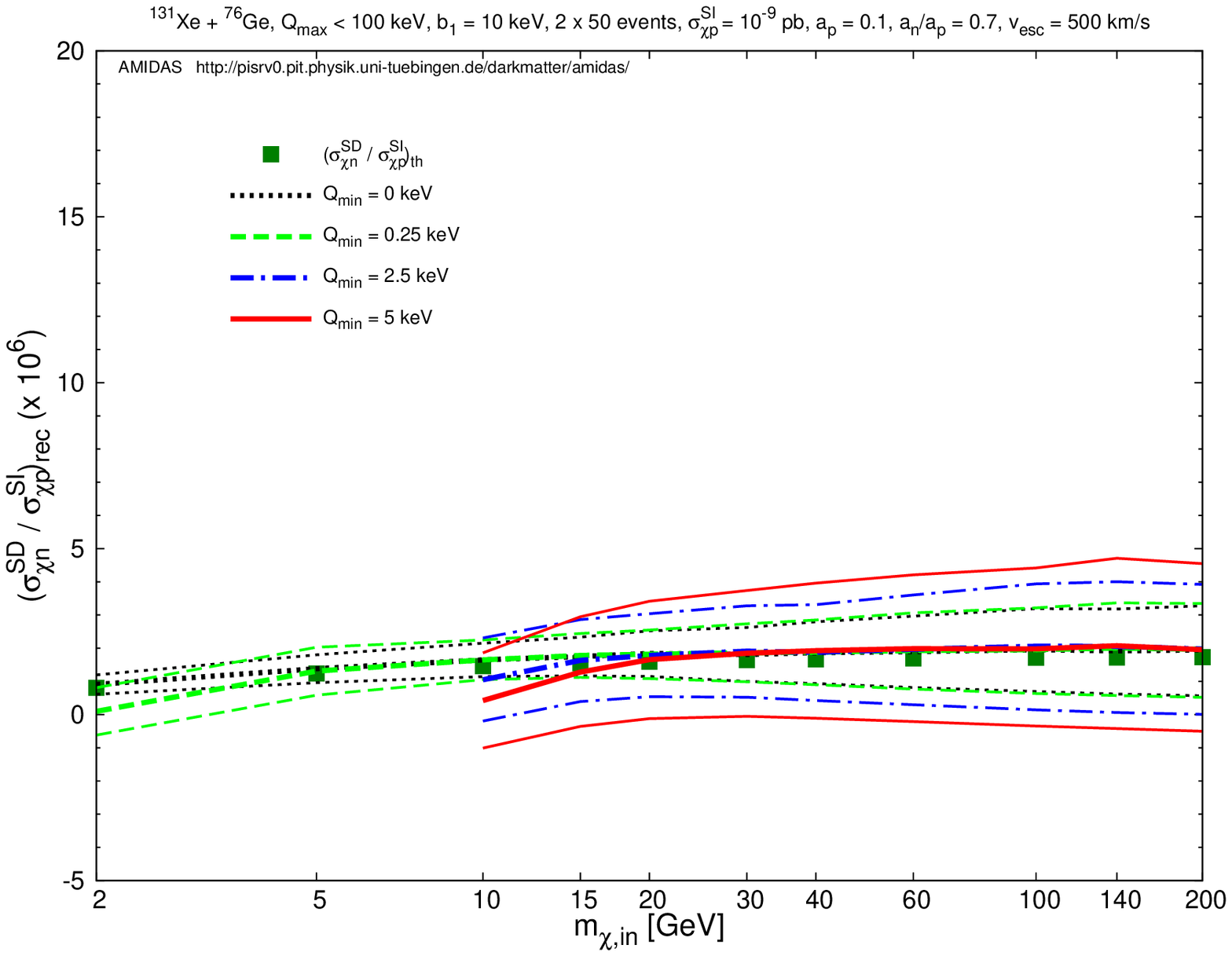}
\end{center}
\caption{
 As in Figs.~\ref{fig:rsigmaSDnSIXYZ-FI-500-100},
 except that
 $\sigmanSD / \sigmapSI$ and $\sigma\abrac{\sigmanSD / \sigmapSI}$
 have been estimated by Eq.~(\ref{eqn:rsigmaSDnSI_even})
 with the $\rmXA{Xe}{131}$ + $\rmXA{Ge}{76}$ target combination.
}
\label{fig:rsigmaSDnSIXY-FI-500-100}
\end{figure}

 Finally,
 in Figs.~\ref{fig:rsigmaSDpSIXYZ-FI-500-100}
 and \ref{fig:rsigmaSDpSIXY-FI-500-100},
 we show
 the reconstructed $\sigmapSD / \sigmapSI$ ratios
 estimated by Eq.~(\ref{eqn:rsigmaSDpSI})
 and the 1$\sigma$ statistical uncertainties
 with the $\rmXA{F}{19}$ + $\rmXA{I}{127}$ + $\rmXA{Si}{28}$
 and by Eq.~(\ref{eqn:rsigmaSDpSI_even})
 with the $\rmXA{Na}{23}$ + $\rmXA{Ge}{76}$ target combinations
 separately.
 As in Figs.~\ref{fig:ranapSISD-FI-500-100},
 in the upper frames of Figs.~\ref{fig:rsigmaSDpSIXYZ-FI-500-100}
 and \ref{fig:rsigmaSDpSIXY-FI-500-100},
 the input WIMP mass
 has been fixed as 20 GeV
 and we have simulated input $\armn / \armp$ ratios between $\pm 4$;
 in the lower frames of two figures,
 we have fixed the input $\armn / \armp$ ratio as 0.7
 and simulated input WIMP masses between 2 and 200 GeV.
 As before,
 four different minimal cut--off energies
 have been considered:
 $\Qmin = 0$ (dotted black),
 0.25 keV (dashed green),
 2.5 keV (dash--dotteded blue),
 and 5 keV (solid red),
 and
 the width of the first energy bin
 (before tuning)
 has been fixed as
 $b_1 = 10$ keV.
 Meanwhile,
 in Figs.~\ref{fig:rsigmaSDnSIXYZ-FI-500-100}
 and \ref{fig:rsigmaSDnSIXY-FI-500-100},
 we show also
 the reconstructed $\sigmanSD / \sigmapSI$ ratios
 estimated by Eq.~(\ref{eqn:rsigmaSDnSI})
 and the 1$\sigma$ statistical uncertainty bounds
 with the $\rmXA{F}{19}$ + $\rmXA{I}{127}$ + $\rmXA{Si}{28}$
 and by Eq.~(\ref{eqn:rsigmaSDnSI_even})
 with the $\rmXA{Xe}{131}$ + $\rmXA{Ge}{76}$ target combinations
 separately.

 These plots demonstrate clearly that,
 first,
 as for reconstructing the other WIMP properties,
 the (pretty) small maximal kinematic cut--off energy
 depending on the mass of the incident WIMPs
 would be the most critical issue
 for the reconstructions of
 the ratios between the WIMP--nucleon cross sections.
 Once WIMPs are so light that
 large parts of the theoretically analyzable energy ranges
 are cut by the threshold energies of the analyzed data sets,
 the ratios between the WIMP--nucleon cross sections
 could be (strongly) over-- or underestimated.
 However,
 as shown in the four lower frames of
 Figs.~\ref{fig:rsigmaSDpSIXYZ-FI-500-100} to
 \ref{fig:rsigmaSDnSIXY-FI-500-100},
 by using different (combinations of) target nuclei,
 one would obtain incompatible results
 for reminding us to reduce
 the experimental threshold energies.

 In contrast,
 once WIMPs are heavier than $\sim 15$ GeV,
 the maximal kinematic cut--off energies
 of our target nuclei
 are (much) higher than
 the threshold energies of the analyzed data sets,
 the results reconstructed with different target combinations
 would match each other pretty well
 and also be pretty precise to the true (input) values.
 Then
 the most serious problem with increasing the threshold energies
 would only be
 the enlargement of the statistical uncertainties
 on the reconstructed cross--section ratios.

\section{Summary and conclusions}

 In this paper,
 we have revisited our data analysis procedures
 developed for reconstructing different WIMP properties:
 the WIMP mass,
 the SI scaler WIMP--nucleon couplings
 as well as
 the ratios between the SI and SD WIMP--nucleon couplings/cross sections
 by taken into account
 non--negligible experimental threshold energies of
 the analyzed data sets.
 All needed expressions
 for the reconstruction processes
 have been checked and modified properly.

 Our simulation results show that,
 firstly,
 the (pretty) small maximal kinematic cut--off energy
 depending on the mass of the incident WIMPs
 would be the most critical issue
 for the reconstructions of WIMP properties.
 For the case that
 WIMPs are as light as $\lsim$ 10 GeV,
 the expected maximal kinematic cut--offs
 for light targets,
 e.g.~Si and Ar,
 would be $\lsim$ 20 keV
 and
 even only a few keV
 for heavy targets,
 e.g.~Ge and Xe.
 It is therefore possible that
 very few or even no WIMP scattering events could be observed
 between the experimental threshold energies
 and the maximal kinematic cut--offs.
 Once we could fortunately observe a number of WIMP signals,
 large parts of the theoretically analyzable energy ranges
 would still be cut by the threshold energies of the analyzed data sets,
 and,
 consequently,
 the reconstructed WIMP properties
 could be (strongly) over-- or underestimated.
 Nevertheless,
 as demonstrated in this paper,
 one could use data sets with different (combinations of) target nuclei
 for the same analyses
 and
 (in)compatible results
 would help us to check
 the reliability of our reconstructions.

 On the other hand,
 once WIMPs are heavier than $\sim 15$ GeV,
 the maximal kinematic cut--off energies
 of our target nuclei
 would be (much) higher than
 the threshold energies of the analyzed data sets,
 our simulation results
 with different target combinations
 could match each other pretty well
 and also be pretty precise to the true (input) values.
 For this case
 the most serious problem with increasing the threshold energies
 would only be
 the enlargement of the statistical uncertainties
 on the reconstructed WIMP properties.

 Moreover,
 for light WIMPs ($\mchi \lsim 10$ GeV),
 since the analyzed energy ranges
 would be very narrow,
 one should take a small width for the first energy bin.
 In contrast,
 once the true WIMP mass is larger than $\sim$ 60 GeV,
 using a larger bin width
 would be helpful for alleviating some systematic deviations.

 In our simulations presented in this paper,
 the Galactic escape velocity
 has been set conservatively as $\vesc = 500$ km/s.
 Our further simulations
 show that,
 firstly,
 with a larger escape velocity and
 thus a larger kinematic cut--off on the velocity distribution
 as well as
 on the recoil spectra,
 the systematic deviations of the reconstructed results
 from the true (input) values
 could be reduced,
 unexpectedly,
 only a little bit.
 On the other hand,
 once the (true) escape velocity
 is larger than our simulation setup,
 for heavy target nuclei,
 the cut--offs of the kinematic energy
 could become higher than
 the experimental threshold energies
 and some WIMP events could thus be observed.
 However,
 since large parts of the theoretically analyzable energy ranges
 would still be cut by the relatively pretty high threshold energies,
 the reconstructed results would be (strongly) deviated from the true values
 and not (very) reliable.

 In summary,
 as a supplement of our earlier works
 on reconstructions of different WIMP properties,
 we modified in this paper all estimators
 for the more general case
 with non--negligible threshold energy.
 Hopefully,
 these modifications
 could not only be more suitable
 for practical data analyses
 in direct detection experiments,
 but also offer preciser information
 about Galactic Dark Matter.

\subsubsection*{Acknowledgments}
 The authors would like to thank
 the Physikalisches Institut der Universit\"at T\"ubingen
 for the technical support of the computational work
 presented in this paper.
 CLS would also appreciate
 the friendly hospitality of
 the Gran Sasso Science Institute
 during the finalization of this paper.
 This work
 was partially supported by
 the Department of Human Resources and Social Security of
 Xinjiang Uygur Autonomous Region
 as well as
 the CAS Pioneer Hundred Talents Program.

\appendix
\setcounter{equation}{0}
\setcounter{figure}{0}
\renewcommand{\theequation}{A\arabic{equation}}
\renewcommand{\thefigure}{A\arabic{figure}}
%
%
%
\section{Formulae for estimating statistical uncertainties}

 Here we list the {\em modified} formulae needed
 for estimating statistical uncertainties
 on the reconstructed WIMP properties
 by using our model--independent methods.
 Detailed derivations and discussions
 can be found in Refs.~%
 \cite{DMDDf1v, DMDDf1v-calN, DMDDmchi, DMDDfp2, DMDDranap}
 (with some necessary modifications).

 First,
 from Eqs.~(\ref{eqn:rmin}), (\ref{eqn:Kn}) and (\ref{eqn:Qsn}),
 the statistical uncertainty
 on the modified estimator $r^{\ast}(\Qmin)$
 defined in Eq.~(\ref{eqn:rmin_ast})
 can be expressed as
\beqn
     \sigma^2(r^{\ast}(\Qmin))
 \=  \bBig{r^{\ast}(\Qmin)}^2
     \cleft{   \frac{1}{N_1}
             + \cleft{   \bbrac{  \frac{1}{k_1}
                                - \afrac{b_1}{2}
                                  \abrac{1 + \coth\afrac{b_1 k_1}{2}} } } }
     \non\\
 \conti ~~~~ ~~~~ ~~~~ ~~~~ ~~~~ ~~~~ ~~~~ ~~ \times 
     \cBiggr{  \cBiggr{  \bbigg{K_1(\Qmin) \~ \Qmin + 1}
                       - \Qmin }^2
              \sigma^2(k_1) }
\~.
\label{eqn:sigma_rmin_ast}
\eeqn
 Meanwhile,
 since all $I_n$ are determined from the same data,
 they are correlated with each other
 as well as
 with $r^{\ast}(\Qmin)$
 through the contribution of the measured recoil energies
 in the first $Q-$bin.
 Following the definition of $I_n$ given in Eq.~(\ref{eqn:In_sum}),
 we have
\beq
     I_{n, 1}(\Qmin, \Qmin + b_1)
  =  r_1
     \int_{\Qmin}^{\Qmin + b_1}
     \bfrac{Q^{(n - 1) / 2}}{F^2(Q)} e^{k_1 (Q - Q_{s, 1})} \~ dQ
 \to \sum_{i = 1}^{N_1} \frac{Q_{1, i}^{(n - 1) / 2}}{F^2(Q_{1, i})}
\~,
\label{eqn:In1_sum}
\eeq
 Hence,
 the correlation between the uncertainties on $r^{\ast}(\Qmin)$
 and on $I_n$ is given by
\beqn
 \conti
     {\rm cov}(r^{\ast}(\Qmin), I_n)
     \non\\
 \=  r^{\ast}(\Qmin) \~ I_{n, 1}(\Qmin, \Qmin + b_1)
     \non\\
 \conti ~~~~ \times
     \cleft{  \frac{1}{N_1}
            + \cbrac{  \bbrac{  \frac{1}{k_1}
                              - \afrac{b_1}{2}
                                \abrac{1 + \coth\afrac{b_1 k_1}{2}} }
                       \bbigg{K_1(\Qmin) \~ \Qmin + 1}
                     - \Qmin } }
     \non\\
 \conti ~~~~ ~~~~ ~~~~ ~~ \times 
     \cright{ \bbrac{  \frac{I_{n + 2, 1}(\Qmin, \Qmin + b_1)}
                            {I_{n    , 1}(\Qmin, \Qmin + b_1)}
                     - Q_1
                     + \frac{1}{k_1}
                     - \afrac{b_1}{2} \coth\afrac{b_1 k_1}{2} }
              \sigma^2(k_1) }
\~.
\label{eqn:cov_rmin_ast_In}
\eeqn

 On the other hand,
 according to the modifications of
 the definitions of $\calR_{n, X}$ and $\calR_{\sigma, X}$
 given in Eqs.~(\ref{eqn:RnX_min}) and (\ref{eqn:RsigmaX_min}),
 the short--hand notation for the six quantities
 introduced in Ref.~\cite{DMDDmchi}
 on which the estimate of $\mchi$ depends
 are now:
\beq
     c_{1, X}
  =  I_{n, X}
\~,
     ~~~~ ~~~~ ~~~~ 
     c_{2, X}
  =  I_{0, X}
\~,
     ~~~~ ~~~~ ~~~~ 
     c_{3, X}
  =  r_X^{\ast}(\QminX)
\~,
\label{eqn:ciX}
\eeq
 and similarly for the $c_{i, Y}$;
 the last element $c_{3, (X, Y)}$
 are now replaced by $r_{(X, Y)}^{\ast}(Q_{{\rm min}, (X, Y)})$.
 Then
 the explicit expressions for the derivatives of
 $\calR_{n, (X, Y)}$ and $\calR_{\sigma, (X, Y)}$
 with respect to $c_{i, (X, Y)}$
 can then be obtained directly
 by replacing $r_{(X, Y)}(Q_{{\rm min}, (X, Y)})$
 by $r_{(X, Y)}^{\ast}(Q_{{\rm min}, (X, Y)})$
 (see Appendices of Refs.~\cite{DMDDfp2, DMDDranap}).

 Finally,
 the derivative of $\sigmapSD / \sigmapSI$
 with respect to $\calR_{m, X}$
 given in Eq.~(A.3) of Ref.~\cite{DMDDranap}
 should be corrected by
\beqn
 \conti
     \pp{\calR_{m, X}} \afrac{\sigmapSD}{\sigmapSI}
     \non\\
 \=  \frac{  \calCpX \FSDQminX \FSIQminY
           - \calCpY \FSDQminY \FSIQminX }
          {\bBig{  \calCpX \FSDQminX
                 - \calCpY \FSDQminY (\calR_{m, X} / \calR_{m, Y}) }^2}
     \afrac{1}{\calR_{m, Y}}
\~,
\label{eqn:drsigmaSDpSI_dRmX}
\eeqn
 i.e.,
 there should be no ``$-$ (minus)'' sign
 before the fraction.


\begin{thebibliography}{99}
%

%
\bibitem{Goodman85}
 {M.~W.~Goodman and E.~Witten,
  {\it ``Detectability of Certain Dark--Matter Candidates''},
  {\it Phys.~Rev.}~{\bf D31}, 3059-–3063 (1985).}
%
\bibitem{Wasserman86}
 {I.~Wasserman,
  {\it ``Possibility of Detecting Heavy Neutral Fermions in the Galaxy''},
  {\it Phys.~Rev.}~{\bf D33}, 2071--2078 (1986).}
%
\bibitem{Drukier86}
 {A.~K.~Drukier, K.~Freese and D.~N.~Spergel,
  {\it ``Detecting Cold Dark Matter Candidates''},
  {\it Phys.~Rev.}~{\bf D33}, 3495--3508 (1986).}
%
\bibitem{Griest88}
 {K.~Griest,
  {\it ``Cross--Sections, Relic Abundance and Detection Rates for Neutralino Dark Matter''},
  {\it Phys.~Rev.}~{\bf D38}, 2357--2375 (1988),
  Erratum: ibid.~{\bf D39}, 3802--3803 (1989).}
%

%
\bibitem{Smith90}
 {P.~F.~Smith and J.~D.~Lewin,
  {\it ``Dark Matter Detection''},
  {\it Phys.~Rept.}~{\bf 187}, 203--280 (1990).}
%
\bibitem{SUSYDM96}
 {G.~Jungman, M.~Kamionkowski and K.~Griest,
  {\it ``Supersymmetric Dark Matter''},
  {\it Phys.~Rep.}~{\bf 267}, 195--373 (1996),
  {\tt arXiv:hep-ph/9506380}.}
%
\bibitem{Lewin96}
 {J.~D.~Lewin and P.~F.~Smith,
  {\it ``Review of Mathematics, Numerical Factors, and Corrections
         for Dark Matter Experiments Based on Elastic Nuclear Recoil''},
  {\it Astropart.~Phys.}~{\bf 6}, 87--112 (1996).}
%

%
\bibitem{Ramachers02}
 {Y.~Ramachers,
  {\it ``WIMP Direct Detection Overview''},
  {\it Nucl.~Phys.~Proc.~Suppl.}~{\bf 118}, 341--350 (2003),
  {\tt arXiv:astro-ph/0211500}.}
%
\bibitem{Jesus04}
 {M.~de Jesus,
  {\it ``WIMP/Neutralino Direct Detection''},
  {\it Int.~J.~Mod.~Phys.}~{\bf A19}, 1142--1151 (2004),
  {\tt arXiv:astro-ph/0402033}.}
%

%
\bibitem{Gaitskell04}
 {R.~J.~Gaitskell,
  {\it ``Direct Detection of Dark Matter''},
  {\it Ann.~Rev.~Nucl.~Part.~Sci.}~{\bf 54}, 315--359 (2004).}
%

%
\bibitem{Cerdeno10}
 {D.~G.~Cerde$\rm \tilde{n}$o and A.~M.~Green,
  {\it ``Direct Detection of WIMPs''},
  contribution to
  {\it ``Particle Dark Matter: Observations, Models and Searches''},
  edited by G.~Bertone,
  Cambridge University Press (2010),
  Chapter 17,
  Hardback ISBN 9780521763684,
  {\tt arXiv:1002.1912 [astro-ph.CO]}.}
%

%
\bibitem{Saab12}
 {T.~Saab,
  {\it ``An Introduction to Dark Matter Direct Detection Searches and Techniques''},
  {\tt arXiv:1203.2566 [physics.ins-det]} (2012).}
%

%
\bibitem{Baudis12c}
 {L.~Baudis,
  {\it ``Direct Dark Matter Detection: the Next Decade''},
  Issue on {\it ``The Next Decade in Dark Matter and Dark Energy''},
  {\it Phys.~Dark Univ.}~{\bf 1}, 94--108 (2012),
  {\tt arXiv:1211.7222 [astro-ph.IM]}.}
%

%
\bibitem{Baudis15}
 {L.~Baudis,
  {\it ``Dark Matter Searches''},
  {\it Annalen Phys.~(Berlin)}, 74--83 (2016),
  {\tt arXiv:1509.00869 [astro-ph.CO]}.}
%
\bibitem{Drees16}
 {M.~Drees and G.~Gerbier,
  contribution to
  {\it ``The Review of Particle Physics 2016''},
  {\it Chin.~Phys.}~{\bf C40}, 100001 (2016),
  {\it 26.~Dark Matter}.}
%
\bibitem{JLiu17}
 {J.~Liu, X.~Chen and X.~Ji,
  {\it ``Current Status of Direct Dark Matter Detection Experiments''},
  {\it Nature Phys.}~{\bf 13}, 212 (2017),
  {\tt arXiv:1709.00688 [astro-ph.CO]}.}
%

%
\bibitem{Green-mchi07}
 {A.~M.~Green,
  {\it ``Determining the WIMP Mass Using Direct Detection Experiments''},
  {\it J.~Cosmol.~Astropart.~Phys.}~{\bf 0708}, 022 (2007),
  {\tt arXiv:hep-ph/0703217}.}
%
\bibitem{Green-mchi08}
 {A.~M.~Green,
  {\it ``Determining the WIMP Mass from a Single Direct Detection Experiment,
         a More Detailed Study''},
  {\it J.~Cosmol.~Astropart.~Phys.}~{\bf 0807}, 005 (2008),
  {\tt arXiv:0805.1704 [hep-ph]}.}
%
\bibitem{Green-mchi12}
 {B.~J.~Kavanagh and A.~M.~Green,
  {\it ``Improved Determination of the WIMP Mass from Direct Detection Data''},
  {\it Phys.~Rev.}~{\bf D86}, 065027 (2012),
  {\tt arXiv:1207.2039 [astro-ph.CO]}.}
%
\bibitem{Green-mchi13}
 {B.~J.~Kavanagh and A.~M.~Green,
  {\it ``Model Independent Determination of
         the Dark Matter Mass from Direct Detection Experiments''},
  {\it Phys.~Rev.~Lett.}~{\bf 111}, 031302 (2013),
  {\tt arXiv:1303.6868 [astro-ph.CO]}.}
%

%
\bibitem{Cannoni10}
 {M.~Cannoni, J.~D.~Vergados and M.~E.~Gomez,
  {\it ``Extraction of Neutralino--Nucleon Scattering Cross Sections from Total Rates''},
  {\it Phys.~Rev.}~{\bf D83}, 075010 (2011),
  {\tt arXiv:1011.6108 [hep-ph]}.}
%

%
\bibitem{Hoferichter16}
 {M.~Hoferichter, P.~Klos, J.~Men\'endez and A.~Schwenk,
  {\it ``Analysis Strategies for General Spin--Independent WIMP--Nucleus Scattering''},
  {\it Phys.~Rev.}~{\bf D94}, 063505 (2016),
  {\tt arXiv:1605.08043 [hep-ph]}.}
%

%
\bibitem{Akrami10a}
 {Y.~Akrami, C.~Savage, P.~Scott, J.~Conrad and J.~Edsj\"o,
  {\it ``Statistical Coverage for Supersymmetric Parameter Estimation:
         A Case Study with Direct Detection of Dark Matter''},
  {\it J.~Cosmol.~Astropart.~Phys.}~{\bf 1107}, 002 (2011),
  {\tt arXiv:1011.4297 [hep-ph]}.}
%
\bibitem{Akrami10b}
 {Y.~Akrami, C.~Savage, P.~Scott, J.~Conrad and J.~Edsj\"o,
  {\it ``How Well Will Ton--Scale Dark Matter Direct Detection Experiments
         Constrain Minimal Supersymmetrh?''},
  {\it J.~Cosmol.~Astropart.~Phys.}~{\bf 1104}, 012 (2011),
  {\tt arXiv:1011.4318 [astro-ph.CO]}.}
%

%
\bibitem{Pato10b}
 {M.~Pato, L.~Baudis, G.~Bertone, R.~Ruiz de Austri, L.~E.~Strigari and R.~Trotta,
  {\it ``Complementarity of Dark Matter Direct Detection Targets''},
  {\it Phys.~Rev.}~{\bf D83}, 083505 (2011),
  {\tt arXiv:1012.3458 [astro-ph.CO]}.}
%
\bibitem{Pato11}
 {M.~Pato,
  {\it ``What Can(not) be Measured with
         Ton--Scale Dark Matter Direct Detection Experiments''},
  {\it J.~Cosmol.~Astropart.~Phys.}~{\bf 1110}, 035 (2011),
  {\tt arXiv:1106.0743 [astro-ph.CO]}.}
%

%
\bibitem{Arina11a}
 {C.~Arina, J.~Hamann and Y.~Y.~Y.~Wong,
  {\it ``A Bayesian View of the Current Status of Dark Matter Direct Searches''},
  {\it J.~Cosmol.~Astropart.~Phys.}~{\bf 1109}, 022 (2011),
  {\tt arXiv:1105.5121 [hep-ph]}.}
%
\bibitem{Arina12}
 {C.~Arina,
  {\it ``Chasing a Consistent Picture for Dark Matter Direct Searches''},
  {\it Phys.~Rev.}~{\bf D86}, 123527 (2012),
  {\tt arXiv:1210.4011 [hep-ph]}.}
%
\bibitem{Arina13a}
 {C.~Arina, G.~Bertone, H.~Silverwood,
  {\it ``Complementarity of Direct and Indirect Dark Matter Detection Experiments''},
  {\it Phys.~Rev.}~{\bf D88}, 013002 (2013),
  {\tt arXiv:1304.5119 [hep-ph]}.}
%
\bibitem{Arina13b}
 {C.~Arina,
  {\it ``Bayesian Analysis of Multiple Direct Detection Experiments''},
  {\it Phys.~Dark Univ.}~{\bf 5--6}, 1--17 (2014),
  {\tt arXiv:1310.5718 [hep-ph]}.}
%

%
\bibitem{Cerdeno13}
 {D.~G.~Cerde$\rm \tilde{n}$o {\it et al.},
  {\it ``Complementarity of Dark Matter Direct Detection: the Role of Bolometric Targets''},
  {\it J.~Cosmol.~Astropart.~Phys.}~{\bf 1307}, 028 (2013),
  {\tt arXiv:1304.1758 [hep-ph]}.}
%
\bibitem{Cerdeno14}
 {D.~G.~Cerde$\rm \tilde{n}$o {\it et al.},
  {\it ``Scintillating Bolometers: A Key for Determining WIMP Parameters''},
  {\it Int.~J.~Mod.~Phys.}~{\bf A29}, 1443009 (2014),
  {\tt arXiv:1403.3539 [astro-ph.IM]}.}
%
\bibitem{Cerdeno18}
 {D.~G.~Cerde$\rm \tilde{n}$o, A.~Cheek, E.~Reid and H.~Schulz,
  {\it ``Surrogate Models for Direct Dark Matter Detection''},
  {\tt arXiv:1802.03174 [hep-ph]} (2018).}
%

%
\bibitem{McDermott11}
 {S.~D.~McDermott, H.~B.~Yu and K.~M.~Zurek,
  {\it ``The Dark Matter Inverse Problem:
         Extracting Particle Physics from Scattering Events''},
  {\it Phys.~Rev.}~{\bf D85}, 123507 (2012),
  {\tt arXiv:1110.4281 [hep-ph]}.}
%

%
\bibitem{Strege12}
 {C.~Strege, R.~Trotta, G.~Bertone, A.~H.~G.~Peter and P.~Scott,
  {\it ``Fundamental Statistical Limitations of
         Future Dark Matter Direct Detection Experiments''},
  {\it Phys.~Rev.}~{\bf D86}, 023507 (2012),
  {\tt arXiv:1201.3631 [hep-ph]}.}
%

%
\bibitem{Newstead13}
 {J.~L.~Newstead, T.~D.~Jacques, L.~M.~Krauss, J.~B.~Dent, F.~Ferrer,
  {\it ``The Scientific Reach of Multi--Ton Scale Dark Matter Direct Detection Experiments''},
  {\it Phys.~Rev.}~{\bf D88}, 076011 (2013),
  {\tt arXiv:1306.3244 [astro-ph.CO]}.}
%

%
\bibitem{Savage15}
 {C.~Savage, A.~Scaffidi, M.~White and A.~G.~Williams,
  {\it ``LUX Likelihood and Limits on
         Spin--Independent and Spin--Dependent WIMP Couplings with LUXCalc''},
  {\it Phys.~Rev.}~{\bf D92}, 103519 (2015),
  {\tt arXiv:1502.02667 [hep-ph]}.}
%

%
\bibitem{Strigari09}
 {L.~E.~Strigari and R.~Trotta,
  {\it ``Reconstructing WIMP Properties in Direct Detection Experiments
         Including Galactic Dark Matter Distribution Uncertainties''},
  {\it J.~Cosmol.~Astropart.~Phys.}~{\bf 0911}, 019 (2009),
  {\tt arXiv:0906.5361 [astro-ph.HE]}.}
%
\bibitem{APeter09}
 {A.~H.~G.~Peter,
  {\it ``Getting the Astrophysics and Particle Physics of Dark Matter
         Out of Next--Generation Direct Detection Experiments''},
  {\it Phys.~Rev.}~{\bf D81}, 087301 (2010),
  {\tt arXiv:0910.4765 [astro-ph.CO]}.}
%
\bibitem{APeter11}
 {A.~H.~G.~Peter,
  {\it ``WIMP Astronomy with Liquid--Noble and Cryogenic
         Direct--Detection Experiments''},
  {\it Phys.~Rev.}~{\bf D83}, 125029 (2011),
  {\tt arXiv:1103.5145 [astro-ph.CO]}.}
%
\bibitem{Pato12}
 {M.~Pato, L.~E.~Strigari, R.~Trotta and G.~Bertone,
  {\it ``Taming Astrophysical Bias in Direct Dark Matter Searches''},
  {\it J.~Cosmol.~Astropart.~Phys.}~{\bf 1302}, 041 (2013),
  {\tt arXiv:1211.7063 [astro-ph.CO]}.}
%

%
\bibitem{Fox10a}
 {P.~J.~Fox, G.~D.~Kribs and T.~M.~P.~Tait,
  {\it ``Interpreting Dark Matter Direct Detection
         Independently of the Local Velocity and Density Distribution''},
  {\it Phys.~Rev.}~{\bf D83}, 034007 (2011),
  {\tt arXiv:1011.1910 [hep-ph]}.}
%
\bibitem{Fox10b}
 {P.~J.~Fox, J.~Liu and N.~Weiner,
  {\it ``Integrating Out Astrophysical Uncertainties''},
  {\it Phys.~Rev.}~{\bf D83}, 103514 (2011),
  {\tt arXiv:1011.1915 [hep-ph]}.}
%
\bibitem{Fox14}
 {P.~J.~Fox, Y.~Kahn and M.~McCullough,
  {\it ``Taking Halo--Independent Dark Matter Methods Out of the Bin''},
  {\it J.~Cosmol.~Astropart.~Phys.}~{\bf 1410}, 076 (2014),
  {\tt arXiv:1403.6830 [hep-ph]}.}
%

%
\bibitem{DelNobile13a}
 {E.~Del Nobile, G.~B.~Gelmini, P.~Gondolo and J.-H.~Huh,
  {\it ``Halo--Independent Analysis of Direct Detection Data for Light WIMPs''},
  {\it J.~Cosmol.~Astropart.~Phys.}~{\bf 1310}, 026 (2013),
  {\tt arXiv:1304.6183 [hep-ph]}.}
%
\bibitem{DelNobile13b}
 {E.~Del Nobile, G.~Gelmini, P.~Gondolo and J.-H.~Huh,
  {\it ``Generalized Halo Independent Comparison of Direct Dark Matter Detection Data''},
  {\it J.~Cosmol.~Astropart.~Phys.}~{\bf 1310}, 048 (2013),
  {\tt arXiv:1306.5273 [hep-ph]}.}
%
\bibitem{Cirelli13}
 {M.~Cirelli, E.~Del Nobile and P.~Panci,
  {\it ``Tools for Model--Independent Bounds in Direct Dark Matter Searches''},
  {\it J.~Cosmol.~Astropart.~Phys.}~{\bf 1310}, 019 (2013),
  {\tt arXiv:1307.5955 [hep-ph]}.}
%
\bibitem{NRopsDD}
 {M.~Cirelli, E.~Del Nobile and P.~Panci,
  {\it ``Tools for Model--Independent Bounds in Direct Dark Matter Searches,
         release 2.0''},
  {\tt http://www.marcocirelli.net/NRopsDD.html} (2013).}
%
\bibitem{DelNobile14a}
 {E.~Del Nobile,
  {\it ``Halo--Independent Comparison of Direct Dark Matter Detection Data: A Review''},
  {\it Adv.~High Energy Phys.}~{\bf 2014}, 604914 (2014),
  {\tt arXiv:1404.4130 [hep-ph]}.}
%
\bibitem{DelNobile14b}
 {E.~Del Nobile, G.~B.~Gelmini, P.~Gondolo and J.-H.~Huh,
  {\it ``Update on the Halo--Independent Comparison of Direct Dark Matter Detection Data''},
  {\it Phys.~Procedia} {\bf 61}, 45--54 (2015),
  {\tt arXiv:1405.5582 [hep-ph]}.}
%

%
\bibitem{Feldstein14a}
 {B.~Feldstein and F.~Kahlhoefer,
  {\it ``A New Halo--Independent Approach to Dark Matter Direct Detection Analysis''},
  {\it J.~Cosmol.~Astropart.~Phys.}~{\bf 1408}, 065 (2014),
  {\tt arXiv:1403.4606 [hep-ph]}.}
%
\bibitem{Feldstein14b}
 {B.~Feldstein and F.~Kahlhoefer,
  {\it ``Quantifying (Dis)agreement
         Between Direct Detection Experiments in a Halo--Independent Way''},
  {\it J.~Cosmol.~Astropart.~Phys.}~{\bf 1412}, 052 (2014),
  {\tt arXiv:1409.5446 [hep-ph]}.}
%
\bibitem{Kahlhoefer16}
 {F.~Kahlhoefer and S.~Wild,
  {\it ``Studying Generalised Dark Matter Interactions with Extended Halo--Independent Methods''},
  {\it J.~Cosmol.~Astropart.~Phys.}~{\bf 1610}, 032 (2016),
  {\tt arXiv:1607.04418 [hep-ph]}.}
%

%
\bibitem{Gelmini14}
 {G.~B.~Gelmini,
  {\it ``Halo--Independent Analysis of Direct Dark Matter Detection Data
         for Any WIMP Interaction''},
  {\it Nucl.~Part.~Phys.~Proc.}~273--275 (2016),
  {\tt arXiv:1411.0787 [hep-ph]}.}
%
\bibitem{Gelmini15b}
 {G.~B.~Gelmini, A.~Georgescu, P.~Gondolo and J.-H.~Huh,
  {\it ``Extended Maximum Likelihood Halo--Independent Analysis of
         Dark Matter Direct Detection Data''},
  {\it J.~Cosmol.~Astropart.~Phys.}~{\bf 1511}, 038 (2015),
  {\tt arXiv:1507.03902 [hep-ph]}.}
%
\bibitem{Gelmini16}
 {G.~B.~Gelmini, J.-H.~Huh, S.~J.~Witte,
  {\it ``Assessing Compatibility of Direct Detection Data:
         Halo--Independent Global Likelihood Analyses''},
  {\it J.~Cosmol.~Astropart.~Phys.}~{\bf 1610}, 029 (2016),
  {\tt arXiv:1607.02445 [hep-ph]}.}
%
\bibitem{Gelmini17}
 {G.~B.~Gelmini, J.-H.~Huh, S.~J.~Witte,
  {\it ``Unified Halo--Independent Formalism from Convex Hulls
         for Direct Dark Matter Searches''},
  {\it J.~Cosmol.~Astropart.~Phys.}~{\bf 1712}, 039 (2017),
  {\tt arXiv:1707.07019 [hep-ph]}.}
%

%
\bibitem{Cherry14}
 {J.~F.~Cherry, M.~T.~Frandsen, I.~M.~Shoemaker,
  {\it ``Halo Independent Direct Detection of Momentum--Dependent Dark Matter''},
  {\it J.~Cosmol.~Astropart.~Phys.}~{\bf 1410}, 022 (2014),
  {\tt arXiv:1405.1420 [hep-ph]}.}
%
\bibitem{Kahn14}
 {Y.~Kahn,
  {\it ``Unbinned Halo--Independent Methods for Emerging Dark Matter Signals''},
  {\tt arXiv:1411.4557 [hep-ph]} (2014).}
%

%
\bibitem{DMDDf1v}
 {M.~Drees and C.-L.~Shan,
  {\it ``Reconstructing the Velocity Distribution of Weakly Interacting Massive Particles
         from Direct Dark Matter Detection Data''},
  {\it J.~Cosmol.~Astropart.~Phys.}~{\bf 0706}, 011 (2007),
  {\tt arXiv:astro-ph/0703651}.}
%
\bibitem{DMDDf1v-Bayesian}
 {C.-L.~Shan,
  {\it ``Bayesian Reconstruction of the Velocity Distribution of
         Weakly Interacting Massive Particles
         from Direct Dark Matter Detection Data''},
  {\it J.~Cosmol.~Astropart.~Phys.}~{\bf 1408}, 009 (2014),
  {\tt arXiv:1403.5610 [astro-ph.HE]}.}
%

%
\bibitem{DMDDmchi}
 {M.~Drees and C.-L.~Shan,
  {\it ``Model--Independent Determination of the WIMP Mass
         from Direct Dark Matter Detection Data''},
  {\it J.~Cosmol.~Astropart.~Phys.}~{\bf 0806}, 012 (2008),
  {\tt arXiv:0803.4477 [hep-ph]}.}
%

%
\bibitem{DMDDfp2}
 {C.-L.~Shan,
  {\it ``Estimating the Spin--Independent WIMP--Nucleon Coupling
         from Direct Dark Matter Detection Data''},
  {\tt arXiv:1103.0481 [hep-ph]} (2011).}
%
\bibitem{DMDDranap}
 {C.-L.~Shan,
  {\it ``Determining Ratios of WIMP--Nucleon Cross Sections
         from Direct Dark Matter Detection Data''},
  {\it J.~Cosmol.~Astropart.~Phys.}~{\bf 1107}, 005 (2011),
  {\tt arXiv:1103.0482 [hep-ph]}.}
%

%
\bibitem{Altmann01}
 {CRESST Collab., M.~Altmann {\it et al.},
  {\it ``Results and Plans of the CRESST Dark Matter Search''},
  {\tt arXiv:astro-ph/0106314} (2001).}
%
\bibitem{Angloher14}
 {CRESST Collab., G.~Angloher {\it et al.},
  {\it ``Results on Low Mass WIMPs Using an Upgraded CRESST-II Detector''},
  {\it Eur.~Phys.~J.}~{\bf C74}, 3184 (2014),
  {\tt arXiv:1407.3146 [astro-ph.CO]}.}
%
\bibitem{Angloher15b}
 {CRESST Collab., G.~Angloher {\it et al.},
  {\it ``Results on Light Dark Matter Particles with a Low--Threshold CRESST-II Detector''},
  {\it Eur.~Phys.~J.}~{\bf C76}, 25 (2016),
  {\tt arXiv:1509.01515 [astro-ph.CO]}.}
%
\bibitem{Petricca17}
 {CRESST Collab., F.~Petricca {\it et al.},
  {\it ``First Results on Low--Mass Dark Matter from the CRESST-III Experiment''},
  {\tt arXiv:1711.07692 [astro-ph.CO]} (2017).}
%
\bibitem{Strauss18}
 {CRESST Collab., R.~Strauss {\it et al.},
  {\it ``A Prototype Detector for the CRESST-III Low--Mass Dark Matter Search''},
  {\it Nucl.~Instrum.~Meth.}~{\bf A845}, 414--417 (2017),
  {\tt arXiv:1802.08639 [astro-ph.IM]}.}
%

%
\bibitem{Aalseth10}
 {CoGeNT Collab., C.~E.~Aalseth {\it et al.},
  {\it ``Results from a Search for Light--Mass Dark Matter
         with a P--Type Point Contact Germanium Detector''},
  {\it Phys.~Rev.~Lett.}~{\bf 106}, 131301 (2011),
  {\tt arXiv:1002.4703 [astro-ph.CO]}.}
%
\bibitem{Aalseth12}
 {CoGeNT Collab., C.~E.~Aalseth {\it et al.},
  {\it ``CoGeNT:
         A Search for Low--Mass Dark Matter Using p-Type Point Contact Germanium Detectors''},
  {\it Phys.~Rev.}~{\bf D88}, 012002 (2013),
  {\tt arXiv:1208.5737 [physics.ins-det]}.}
%

%
\bibitem{WZhao13}
 {CDEX Collab., W.~Zhao {\it et al.},
  {\it ``First Results on Low--Mass WIMP from the CDEX-1 Experiment
         at the China Jinping Underground Laboratory''},
  {\it Phys.~Rev.}~{\bf D88}, 052004 (2013),
  {\tt arXiv:1306.4135 [hep-ex]}.}
%
\bibitem{QYue14}
 {CDEX Collab., Q.~Yue {\it et al.},
  {\it ``Limits on Light WIMPs from the CDEX-1 Experiment
         with a P--Type Point--Contact Germanium Detector
         at the China Jingping Underground Laboratory''},
  {\it Phys.~Rev.}~{\bf D90}, 091701 (2014),
  {\tt arXiv:1404.4946 [hep-ex]}.}
%
\bibitem{WZhao16}
 {W.~Zhao {\it et al.},
  {\it ``A Search of Low--Mass WIMPs with P--Type Point Contact Germanium Detector
         in the CDEX-1 Experiment''},
  {\it Phys.~Rev.}~{\bf D93}, 092003 (2016),
  {\tt arXiv:1601.04581 [hep-ex]}.}
%
\bibitem{HJiang18}
 {CDEX Collab.,H.~Jiang {\it et al.},
  {\it ``Limits on Light WIMPs from the First 102.8 kg--days Data of the CDEX-10 Experiment''},
  {\it Phys.~Rev.~Lett.}~{\bf 120}, 241301 (2018),
  {\tt arXiv:1802.09016 [hep-ex]}.}
%

%
\bibitem{Agnese13d}
 {SuperCDMS Collab., R.~Agnese {\it et al.},
  {\it ``CDMSlite: A Search for Low--Mass WIMPs
         Using Voltage--Assisted Calorimetric Ionization Detection
         in the SuperCDMS Experiment''},
  {\it Phys.~Rev.~Lett.}~{\bf 112}, 041302 (2014),
  {\tt arXiv:1309.3259 [physics.ins-det]}.}
%
\bibitem{Agnese14a}
 {SuperCDMS Collab., R.~Agnese {\it et al.},
  {\it ``Search for Low--Mass WIMPs with SuperCDMS''},
  {\it Phys.~Rev.~Lett.}~{\bf 112}, 241302 (2014),
  {\tt arXiv:1402.7137 [hep-ex]}.}
%
\bibitem{Agnese15b}
 {SuperCDMS Collab., R.~Agnese {\it et al.},
  {\it ``WIMP--Search Results from the Second CDMSlite Run''},
  {\it Phys.~Rev.~Lett.}~{\bf 116}, 071301 (2016),
  {\tt arXiv:1509.02448 [astro-ph.CO]}.}
%
\bibitem{Agnese17a}
 {SuperCDMS Collab., R.~Agnese {\it et al.},
  {\it ``Low--Mass Dark Matter Search with CDMSlite''},
  {\it Phys.~Rev.}~{\bf D97}, 022002 (2018),
  {\tt arXiv:1707.01632 [astro-ph.CO]}.}
%

%
\bibitem{Amole15a}
 {PICO Collab., C.~Amole {\it et al.},
  {\it ``Dark Matter Search Results from the PICO-2L C$_3$F$_8$ Bubble Chamber''},
  {\it Phys.~Rev.~Lett.}~{\bf 114}, 231302 (2015),
  {\tt arXiv:1503.00008 [astro-ph.CO]}.}
%
\bibitem{Amole16}
 {PICO Collab., C.~Amole {\it et al.},
  {\it ``Improved Dark Matter Search Results from PICO-2L Run-2''},
  {\it Phys.~Rev.}~{\bf D93}, 061101 (2016),
  {\tt arXiv:1601.03729 [astro-ph.CO]}.}
%

%
\bibitem{Agnes18a}
 {DarkSide Collab., P.~Agnes {\it et al.},
  {\it ``Low--Mass Dark Matter Search with the DarkSide-50 Experiment''},
  {\tt  arXiv:1802.06994 [astro-ph.HE]} (2018).}
%

%
\bibitem{DMDDf1v-calN}
 {C.-L.~Shan,
  {\it ``Reconstructing the WIMP Velocity Distribution
         from Direct Dark Matter Detection Data
         with a Non--Negligible Threshold Energy''},
  {\it Int.~J.~Mod.~Phys.}~{\bf D24}, 1550090 (2015),
  {\tt arXiv:1503.04930 [astro-ph.HE]}.}
%

%
\bibitem{Tovey00}
 {D.~R.~Tovey {\it et al.},
  {\it ``A New Model--Independent Method
         for Extracting Spin--Dependent Cross Section Limits
         from Dark Matter Searches''},
  {\it Phys.~Lett.}~{\bf B488}, 17--26 (2000),
  {\tt arXiv:hep-ph/0005041}.}
%

%
\bibitem{Giuliani05}
 {F.~Giuliani and T.~A.~Girard,
  {\it ``Model--Independent Limits
         from Spin--Dependent WIMP Dark Matter Experiments''},
  {\it Phys.~Rev.}~{\bf D71}, 123503 (2005),
  {\tt arXiv:hep-ph/0502232}.}
%
\bibitem{Girard05}
 {T.~A.~Girard and F.~Giuliani,
  {\it ``On the Direct Search for Spin--Dependent WIMP Interactions''},
  {\it Phys.~Rev.}~{\bf D75}, 043512 (2007),
  {\tt arXiv:hep-ex/0511044}.}
%

%
\bibitem{Freese88}
 {K.~Freese, J.~Frieman and A.~Gould,
  {\it ``Signal Modulation in Cold--Dark--Matter Detection''},
  {\it Phys.~Rev.}~{\bf D37}, 3388--3405 (1988).}
%

%
\bibitem{RPP16AP}
 {C.~Patrignani {\it et al.} (Particle Data Group),
  {\it ``The Review of Particle Physics 2016''},
  {\it Chin.~Phys.}~{\bf C40}, 100001 (2016),
  {\it 2.~Astrophysical Constants and Parameters}.}
%

%
\end{thebibliography}
\end{document}